# NC-AFM AND XPS INVESTIGATION OF SINGLE-CRYSTAL SURFACES SUPPORTING COBALT (III) OXIDE NANOSTRUCTURES GROWN BY A PHOTOCHEMICAL METHOD

By

DAVID JOSEPH MANDIA, B.Sc. (Hons)

A THESIS

SUBMITTED TO THE FACULTY OF GRADUATE STUDIES

IN PARTIAL FULFILLMENT OF THE REQUIREMENTS FOR THE

DEGREE OF MASTER OF SCIENCE (M.Sc.)

Department of Chemistry

University of Ottawa

April, 2012

The work has been carried out under the supervision of Dr. Javier B. Giorgi

# ABSTRACT

The work of this thesis comprises extensive Noncontact Atomic Force Microscopy (NC-AFM) characterization of clean metal-oxide (YSZ(100)/(111) and MgO(100)) and graphitic (HOPG) supports as templates for the novel, photochemically induced nucleation of cobalt oxide nanostructures, particularly Cobalt (III) Oxide. The nanostructure-support surfaces were also studied by X-ray Photoelectron Spectroscopy (XPS) to verify the nature of the supported cobalt oxide and to corroborate the surface topographic and phase NC-AFM data. Heteroepitaxial growth of $Co_2O_3$ nanostructures proves to exhibit a variety of different growth modes based on the structure of the support surface. On this basis, single-crystal support surfaces ranging from nonpolar to polar and atomically flat to highly defective and reactive were chosen, again, yielding numerous substrate-nanostructure interactions that could be probed by high-performance surface science techniques.

# TABLE OF CONTENTS

# LIST OF TABLES

# LIST OF FIGURES

doped with $Y_2O_3$). Red and white balls are for oxygen and any metal, respectively. The Oxide-Metal-Oxide (OMO) layers make up the bulk material and often reconstruct to stabilize the surface energy.

**Figure 3.1.4** High resolution NC-AFM images of the clean, pre-growth YSZ (111) single-crystal surface **A)** 2.0 x 2.0 µm$^2$ micrograph of as-received single crystal (nominal z-range=5.221 Å). Micrographs **B)-D)** depict sample after ambient thermal cycle in air at 1000 $^oC$

for 1 hr. **B)** 1.6 x 1.6 µm$^2$ micrograph of post-annealed sample (nominal $z_{range}$=4.485 Å). **C)** 1.0 x 1.0 µm$^2$ micrograph of post-annealed sample (nominal $z_{range}$=4.565 Å). **D)** 500 x 500 nm$^2$ micrograph of post-annealed sample (nominal $z_{range}$= 3.727 Å)

**Figure 3.1.5** NC-AFM height profile for YSZ(111) surface (also in Figure 3.1.3d), revealing a high density of terraces with minimal edge defects. This step-edge morphology is indicative of a surface flattening or "smoothing" process that occurs on the single-crystal surface after a thermal annealing treatment

**Figure 3.2.1** 2.0 x 2.0 µm$^2$ NC-AFM micrograph of a Type I surface (post-anneal, 1000$^o$C, 1 hr) with associated height profile on the right, corresponding to the grey trace line (note that the profile begins at the top of the trace). Parallel steps and terrace features dominate this and frequently contain defects in the form of small pits or kinks. RMS roughness was 0.7380Å

**Figure 3.2.2** Ambient NC-AFM images of clean YSZ(100) (post-anneal, 1000 $^o$C, 1 hr) surface depicting a roughly Type II surface orientation with defect structures (pits or clusters) forming preferentially at the step edges. **A)** 2.0 x 2.0 µm$^2$ scan (z-range=4.697Å, RMS=1.681 Å, scan speed=0.2 line/s) **B)** 819 x 819 nm$^2$ scan (z-range=3.754 Å, RMS=1.265 Å, scan speed=0.5 line/s) revealing defect pits of ~1**a** (recall a=5.124Å) length with nearly flat bottoms in the line profile for I in D). 500 x 500 nm$^2$ scan (z-range=5.372Å, RMS=1.003 Å, scan speed=~1.0 line/s) showing terrace step length of ~1/2**a**, as shown in height profile for II in **D)**, and spherical defect clusters formed at the step features

**Figure 3.2.3** High resolution NC-AFM images of the clean, pre-growth YSZ (100) sample. **A)** 2.0 x 2.0 µm$^2$ micrograph of as-received single crystal (nominal z-range=9.762 Å). Micrographs **B)-D)** depict sample after ambient thermal cycle in air at 1000 $^o$C for 1 hr. **B)** 2 x 2 µm$^2$ micrograph of single crystal, post-anneal (nominal $z_{range}$=7.673 Å). **C)** 1.0 x 1.0 µm$^2$ micrograph of single crystal, post-anneal (nominal $z_{range}$=5.993 Å). **D)** 500 x 500 nm$^2$ micrograph of single crystal, post-anneal (nominal $z_{range}$= 4.759 Å). Images required 2$^{nd}$-order polynomial flattening during acquisition

**Figure 3.2.4** 2.0 x 2.0 µm$^2$ NC-AFM phase image of clean YSZ(100) surface (corresponding topography micrograph in Figure 3.2.3b) containing clusters of terrace-specific defects. Saturation of the signal and local contrast in the phase image suggest that the defects are non-stoichiometric with respect to the surface.

**Figure 3.3.1** Atomic "checker-board" surface structure of MgO(100) with $O^{2-}$ ions represented by large grey circles and $Mg^{2+}$ ions by the small black circles. The ideal <100> orientation allows the MgO(100) surface to be ionic neutral (nonpolar) without requirements for charge stabilization.

envelope. Spectrum was charge-corrected with respect to the $Zr^{4+}$ $3d_{5/2}$ peak envelope which occurs at 182.6 eV. Spectrum obtained the Specs/RHK system.*Note: The Co 2p region had the highest resolution in the survey scan so this is not the fit for the component region

**Figure 4.4.1** High resolution NC-AFM image set of the $Co_2O_3$ nanostructures grown onto the clean MgO(100) single-crystal surface. A) 2.0 x 2.0 $\mu m^2$ topographic image of post-growth $Co_2O_3$-MgO(100) system B) 2.0 x 2.0 $\mu m^2$ phase image corresponding to A) showing high local contrast of the nanoclusters with the underlying surface substrate. C) 1.0 x 1.0 $\mu m^2$ topographic image of the selected region in A) (scan speed=0.2-0.4 line/s, tip: SSS-NCL, RF=167 kHz).D) Height distribution histogram for the $Co_2O_3$ nanostructures.

**Figure 4.4.2** XP-spectra recorded following photochemical growth of $Co_2O_3$ on the clean MgO(100) single-crystal surface. A) Co 2p envelope with peak behavior characteristic of the $Co_2O_3$ species. B) Fit Mg 2p band with broadening due to $MgCO_3$ and $Mg(OH)_2$ C) O 1s envelope showing broadening and splitting due to $MgCO_3$ and $Mg(OH)_2$ species. Spectra were collected on the Kratos Axis Ultra DLD system (CCRI), fit with Shirley backgrounds, and calibrated with respect to the Mg 2s peak (nominal position of 89 eV)

**Figure 4.5.1** NC-AFM imaging of the $Co_2O_3$-HOPG system. A) 3.0 x 3.0 $\mu m^2$ topographic image of spherical $Co_2O_3$ NPs grown on the pristine HOPG surface. Clusters actually form both along cleavage steps and atop the pristine islands of freshly cleaved HOPG B) Particle height histogram for the $Co_2O_3$ nanostructures. C) 4.0 x 4.0 $\mu m^2$ phase image of the nanostructures and clusters forming preferentially at cleavage steps. (scan speed=0.3-0.5, DLC tip, RF=~150 kHz).

**Figure 4.5.3** XP-spectrum of the Co 2p region of the Co $2p_{3/2}$ band fit for $Co_2O_3$NP species (major) at 779.5 eV. CoO was also a likely minor species causing the broadband. There is likely a higher contribution from the surface contamination of other cobalt oxides on this support, evident from the smaller fit for $Co_2O_3$. Spectrum was acquired on the Kratos Axis DLD Ultra system (CCRI).

**Figure 4.5.4** Fit component XP-spectrum of O 1s envelope corresponding to the $Co_2O_3$ and other surface species attributed to hydroxyl contaminants ($Co(OH)_2$/CoO(OH)) and the presence of higher oxides of cobalt.[75] Spectrum acquired on Kratos Axis DLD Ultra (CCRI).

# LIST OF ABBREVIATIONS AND SYMBOLS

| | |
|---|---|
| $A$ | Hamaker constant for inorganic materials |
| $A_0$ | Resonant amplitude |
| $\alpha,\beta$ | Damping parameters |
| $A_f$ | Damped amplitude response |
| AAC | Acoustic Alternating Current |
| AFM | Atomic Force Microscopy |
| AM-AFM | Amplitude-Modulated Atomic Force Microscopy |
| $C$ | Collapsed potential energy term or capacitance |
| CPD | Contact Potential Difference |
| $d_{hkl}$ | d-spacing parameter |
| $d_{ts}$ | Distance between tip and sample surface |
| $d_z$ | Distance of z-range motion of AFM within meniscus |
| $E_b$ | Binding energy |
| $E_f$ | Final state energy of electron |
| EFM | Electrostatic Force Microscopy |
| $E_i$ | Initial state energy of electron |
| $E_k$ | Kinetic energy of photoelectron |
| $f$ | Driving frequency |
| $f_0$ | Resonant (natural) frequency |
| $F_{cap}$ | Capillary force |
| fcc | Face-centered Cubic |
| $F_{chem}$ | Chemical force |
| $F_{cpd}$ | Contact potential difference force |
| FdM | Frank-van der Merve growth |
| $F_{electr}$ | Electrostatic force |
| FFT | Fast Fourier Transform |
| $f_m$ | Modified frequency |
| FM-AFM | Frequency-Modulated Atomic Force Microscopy |
| $F_{ts}$ | Tip-surface forces |
| $F_{vdW}$ | van der Waals force term |
| FWHM | Full Width At Half Maximum |
| GL | Gaussian-Lorentzian peak |
| $\gamma$ | Surface tension of meniscus |
| $\gamma_{sub}$ | Substrate energy |
| $\gamma_{metal}$ | Metal (oxide) nanocluster energy |
| $\gamma_{int}$ | Interface (substrate-metal) energy |
| {h,k,l} | Miller index plane |
| $h\upsilon$ | Photon (X-ray) |
| $I_t$ | STM tunneling current |
| $k$ | Boltzmann's constant |
| $m$ | Effective mass |
| MAC | Magnetic Alternating Current |
| ML | Monolayer |
| NC | Nanocluster (or 3D-nanocluster/cluster) |
| NC-AFM | Noncontact Atomic Force Microscopy |
| NP | Nanoparticle |
| OMO | Oxide-Metal-Oxide layer |
| $r$ | atom-to-atom separation distance |
| % At | % atomic composition |
| $P_L$ | Laplace pressure |
| PPP | Point Probe Plus |
| $Q$ | Quality factor |
| $R$ | Radius of (spherical) AFM tip |
| $R_q$ | RMS Roughness |
| RF | Resonant Frequency |
| $R_{avg}$ | Average roughness parameter |
| $R_{ku}$ | Kurtosis Parameter |
| $R_p$ | Peak (max) roughness parameter |
| $R_v$ | Valley (min) roughness parameter |
| $R_{MPV}$ | Max-peak-to-valley roughness (highest feature) height |
| $r_1$, $r_2$ | Radii of tip-surface meniscus |
| $r_K$ | Kelvin radius |
| RSF | Relative Sensitivity Factor |
| $\rho_t$ | Atomic density of AFM tip |
| $\rho_t$ | Atomic Density of the surface |
| $\sigma$ | Equilibrium bond distance |
| SK | Stranski-Krastanov growth |
| SPM | Scanning Probe Microscopy |
| STM | Scanning Tunneling Microscopy |
| $t$ | Amplitude response time |
| UHV | Ultra-High Vacuum |
| $U(r)$ | Interaction Potential |
| $V_s$ | Integration volume of the surface |
| $V_t$ | Integration volume of the AFM tip |
| $V_{cpd}$ | Contact potential difference potential energy |
| $V_{morse}$ | Morse potential |
| $V_{Lennard\text{-}Jones}$ | Lennard-Jones potential |
| $V_{ts}$ | Tip-surface potential energy |
| vdW | van der Waals |
| VW | Volmer-Weber growth |
| SSS | Super-Sharp Silicon |
| $\varphi$ | Work function of solid |
| $\phi$ | Phase differential (contrast function) |
| $X_i$ | Atomic composition |
| XPS | X-ray photoelectron spectroscopy |
| YSZ | Yttria-stabilized Zirconia |
| $\bar{z}$ | Arithmetic mean height |
| $z$ | z-range of AFM tip or atom-atom separation distance |

# ACKNOWLEDGEMENTS

First and foremost, I want to acknowledge my supervisor, Dr. Javier B. Giorgi for his unwavering patience and undeniable surface science wizardry. If it weren`t for the former, I wouldn`t have been able to strive for the latter. Finally, my studies under your guidance are the reason why I am still excited about chemistry and, ultimately, the reason why I am pursuing doctoral studies now.

The other guys (and girl!): Thanks to John “Bon Jovi” Selwyn for the everyday musings about chemistry-related TV programs and how you need to patent your trademarked “cal-sintering” process. Also, thanks for being a partner in all the agony that was calibrating the big machine, daily. Will, I only just met you while writing, but thanks for the moral support, coffee, jokes and general affability. Julie, thanks for all the chocolate and the advice on how to tackle one`s master`s degree. Arif, thanks for insightful discussions that mostly pertained to chemistry.

Other honourable mentions: Sander and Yun, formidable members of the CCRI, thanks for help with XPS and AFM, respectively. Richard Green, thanks for spending lots of your time helping us calibrate the UHV-STM/XPS/LEED/Heating and with sample mounting procedures. You really can`t drink coffee before mounting samples; it`s just not an option!

Finally, I need to thank my mom, dad, and sister, since if it were not for their patience and support, I would not have made it even remotely this far. Thanks to Meaghan for being my favourite person in the world and so patient while I was writing. Want to play Sega some time?

# CHAPTER 1 INTRODUCTION

## 1.1 SCOPE OF THE THESIS

### 1.1.1 VICINAL SURFACES AND MOTIVATION

Metal oxides and vicinal single-crystal surfaces thereof offer a broad spectrum of applications including technological ones such as in thin film coatings for anti-corrosion[7-9] and catalyst supports in heterogeneous catalysis studies.[7,8]Especially in the latter case, a fundamental understanding of the support surface is critical in exploring its role in any model catalyst system. Metal oxides have many interesting properties that would be of interest to the surface scientist and run the gamut of insulating, high band gap materials such as Yttria Stabilized Zirconia (YSZ) to low band gap (< 4 eV) semiconducting or conducting materials such as Highly Ordered Pyrolytic Graphite. At the atomic scale, many interesting features arise in the surfaces of metal oxides that are not observed in simpler metallic films largely due to the partly covalent nature of the oxide. Because of the varying bulk stoichiometry of metal oxides, cleavage planes can often form in the material to generate a wide range of surfaces. It is expected that the surface tends towards an equilibrium structure, which is observed through the formation of the lowest energy, most stable planes. This work deals with a range of different surfaces whose stability can be described by Tasker's rules for ionic or partially ionic crystals:[9,10]

**Type 1**: These surfaces involve a sequence of neutral ionic planes with equal numbers of $m^+$ and $X^-$ ions. Generally these rocksalt-like surfaces are completely nonpolar and in the context of this work, MgO(100) adopts this surface type.

**Type 2:** These surfaces are characterized as a sequence of ionic planes, often in the form of alternating oxide-metal-oxide ($O^{2-}$—$m^{+}$-$O^{2-}$ or OMO) layers which comprise no net dipole within the unit cell and no overall charge. Typically the (111) plane, like in the YSZ(111) surface studied in this work, is the most stable and natural cleavage plane and renders the surface relatively nonpolar in most cases.

**Type 3:** These surfaces are made up of charged ionic planes and often render the surface unstable and highly charged with a nonzero net dipole. The polar nature of these surfaces means they must undergo extensive reconstruction to stabilize their surface energy. The YSZ(100) single-crystal surface explored in this work is both polar and highly susceptible to reconstruction. The unstable nature of this particular surface means it often requires surface cleaning via thermal treatments before the (100) surface is achieved. Holding all conditions constant, there are various surface reconstructions that will occur based on the oxygen concentration in the crystal.[11,12]

Since the single-crystal surfaces, with the exception of HOPG, studied herein are insulating materials, noncontact atomic force microscopy (NC-AFM) is the most relevant surface science technique that can probe the structure of these materials in their native and complex state. The insulating nature of the materials means spectroscopic techniques involving high-energy electrons, ions, or photons are quite difficult to employ because of the issue of various charging phenomena.

Despite the multitude of applications of nanoparticles in heterogeneous catalysis, the importance of creating well-defined, nanostructured arrays for growth of thin films or nanoparticles in model catalyst systems is often overlooked. Exploiting the surface structure of the support prior to growth of nanoparticle arrays can aid in understanding the growth behavior of the nanoparticles and the mechanism of growth itself, which is, again, highly dependent on the surface features whether they are kinks, steps, vacancies, or other such defects.[13-15] The principal motivation for this work is to employ *ex-situ* NC-AFM and X-ray Photoelectron Spectroscopy (XPS) to investigate the four different single-crystal surfaces mentioned (YSZ(111), YSZ(100), MgO(100), HOPG) after extensive surface cleaning both before and after Cobalt (III) Oxide ($Co_2O_3$) nanostructures are heteoepitaxially grown by a novel, photochemical growth process. This process is reported extensively by collaborators Tse-Luen Wee and Juan C. Scaiano for the nanocrystalline diamond (NCD) support,[16] but thus far the literature has no reports of extensive ambient NC-AFM imaging of $Co_2O_3$ nanostructures on any of the single-crystal surfaces studied in this thesis. The recent interest of cobalt as a water-splitting catalyst, among many other photocatalytic applications, is a future direction for exploring these $Co_2O_3$-support systems in the first place; however, this work serves as a springboard or starting point in probing the nucleation behavior of the $Co_2O_3$ nanostructures as they decorate surface features/defects of various well-defined and well-prepared single-crystal supports. As mentioned, NC-AFM and XPS are the primary surface science techniques that are both employed and explored as highly-sensitive tools for a rigorous characterization of both insulating metal oxide and conductive systems. The work presented is therefore an instrumental pursuit as well, highlighting the versatility and efficacy of these tools.

## 1.2 ASPECTS OF ATOMIC FORCE MICROSCOPY

### 1.2.1 HISTORICAL BACKGROUND

Atomic force microscopy (AFM) is part of the rapidly growing library of Scanning Probe Microscopy (SPM) tools critical in surface analysis from the micro right down to the molecular and atomic scale. The advent of SPM saw the discovery of Scanning Tunneling Microscopy (STM) in 1981 by Binnig, Rohrer, Weibel, and Gerber, eventually earning its inventors the Nobel Prize in 1986.[17,18] In fact, as per Figure 1.2.1, IBM R & D is largely responsible for many of the most important discoveries in nanoscience and the ever-growing surface science toolkit.

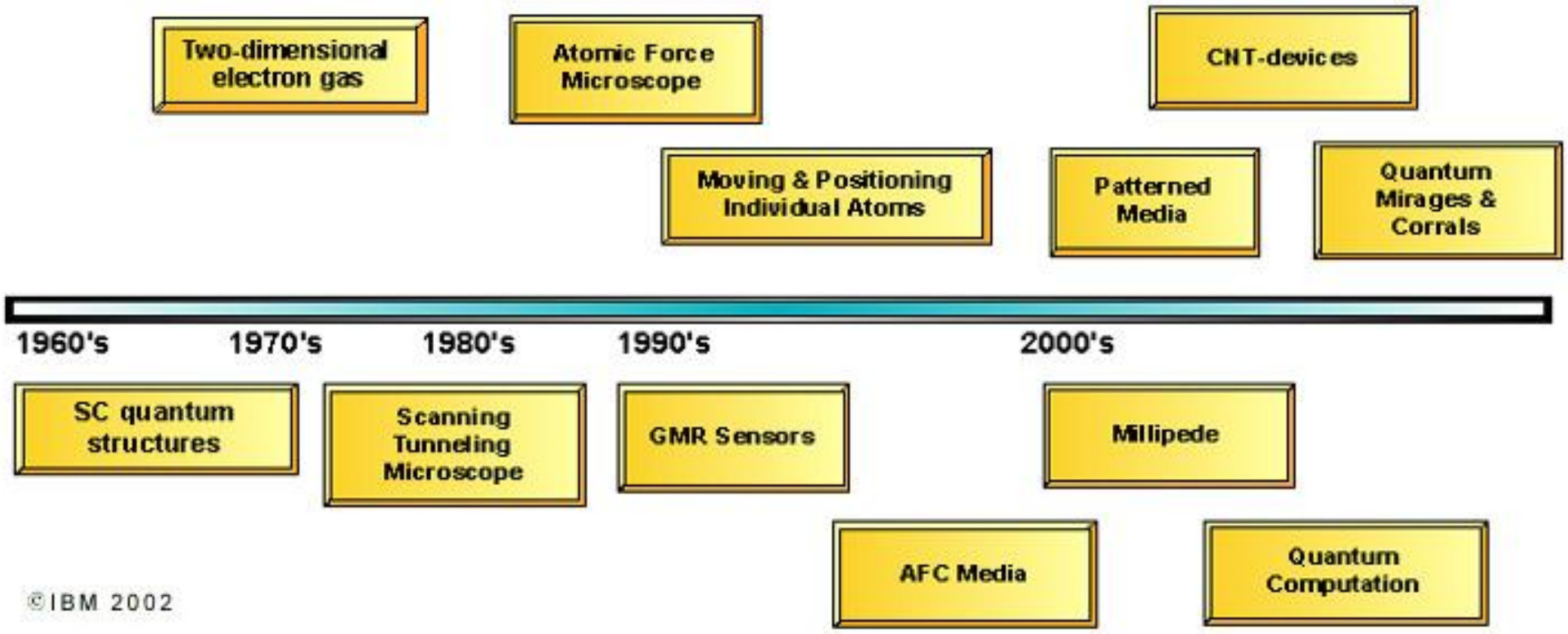

**Figure 1.2.1** Timeline of SPM and other key events in the history of IBM R & D.[1]

As evident from the timeline for IBM R & D in Figure 1.2.1, they are responsible for some of the most important contributions to surface science. From two-dimensional electron gases, which are transistor-like structures with electron motion constrained to two dimensions, to CNT devices and quantum computing, this critical research has led to many high-performance inventions, particularly STM and AFM.

Naturally, though, every instrument has limitations and because STM relies on the tunneling of current into conductive samples, this limits its applications to metal and semiconductor materials. Also, a tip-surface distance component of

$$I_t = e^{-kd_{ts}} \quad (1.1)$$

where $I_t$ is the tunneling current and $d_{ts}$ is a the distance between tip and sample ($k$ being Boltzmann's constant) shows an exponential sensitivity with respect to tip-surface distance and short-range forces in STM such that a change of just 0.1 Å in $d_{ts}$ can induce a full order of magnitude change in $I_t$.[19] In fact, this achievement was concurrent with the publishing of a veritable benchmark paper in the field surface science: the Si(111)-(7 x 7) reconstruction.[20,21] Despite a marvelous library of atomic-scale resolution images, the invention of AFM could only be realized when the limitations that beset STM were elucidated. Thus, Binnig invented the atomic force microscope in 1986, to address the challenge, among many others that belied STM, of imaging insulating substrates was over.[22] A principal motivation for its discovery was to take advantage of the forces that seemed to act collaterally with the tunneling current in the STM process.[23] Ultimately, the goal with AFM was to achieve atomic resolution of virtually any surface, but unlike the fleeting overnight success of STM, this took an additional 5 years from the point of inception when Binnig, Giessibl, and Ohnesorge revealed the first atomic resolution images of inert, well-defined surfaces.[24,25] Another 3-4 years elapsed before true atomic resolution of dynamic, reconstructive surfaces like that of the Si(111)-(7 x 7) surface could be achieved.[26,27] Indeed, the initial work of Kitamura and Iwatsuki[28] as well as Giessibl[26] paved the way for achieving STM-like resolution for NC-AFM, making it one of the most called upon and versatile techniques in surface science. A proper rigorous survey of current AFM literature would confirm many of the countless achievements in AFM, which include: true atomic resolution; 3D measurement of atomic forces; imaging of insulating materials; manipulation of atomic forces; and even mechanical manipulation/assembly of individual atoms.

### 1.2.2 AFM: WORKING PRINCIPLE AND TIP-SURFACE FORCES

Referring to Figure 1.2.2, AFM involves approaching the sample surface with a sharp tip (probe), usually composed of Si or SiN. The tip is mounted to the cantilever and often the whole unit is referred to as “cantilever”, “probe”, “cantilever-tip” and other variations. The reflective coating on the cantilever causes the incident laser beam to reflect off of it and get subsequently detected by a position-sensitive photodiode "quadrant" detector. As the tip scans the surface in a raster pattern, the cantilever will bend, which causes instantaneous changes in the laser deflection signal (usually measured as a change in setpoint or voltage amplitude). In scenarios when the tip is oscillating at a known free-standing resonant frequency over the surface (intermittent contact or "tapping" mode), contact with the surface induces a detuning signal and the changes in the tip-surface forces are interpreted as topographic data.[29] The optical beam deflection method is essentially the driving force behind acquisition of topographic images in AFM and is the basis for plenty of the various imaging modes including the "noncontact" AFM (NC-AFM) regime employed in this work.

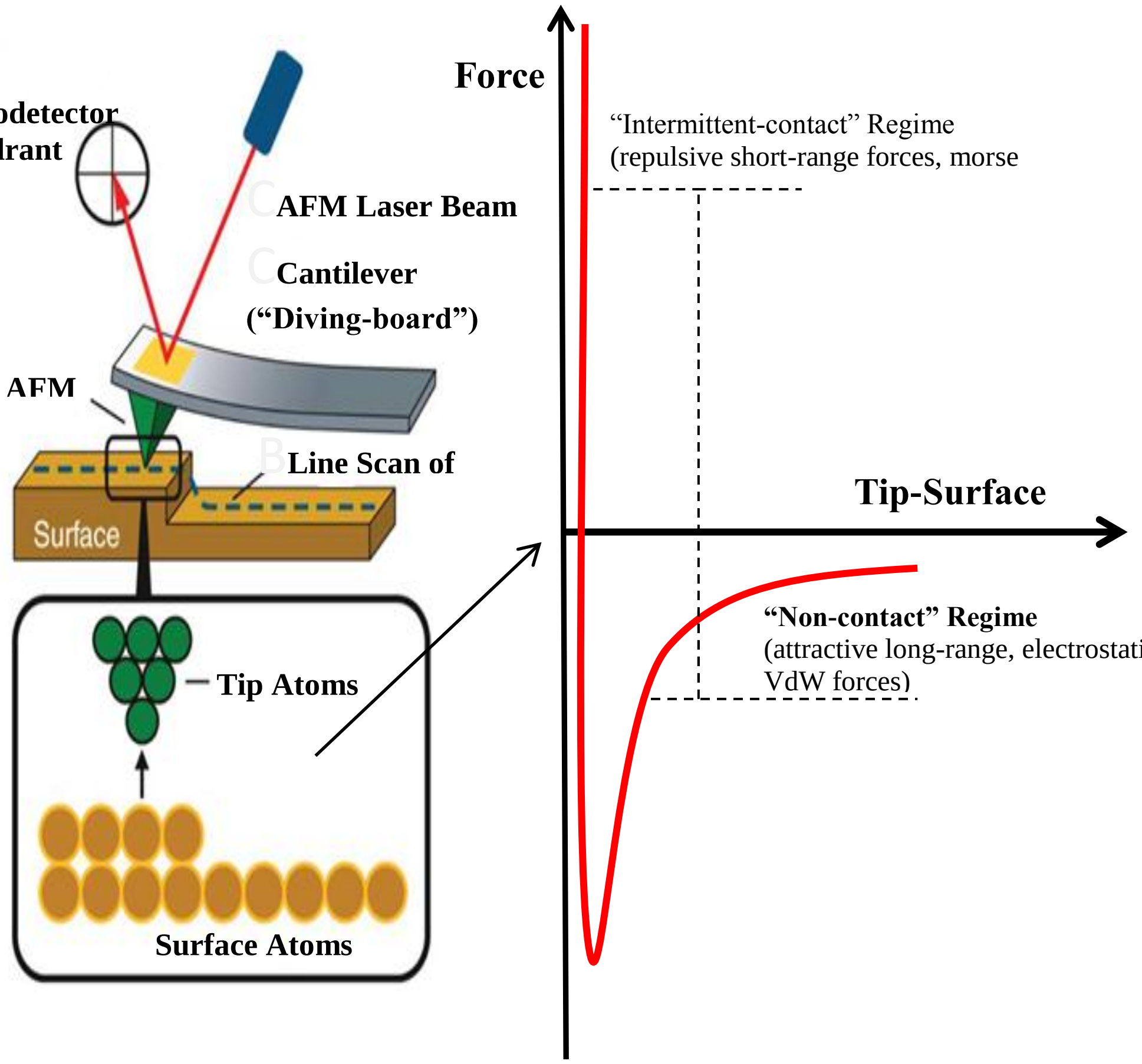

**Figure 1.2.2** General principle of AFM with associated generic force profile, indicating the zones of tip-surface interaction as the tip approaches the surface.

Tip-surface forces, as we will define, often complicate the imaging process and some notable ones for ambient imaging are the strong adhesion or capillary forces induced by water when it wicks around the tip. Ideally, the AFM signal is the resulting sum of all force contributions due to capillary, adhesion and Van der Waals (VdW) forces. Of course, the types of the forces and their contributions to the tip-surface interaction depends on the imaging environment, so the adhesion layers observed in ambient AFM imaging are a non-issue during UHV-STM or AFM. The AFM image signal, as mentioned, should be a cumulative function of tip-surface forces, $F_{ts}$, and more explicitly, a change in the tip-surface potential energy, $V_{ts}$, with respect to a change in the z-component (axial response of the cantilever) value[23]:

$$F_{ts} = -\frac{\partial V_{ts}}{\partial z} \qquad (1.2)$$

Present both in ambient and *in vacuo* conditions, VdW forces are always a consideration in understanding tip-surface interactions. The interaction potential, *U*, between any two neutral atoms is a function of their mutual propensity to polarize.[30] Separated by a radius, *r*, the potential, *U(r)*, is given by:

$$U(r) = -\frac{C}{r^6} \qquad (1.3)$$

where *C* is a collapsed term representing the specific interactions between the two atoms depending on their intrinsic properties. Note the interesting proportionality to $r^{-6}$, which intuitively implies a strong interaction potential for $r \rightarrow 0$ and, in the context of AFM, when the tip atoms are in close proximity to the surface atoms. Since the VdW forces work over a short and long range, there are conceivably innumerable dipole-dipole interactions between atoms, making even the approximation of total VdW forces a non-trivial task. One such approximation is perhaps also the simplest—a term used very loosely— approach in the Hamaker approximation of the London-VdW forces between spherical particles.[31] The model states a few assumptions about the tip (spherical) and surface including that of relative inertness of material properties and a continuous interaction. Assumptions for the VdW force parameter *C* between atoms first must be delineated by dividing the force into 3 distinct interactions:[32]

i) Dipole-dipole force: Molecules with permanent dipoles will always interact to some degree .

ii) Dipole-induced dipole forces: A dipole moment between inert atoms or molecules may be induced by neighbouring dipole field.

iii) (London) Dispersion forces: Non-polar atoms will have instantaneous attractions towards each other due to charge fluctuations that exist in any material. These less prominent interactions are, in essence, instantaneous dipole-induced dipole forces.

Thus, the VdW force, $F_{vdW}$, between any two bodies as defined by the *Hamaker Summation Method* is as follows:

$$F_{vdW} = \rho_t \rho_s \int_{V_s}^{0} \int_{V_t}^{0} -\nabla\left(\frac{-C}{r^6}\right) dV_t dV_s \qquad (1.4)$$

where $\rho_t$ and $\rho_s$ are the atomic densities and $V_t$ and $V_s$ are the integration volumes of the tip and surface, respectively.

The Hamaker constant unique to all materials[33] is rearranged for the tip-surface atomic density:

$$A = \pi^2 C \rho_t \rho_s \quad \Rightarrow \quad \rho_t \rho_s = \frac{A}{\pi^2 C} \qquad (1.5)$$

Substituting the Hamaker constant (1.5), A, into equation 1.4 yields:

$$F_{vdW} = A\int_{V_s}^{0}\int_{V_t}^{0} -\nabla\left(\frac{-C}{r^6}\right)\frac{1}{\pi^2}\,dV_t dV_s \qquad (1.6)$$

This comprehensive term for vdW force actually modifies our original vdW potential term of equation 1.3 in that if the spherical tip of radius, *R*, is close to an atomically flat surface defined by distance, *z*, which is the distance between the apex tip atom (see Figure 1.2.2) and nearest surface atom, the vdW potential is then[32]:

$$V_{ts} = -\frac{AR}{6z} \qquad (1.7)$$

Note that equation 1.7 is the simplest possible analytical expression for the vdW potential and one should take heed of the fact that the complexity of tip-surface geometry rarely warrants an accurate empirical calculation of the Hamaker integral; however, qualitative conclusions about AFM resolution may certainly be drawn from equation 1.7, especially if *z* is constrained as close to 0 as possible.

Another long-range force that needs to be considered for conductive tip-surface systems is electrostatic force. The tip and surface, unless made of identical materials, have a non-zero distribution of charges on their respective surfaces and the mechanism behind the electrostatic force is electron flow between these materials when they are proximal. A potential difference brought about by the electric contact of, for example, the AFM tip and sample surface is intimately related to the work function of each material. A parabolic or spherical tip apex of radius, *R*, separated from the sample surface by distance, *z*, has the general electrostatic force, $F_{cpd}$, of:

$$F_{cpd} = -\frac{\pi\varepsilon_0 R V_{cpd}}{z} \qquad (1.8)$$

where the subscript "cpd" stands for contact potential difference and $V_{cpd}$ is the electrostatic potential. Once established, the electric contact essentially creates a capacitor with a total electrostatic force that can be written in terms of the electrostatic potential:[34]

$$F_{electr.} = \frac{1}{2}\frac{\partial C}{\partial z}V_{cpd}^2 \qquad (1.9)$$

where $C$ is the capacitance between the sample surface and tip. Upon inspection of equation 1.9, it is evident that the electrostatic contact potential can be determined experimentally. Empirical values for the $F_{ts}$ can be extracted from the use of a voltage sweep and the minimum value would correspond to the $V_{cpd}$. Electrostatic force corrections are modulated by the change in tip-surface distance, $z$, and the $\frac{\partial C}{\partial z}$ can be inferred from the shape of the distance-voltage trace. This is actually the driving force behind the Electrostatic Force Microscope (EFM) and, in principle, governs the contrast and resolution limits of amplitude-modulated NC-AFM, which is the central method of characterization this work.

In ambient conditions, NC-AFM image signals are frequently influenced by water vapour that is present and this can happen in a few ways including coverage of the sample surface or by forming a meniscus between the tip and the sample. When the meniscus forms, it is dominated largely by the Laplace pressure[35]:

$$P_L = \gamma\left(\frac{1}{r_1} + \frac{1}{r_2}\right) = \frac{\gamma}{r_K} \qquad (1.10)$$

where $\gamma$ is the surface tension of the meniscus and $r_1$, $r_2$ are the radii of the meniscus. These radii are collapsed into the Kelvin radius term, $r_k$, that describes the radius of a droplet or bubble formed (note that its form is analogous to that of reduced mass). Since any pressure term is a function of some force acting on some unit area, the capillary force,[36] $F_{cap}$, can be derived by finding the product of the Laplace pressure and contact area of the meniscus that is formed:

$$F_{cap} = \frac{\gamma A}{r_k} = \frac{\gamma 2\pi R d_z}{r_k} \tag{1.11}$$

where equation 1.11 indicates the $F_{cap}$ for AFM tip of radius $R$ of penetration depth into the meniscus defined by $d_z$ where subscript $z$ is the z-range motion of the tip. Likewise for the tip-surface forces heretofore described, this particular force contributes to the to the topographic AFM signal and its accurate calculation requires extensive characterization of the actual tip geometry with the surface and the precise shape of the meniscus. To remedy such calculations, meniscus reconstruction models can be developed using a variety of interesting methods but this requirement is beyond the scope of this particular work.

Final considerations, particularly concerning the short-range force region of any tip-surface force profile (Figure 1.2.2) include contributions from chemical forces, which are perhaps the most complex forces to quantify since they deal with the chemical nature of the tip-surface system. Briefly, the Morse and Lennard-Jones potential models are well-known empirical models[23] which provide, at the very least, qualitative insights in to the character of chemical bonds. The former is typically described in quantum chemistry literature as the analytic solution to the exact energy of the $H_2^+$ ion's covalent bond[32] while the latter model is used in a purely qualitative context and has the tip-surface distance dependency $\propto r^{-6}$ of the $F_{vdW}$ described by equations 1.3-1.4 and 1.6-1.7. The empirical models of the Morse[32] and Lennard-Jones[32,41] potentials appear as follows:

$$V_{Morse} = -E_{bond}\left[2e^{-\kappa(z-\sigma)} - e^{-2\kappa(z-\sigma)}\right] \qquad (1.12)$$

$$V_{Lennard-Jones} = -E_{bond}\left(2\frac{z^6}{\sigma^6} - \frac{z^{12}}{\sigma^6}\right) \qquad (1.13)$$

Both potentials model a chemical bond with: bonding energy, $E_{bond}$; equilibrium distance, $\sigma$, which is the most stable average bond length of molecules in their ground-state (this could also be for any atom-atom separation distance, $r$); decay length, $\kappa$, which describes the maximum length that bond cohesion is sustained; and the atom-atom or molecule-molecular separation, $z$, which is, for all intents and purposes, the AFM tip-surface distance previously shown. It is worth noting that the Morse potential model is also a qualitative approach to underscoring the chemical forces between the tip and surface atoms, $F_{chem}$, but lacks a term for both bond angle and strength.[32,42]

To summarize, AFM imaging comprises several different force contributions that, additively, generate the topographic AFM signal of the total force:

$$F_{tot} = F_{vdW} + F_{electr} + F_{cap} + F_{chem} + \ldots(F_i) \qquad (1.14)$$

where $F_i$ is any other force "i" (i.e. magnetic) that may contribute to the AFM signal.

and, though it's clearly not a linear combination of these individual components, a topographic AFM image of any resolution will always serve to be a convolution or *ad hoc* fingerprint encoding all of the tip-sample interactions.

### 1.2.3 PRINCIPLES OF NONCONTACT ATOMIC FORCE MICROSCOPY IN AMPLITUDE MODULATION MODE

Falling under the umbrella of AFM operation modes is NC-AFM. Other modes, where the tip is engaged with the sample within the repulsive region (Figure 1.2.2), include the static and contact modes. These more aggressive imaging modes often lend themselves to tip-surface modification, deformation of the tip, or damage to the sample. Due to the reactive nature of the samples in this work, coupled with the insurmountable problem of tip-surface chemical bonding observed in contact AFM,[43,44] NC-AFM is the principal mode of characterization, hence, it is hereafter the only mode described. In this particular mode, the tip-surface interactions occur in the attractive region of the force profile (Figure 1.2.2) and deformation or chemical modification of the sample surface is avoided. It should be noted that an intermediate AFM mode exists between contact and noncontact referred to as "tapping" or intermittent contact mode, also is also often used for ambient AFM imaging for a myriad of purposes and sample types.[45–50] In principle, this mode is identical to the amplitude modulation AFM (AM-AFM) for ambient NC-AFM  wherein a stiff, high-Q cantilever is oscillated at its resonant frequency over the surface, returning a feedback signal containing topographic, phase, and amplitude data[22]; however, in this mode the tip is operating at exactly the resonant frequency or in what is coined the jump-to-contact region of the force profile, between the attraction and repulsion regimes of Figure 1.2.2. To aid in understanding, the following figure depicts the oscillation amplitude of an AFM tip as a function of the excitation or drive frequency. The harmonic, Gaussian-like shape of the resonance curve is typical of the cantilevers (usually coated or uncoated Si/SiN) used for NC-AFM such as that performed in this work. The type of dynamic AFM performed in this investigation is, indeed, *true* NC-AFM (attractive region of Figure 1.2.3) in amplitude modulation mode or, with respect to our instrument, in acoustic alternating current (AAC) mode.

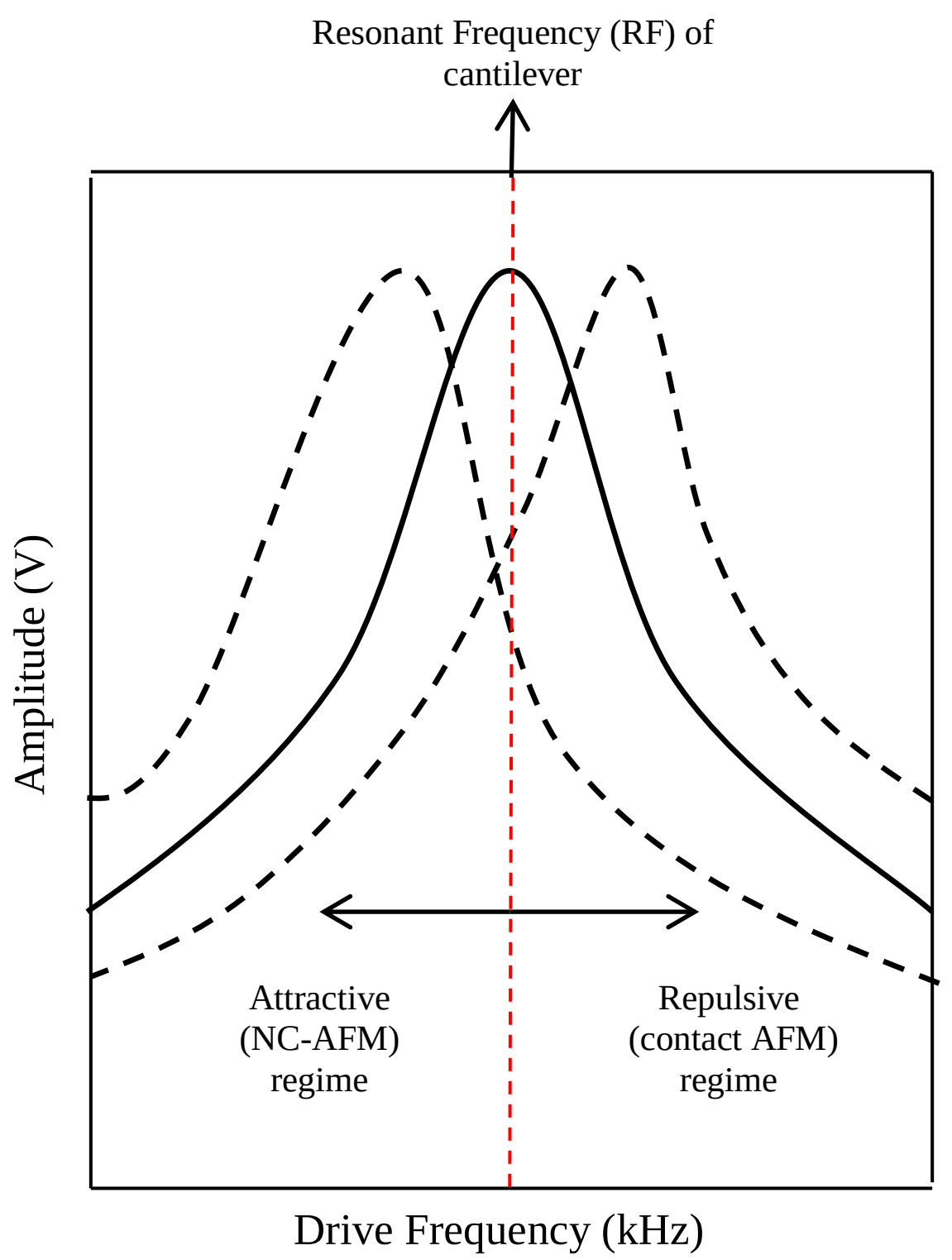

**Figure 1.2.3** Tuning curves of an AFM tip after it is excited to its resonant frequency (centered about dashed red line) and tip response (RF shift) in relation to tip-surface distance. Left of the RF curve (tapping-mode frequency) is the attraction-dominant region and right of the RF curve is a the repulsive-dominant region for NC-AFM and contact-AFM, respectively.

The remainder of this section is dedicated to a description of the relevant properties of ambient NC-AFM as performed in the dynamic, AM-AFM mode.

In dynamic AFM, there are two primary modes of operation, namely, frequency modulation (FM) and, as mentioned, amplitude modulation (AM). AM-AFM involves excitation of the tip-sample ensemble at a constant frequency close to the resonant frequency, $f_0$, which corresponds to some fixed amplitude signal at or near the resonant amplitude, $A_0$.[36,51] As the gap closes between the tip and surface, elastic and inelastic interactions induce changes in amplitude and phase channels of the oscillating cantilever.[51] While for the most part AM-AFM is used for ambient NC-AFM mode, it is also reported to have been used at closer tip-surface distance (repulsive) regimes (intermittent contact mode) as well.[52,53] The oscillation of the AFM tip has a motion that is roughly analogous to that of a simple harmonic oscillator.[54] To a first approximation, the AM-AFM mode adheres to a relatively simple and widely used[55–61] model, the point-mass model,[53] which is analytically solvable. The motion of the cantilever in free oscillation is initially defined by the expression in the form of a 2nd-order differential ($z$):

$$F_{ts}(d) = mz'' + kz + \frac{mf_0}{Q} z' - A_0 \cos ft \qquad (1.15)$$

where $f_0$, $A_0$, $F_{ts}$, and $d$ are the familiar resonant frequency, near-resonant amplitude, tip-surface force, and tip-surface distance components, respectively, while $t$, $Q$, $m$, $f$, and $k$ represent, respectively, amplitude response time, the quality factor, effective mass (sometimes $m_{eff}$), driving frequency, and spring (force) constant of the cantilever. An assumption exists in many AM-AFM models that equates effective mass, $m$, ~$0.25m_c$ where $m_c$ is the total cantilever mass, implying that the tip is typically 25% of the cantilever's mass.[54] Although the nonlinearity of tip-surface forces, particularly any contribution to the $F_{ts}$ term, makes analytic solutions to equation 1.15 difficult, an important relationship with respect to the free resonant frequency, $f$ ($f$ is $f_0$ in this case) and effective mass of the cantilever, $m$, arises (as one may recall from introductory physics!)[62]:

$$f = \sqrt{\frac{k}{m}} \tag{1.16}$$

Next, forcing the condition of $F_{ts}=0$, equation 1.15 becomes:

$$mz'' = -kz - \frac{mf_0}{Q} z' - A_0 \cos ft \tag{1.17}$$

where tip-surface interaction is constrained to a amplitude response timescale of $t \approx \frac{2Q}{f_0}$ meaning that the feedback signal is generally slow due to the high Q-factor[62] (>$10^2$ for ambient conditions) of most AFM cantilevers, especially those used in this work involving ambient NC-AFM. Qualitatively, this is fairly obvious when comparing the tip-surface response and amount of low-frequency noise in the AFM tip's amplitude response during tuning cycles for the longer (lower $f$) versus shorter (higher $f$) cantilevers. Furthermore, the AAC mode AFM controller used in this study actually provides a bandpass filter with a center frequency on the order of the resonant frequency, making the lower $1/f$ (flicker) noise and higher frequency mechanical noise very discriminant in the resulting topographic image, allowing for easy filtering during image processing (see Chapter 2 for example of FFT filtering technique). When the AFM tip experiences damping (i.e. when it is in close proximity to the surface) the steady motion of the simple harmonic is perturbed and $f$ deviates from the $f_0$.[54,62] Because of this deviation, the AFM The modified resonant frequency, $f_m$, rendered in terms of the free resonant frequency, $f_0$,:

$$f_m = f_0\left(1-\frac{1}{4Q^2}\right)^{1/2} \qquad (1.18)$$

and substituted into the general solution (z''→z) for the 2nd-order differential in equation 1.17 as follows[53]:

$$z = A\cos(ft-\phi)+B\exp\left(-\frac{\alpha}{2}t\right)\cos(f_m t-\beta) \qquad (1.19)$$

where the z-motion of the cantilever is now a function of the general solution for the damped harmonic oscillator system with linear combination coefficients *A* and *B* capturing the phase difference, *Φ,* sustained as the cantilever motion and driving frequency, *f*, constantly fluctuate throughout a typical NC-AFM image cycle. Of course, since the *f* is now somewhat dampened and therefore not the same as its RF, $f_0$, the resonant amplitude response, $A_0$, now varies and is effectively some other amplitude, $A_f$* forming the Lorentzian-like function :[53]

$$A_f = \frac{A_0/m}{\left(\left(f_0^2-f^2\right)^2+\left(f\,f_0/Q\right)^2\right)^{1/2}} \qquad (1.20)$$

with a corresponding phase shift, *Φ*, that is the basis for local contrast in the NC-AFM image described as:

*Please excuse the unfortunate inconsistency for the notation of linear combination coefficient, *A*, in equation 1.19 and symbol for amplitude (free resonant in $A_0$ and instantaneous in $A_f$ amplitudes)

$$\phi = \tan^{-1}\left(\frac{ff_0/Q}{\left(f_0^2-f^2\right)}\right) \qquad (1.21)$$

It should be highlighted that the damping constants ($\alpha$, $\beta$) in the second term of equation 1.20 are related to what I have referred to as the amplitude response timescale by $t \approx \frac{2Q}{f_0} \approx \frac{1}{\alpha}$, imposing a limits on the NC-AFM (AAC mode herein) bandwidth (used colloquially here as the number of bits rendered per second).[54] Alas, in UHV conditions[62] where Q-factors are often ~$10^5$-$10^6$ the story is much worse and AM-AFM must be abandoned for more complex ensembles involving a series of phase shifters, RMS to DC converters, and constant oscillation amplitude controllers.[62] This report exclusively involves the NC-AFM imaging in AM-AFM mode, so a detailed workup of the FM-AFM mode is omitted. However, the preceding workup of generic tip-surface force calculations involved in NC-AFM should serve as a benchmark for the complexity of NC-AFM imaging in any operating mode, whether it is AM-AFM, FM-AFM, or otherwise, and this cannot be understated when interpreting the experimental results. There is, on the other hand, a rich phenomenology or "art" to the interpretation of NC-AFM data that lends itself to a solid foundation of theoretical discussion. The following sections will define contrast formation, the nature of topographic images, and common imaging artifacts observed with NC-AFM imaging. Finally, the roughness parameters used to perform quantitative analysis on NC-AFM data are addressed.

### 1.2.4 NONCONTACT ATOMIC FORCE MICROSCOPY: TOPOGRAPHIC IMAGES AND ASSOCIATED ARTIFACTS

The local contrast generated in NC-AFM images is derived from all the tip-surface forces that exist during the raster-like oscillations of the cantilever over the sample surface. While both long and short range interactions are dominant in ambient NC-AFM scenarios, the shorter range chemical forces are predominantly responsible for the local contrast formation; sometimes even yielding atomic resolution.[63] In this study where nanostructures sometimes in the form of 3D – clusters are supported on corrugated and other complex surfaces thereof, the need for local contrast formation is imperative for accurate interpretation of topographic data. Because so many variables go into the acquisition of a correct image (see Chapter 2), it is hard to describe topographic image data solely on the basis of a specific tip-cantilever system, even if extensively characterized prior to imaging. The tip's curvature in relation to supported nanostructure features might induce changes in surface composition, thus altering rigorous characterization. All kinds of experimental factors affect the acquired NC-AFM images and since local contrast formation and tip-surface interactions are rarely commensurate, many considerations must be made during image processing where grain analysis, roughness analysis, and other surface statistical methods are carried out. These considerations include some major artifacts, which, if not corrected, skew the theoretical integrity of the aforementioned analyses.

In the noncontact (attractive) AFM mode, the effect of van der Waals tip-surface forces contribute to the local contrast within a range of ~100 nm,[64] so for any surface feature, whether it is a protruding or step-edge site, the contrast formation will be a measure of the tip-surface convolution. In essence, every single bit that gets digitally rendered to a pixel contains information in equal parts from the tip and the surface, especially for AFM in the contact regime.

The tip-convolution or tip-broadening effect is a common, unavoidable phenomenon that is also the underlying reason for the relatively lower lateral resolution (xy) than axial (z) resolution for all of the modes of AFM. The broadening is often observable when the tip size is on the order of the surface feature size (Figure 1.2.4) and, in cases where the surface feature is smaller, one can accidentally image the tip (Figure 1.2.5a). Although this would inadvertently be an effective way of characterizing the tip, it is sometimes not so unequivocally clear that it is the actual tip image, so other relevant effects including strange repeating patterns or multiple-tip (“twinning”) patterns may arise (Figure 1.2.5b).

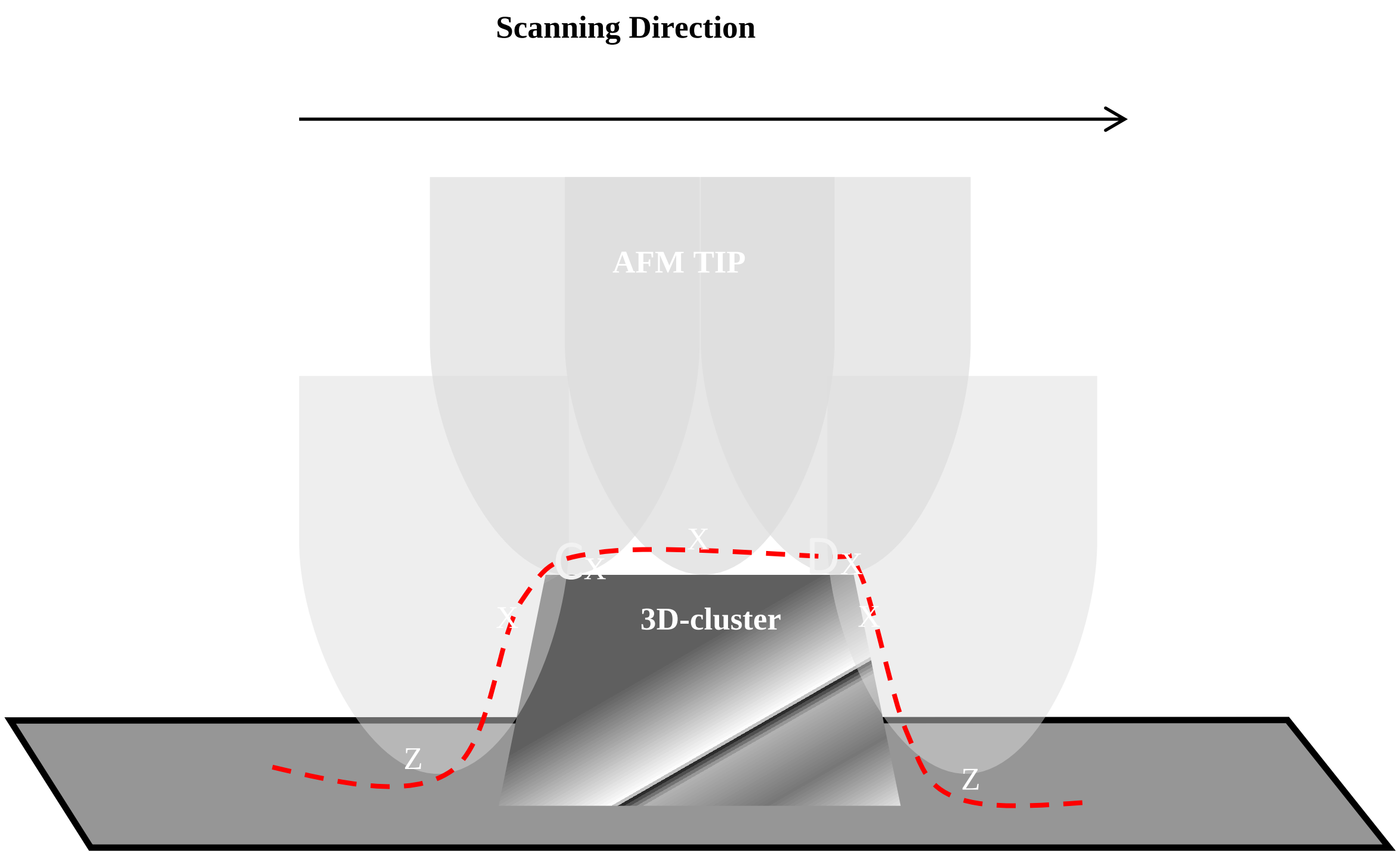

**Figure 1.2.4** Schematic of the tip-surface convolution or tip-broadening effect in NC-AFM wherein the tip may over shoot nanostructures such as a 3D-clusters protruding from the surface. Actual points of tip-surface signals ("contact"), x, are often different from the apparent signal, z, when the tip apex is larger than the surfaces features being imaged.

To elucidate this problem, in some cases the use of (expensive) super-sharp AFM tips ($r_{tip}$~2 nm) can yield optimal, even atomic, resolution. As it is evident from Figure 1.2.4, a convolution of the tip-surface interactions make it impossible to accurately characterize the top facet in the nanocluster, however, with the loss in lateral resolution accounted for, inferences about the morphology of individual nanostructures within the nanocluster can be made without the exhaustive characterization of the tip-cantilever ensemble that would be required to warrant accurate tip-surface deconvolution calculations.

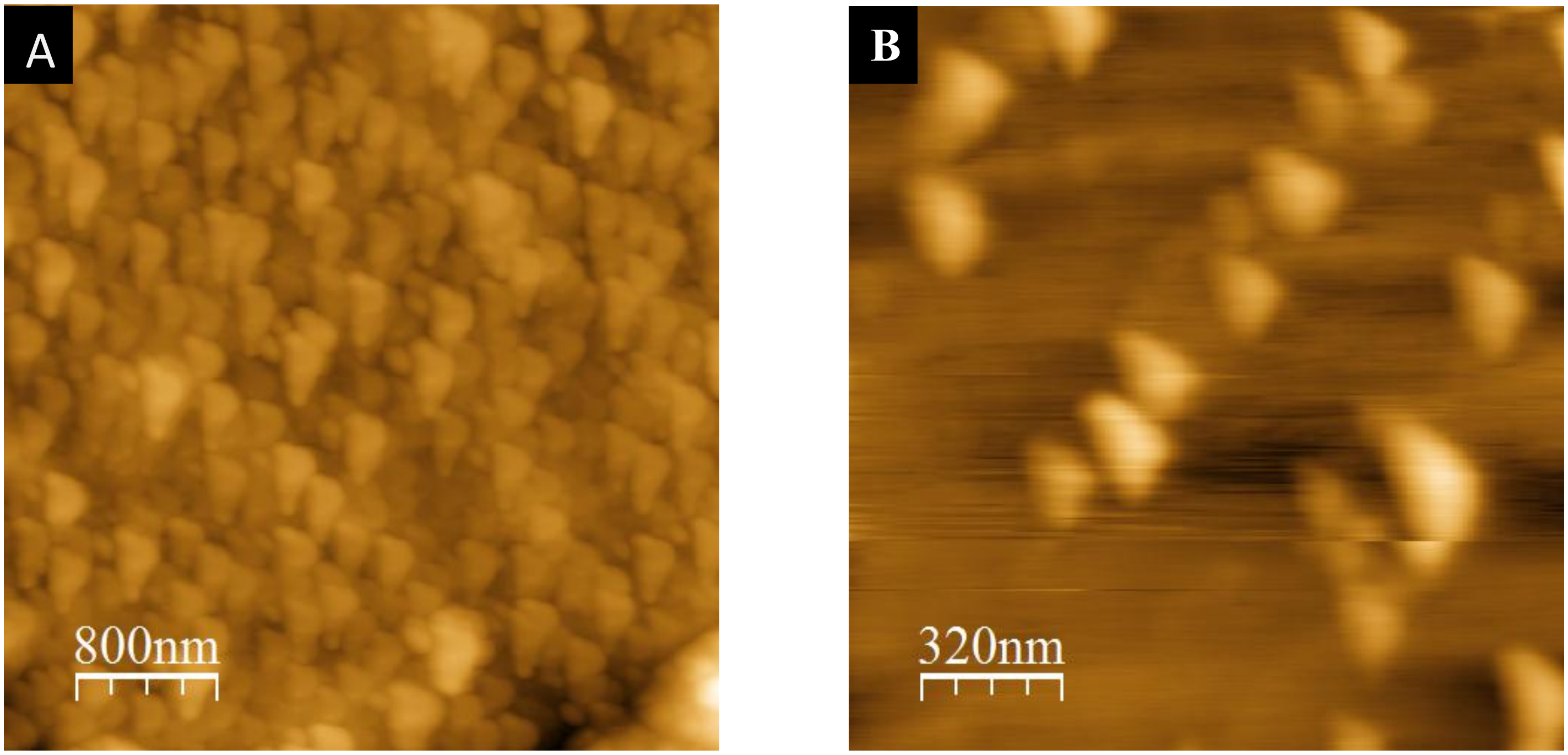

**Figure 1.2.5** A) and B) depict 4.0 µm x 4.0 µm and 1.6 µm x 1.6 µm NC-AFM image artifacts of cobalt oxide-loaded MgO(100) and YSZ(100) surfaces, respectively. It is clear in A) that the features are on the order of size and shape for the Si cantilever-tip system used for imaging, meaning an image of the tip itself was obtained. Similarly, in B) the "twinning" effect of the triangular–shaped Si tip is observed.

Precise grain analysis methods are described at length in chapter 2, but it needs clarifying now and hereafter that the upper limit in terms of the lateral resolution for the nanostructures/3D-clusters was adequate enough that tip-surface deconvolutions would not add any further physical meaning to the interpretation of NC-AFM data. Instead, super-sharp Silicon probes are used in the event of difficulties imaging the top of the surface feature facets on the flat supports.

Height measurements in NC-AFM images rely on the sensitive calibration of the scanner in the z-direction. Parabolic backgrounds (bow/tilt) are a major NC-AFM image artifact (Figure 1.2.6), especially in this work where our pendulum style AFM setup is on the 4$^{th}$ floor of a building, but can be handily rectified by regular z-calibration of xyz-piezo response.[64] This is often correctable in real-time during NC-AFM image cycles via a polynomial fitting feature, which fits each tilted image line to a polynomial function of $n^{th}$-order (usually $2^{nd}$) and then subtracts out the initial tilted image line.

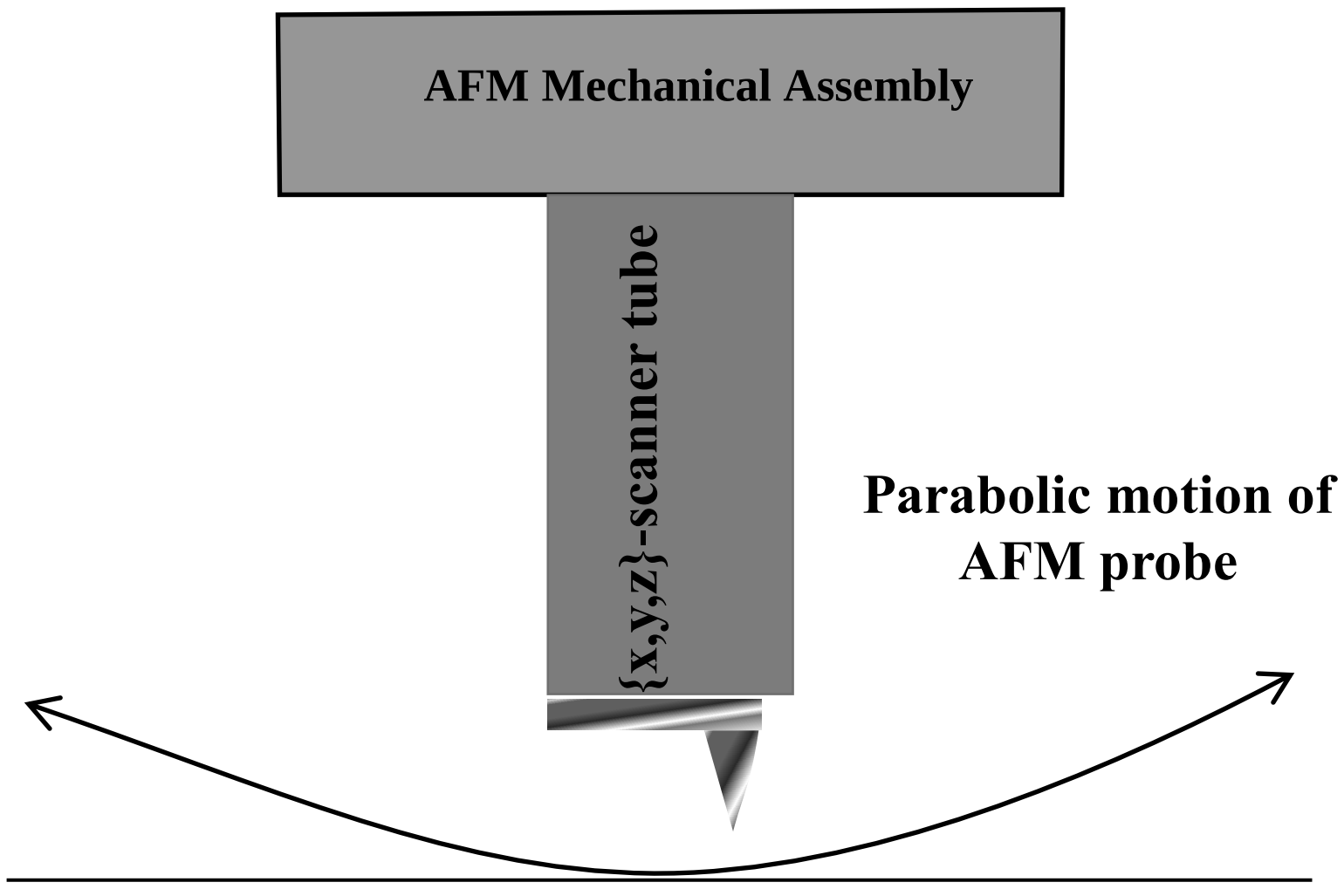

**Figure 1.2.6** Background "bow" or "tilt" artifact in NC-AFM. Since the scanner tube is suspended from the mechanical assembly, the z-range motion is often nonlinear as the probe scans the surface.

In the case of flat surfaces, possibly with terrace features, further flattening can be applied by selecting a background region on the image that is substantially lower than the adjacent feature height and leveling the whole image relative to said feature height. Care must be taken in this method of leveling as background corrections with respect to a feature having a saturated (beyond z-range or setpoint amplitude of cantilever) contrast formation, usually observed as white dots, would be inexact. Almost never is image flattening a sufficient treatment when carrying out a rigorous and correct NC-AFM image analysis, which often involves both an intuition of, in this report a nanostructure-support system, surface morphology and the ability to filter out common types of analytical noise that may hamper the image including, but not limited to: irregularities in surface roughness, contamination of tip-surface ensemble, general wear of components, scanner or sample drift, coupling of electronic noise, and mechanical instabilities. Matrix, fast Fourier Transform (FFT), and other filtering processes carried out in this study's NC-AFM image analysis are described in Chapter 2.

.

### 1.2.5 QUANTITATIVE ANALYSIS OF NONCONTACT ATOMIC FORCE MICROSCOPY: DESCRIPTION OF ROUGHNESS AND STATISTICS PARAMETERS

The amplitude signal in the AM-AFM mode of NC-AFM in this thesis corresponds to the height parameters of the surface topography. In this vein, surface profile parameters are typically expressed through the rendering of 2D-topographic, phase, and amplitude image channels during NC-AFM imaging. The morphology of the single-crystal substrate-supported nanostructures can be accurately quantified solely from the topography images. In cases where the phase imaging channels offer sharper local contrast with respect to lateral resolution, quantification of surface morphology parameters can be carried out using these images. The principal statistical roughness and height parameters[65] calculated for clean, single-crystal support surfaces (YSZ(100)/(111), MgO(100), and HOPG) as well as the morphological data of the single crystal-supported metal oxide nanostructures ($Co_2O_3$ nanostructures) or nanoclusters are summarized succinctly in the following forms:

**Arithmetic Mean Height ( $\bar{z}$ ):**

$$(2D) \qquad \bar{z}(N,M) = \frac{1}{N}\sum_{x=1}^{N} z(x,y) \tag{1.22}$$

$$(3D) \qquad \bar{z}(N,M) = \frac{1}{NM}\sum_{x=1}^{N}\sum_{y=1}^{M} z(x,y) \tag{1.23}$$

**Mean Roughness ( $R_{avg}$ ):**

$$R_{avg}(N,M) = \frac{1}{NM}\sum_{x=1}^{N}\sum_{y=1}^{M}\left(z(x,y) - \bar{z}(N,M)\right) \tag{1.24}$$

**Root Mean Square (RMS) Roughness, $R_q$:**

$$R_q = \sqrt{\frac{1}{NM}\sum_{x=1}^{N}\sum_{y=1}^{M}\left(z(x,y) - \bar{z}(N,M)\right)^2} \tag{1.25}$$

**Maximum Peak-to-Valley Height (MPV), $R_{mpv}$:**

$$\begin{aligned} R_p &= \max(z_i - \bar{z});\ 1 < i < N \\ R_v &= \left|\min(z_i - \bar{z})\right|;\ 1 < i < N \\ &\therefore \\ R_{mpv} &= R_p + R_v \end{aligned} \tag{1.26}$$

**Kurtosis Parameter, $R_{ku}$:**

$$R_{ku} = \frac{1}{NR_q^4}\sum\left(z_i - \bar{z}\right)^4 \tag{1.27}$$

Since the NC-AFM data in this thesis is from amplitude modulation, the corresponding topographic data is all derived from amplitude data, which in turn corresponds to the z-component(axial) height of features. Thus, the preceding statistical roughness parameters are all as a function of the amplitude/height, *z*, and its mean height, $\bar{z}$ at any image cross section/point (x,y). The variables *N* and *M* correspond to the number of height profiles and number of points per height profile, respectively. Since the surfaces, especially the clean, nearly atomically flat single-crystal surfaces often have stepped features, the statistical descriptor, kurtosis, was often used to describe peakedness [66]or "waviness" of the surfaces. Kurtosis parameters are centered around 3 (Mesokurtic[65]), which is designated for a Gaussian distributed peakedness of any surface, while values less than 3 indicate a flat (Platykurtic[65]) surface, and anything above 3 indicates a very rough surface with high peakedness (i.e. not many valleys). Corrugation of the surfaces is defined as a function of the RMS roughness, $R_q$, while the maximum feature heights are correlated with the MPV amplitude, $R_{MPV}$. Overall, equations 1.22-1.27 capture all of the surface morphological data necessary for the complete and rigorous characterization of single crystal-supported nanostructures investigated in this work.

In summary, this section on NC-AFM imaging addresses all aspects of the history, instrumentation, working principles and theory behind the tip-surface interactions responsible for the AFM signal, and all of the relevant data treatment for the NC-AFM work performed in this thesis.

## 1.3 PRINCIPLES OF CLEAN SURFACE STRUCTURES AND SUPPORTED METAL NANOSTRUCTURES

Since a substantial amount of this work is dedicated to characterizing metal oxides supported on single-crystal surfaces, a theoretical basis relevant to the data presented in chapters 3 and 4 is necessary. Nanostructure and cluster shape of the metal oxide is relevant to the nature of the substrate, so the various solid-solid growth modes are addressed.

Perfect single-crystal surfaces are cut along any arbitrary angle and the resulting lattice always has associated direction vectors that are defined as the Miller Indices, {*h*,*k*,*l*}.[67,68]. The supports in this work, YSZ(100)/(111) and MgO(100), bear face-centered cubic (FCC) crystal structures and the general FCC surface structures or orientations look like the following:

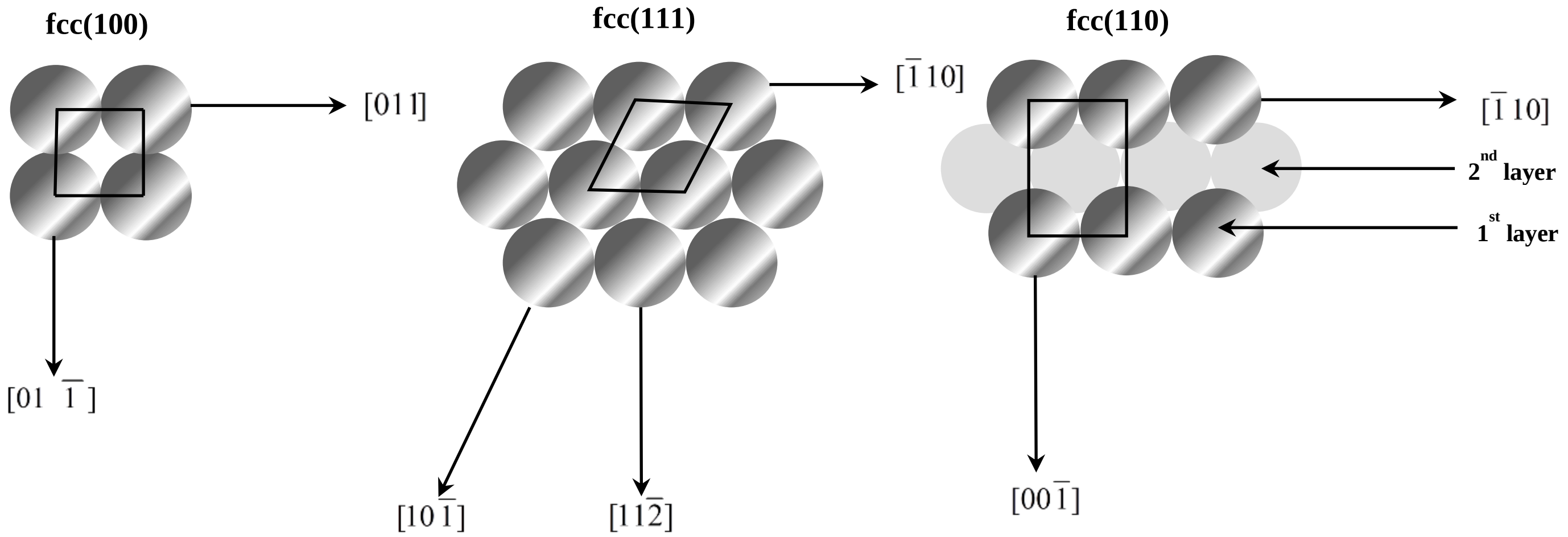

**Figure 1.3.1** Representations of the face-centered cubic surface with low index surface planes (left to right) of {100}, {111}, and {110}. The perpendicular direction planes are drawn to aid in visualization.

The various ideal structures show varying surface orientations with varying influence on structure-support interactions. For instance, on the (100) surface (Figure 1.3.1), there is a one-fold (on top of the center of the atom), 2-fold (bridging two atoms) and 4-fold coordination (in the hollow sites) and since the other surfaces have various different coordination sites this lends itself to a heterogeneity in the adsorbate interactions with clean, ideal surfaces of different surface atom densities. The varying surface atom densities often have different forms that depend on their respective unit cell parameters and Miller indices. Surface atom density is intrinsically related to surface structure and strongly influences surface properties such as the RMS roughness discussed in the last section.

Referring back to the opening statement, the 3 different Miller indices form planes in reciprocal space that correspond to a particular Bravais lattice.[67] There are 14 Bravais Lattices, which are the unit cells expressing the geometry between different planes defined by the Miller indices. Typically, the interplanar d-spacing, $d_{hkl}$, can be calculated for any unit cell.

The separation of planes is defined by:

$$\frac{1}{d_{hkl}^2} = \frac{h^2}{a^2} + \frac{k^2}{b^2} + \frac{l^2}{c^2} \tag{1.28}$$

where $d_{hkl}$ is the d-spacing between {h,k,l} planes and (a,b,c) are unit cell vectors associated with the length of each side. Consider the following relavant example where, by extension of equation 1.28, the interplanar distance of $ZrO_2$(111) (it would work for YSZ(111), too, but there is less fcc character) can be calculated as such:

$$\begin{aligned} d_{hkl}^2 &= \frac{a}{\sqrt{h^2 + k^2 + l^2}} \\ &(a = b = c = 5.184\,\text{Å for cubic lattice of } ZrO_2) \\ d_{\{111\}}^2 &= \frac{5.184\,\text{Å}}{\sqrt{1^2 + 1^2 + 1^2}} = \sim 3\text{Å} \end{aligned} \tag{1.29}$$

This value could be useful when characterizing well-defined parallel steps found on surfaces of Yttria-stabilized Zirconia (YSZ). The varying surface geometries of the single-crystal supports studied in this work should influence varying growth modes and morphology of the nucleated nanostructures. The following section briefly addresses various modes of metal nanostructure nucleation on a substrate in terms of

## 1.3.2 NUCLEATION OF METAL NANOSTRUCTURES ON SUPPORTS

Nucleation of a metal nanostructures or nanoclusters atop the well-defined surfaces (Figure 1.3.2) in this study can happen following three general growth modes:[68]

i) Volmer-Weber (VW) Growth where metal nanostructures on a surface will nucleate and if there exists mobility, will aggregate to form 3D-islands or nanoclusters. Aggregation can also occur if nucleation happens in adjacent nucleation or defect sites.

ii) Stranski-Krastanov (SK) Growth where first a monolayer or self-limiting layer of metal grows, followed by growth of 3D-islands on top of the formed layers. A lattice mismatching or strain between the adsorbed metal and substrate often limits the number of wetting layers. The process favours wetting of the substrate by the metal over-layer.

iii) Frank-van der Merve (FdM) growth or layer-by-layer growth wherein, as the latter's name suggests, growth of the metal over-layer along with precise lattice matching between the metal and surface.

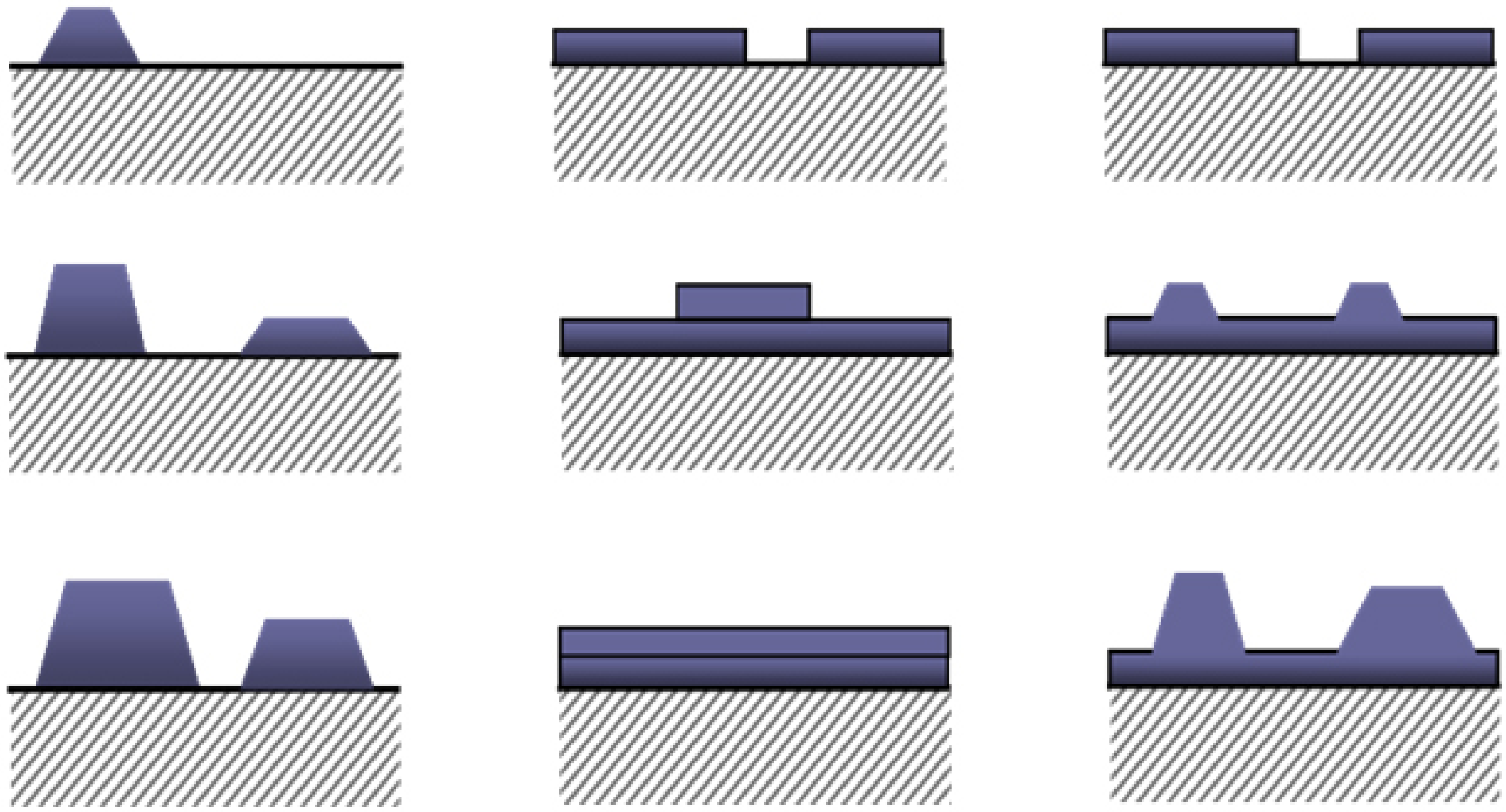

**Figure 1.3.2** Three thermodynamic solid-on-solid growth modes of  solid over-layer (purple) on solid substrates (grey lines). (Going from left to right): i) VW Growth (3D-island formation), ii) FdW Growth (layer-plus-layer growth), iii) SK Growth (layer-plus-island growth).

The present study considers, however, VW growth to be the only feasible growth mode that can occur while, though SK growth might be localized in some cases, the growth modes of SK and FdM are not possible or expected in this particular capacity. The growth process, as will be discussed, is preferential with respect to defect features on the surface, which act as nucleation sites.

VW growth is expected when interaction of the metal nanostructures (particle, etc) with other metal nanostructures is stronger than interaction with the actual substrate.[69] This type of growth has been reported on oxide supports similar to those used in this study[70] where weak metal-substrate is observed. In some cases, the interaction is weak enough that NC-AFM can damage the nanostructure. The VW growth with respect to strength of interaction with support can be distinguished from the other growth modes, particularly FdM, using qualitative thermodynamic arguments:

$$\gamma_{sub} < \gamma_{metal} + \gamma_{int} \quad \text{VB Growth Mode} \tag{1.30}$$

$$\gamma_{sub} \geq \gamma_{metal} + \gamma_{int} \quad \text{FdW Growth Mode} \tag{1.31}$$

where $\gamma_{sub}$, $\gamma_{metal}$, $\gamma_{int}$ are the surface free energy terms for substrate, metal, and substate-metal interface, respectively. Typically for metal-oxide supports, the surface free energy of the substate-metal interface and surface free energy of the metal are greater than that of the support. This, by thermodynamic consequence, means the effect of wetting at the metal-substrate interface is disallowed.[68] Instead, island growth into 3D-nanoclusers whose exact equilibrium and wetting path (if applicable) could be derived from Wulff construction methods, but these are not relevant data for this particular study and so they will not be addressed. However, growth modes of single crystal-supported nanostructures are inferred on the basis of NC-AFM topographic data.

## 1.4 X-RAY PHOTOELECTRON SPECTROSCOPY

An important surface-sensitive technique falling under the blanket of electron spectroscopies is X-ray Photoelectron Spectroscopy (XPS). The technique is built upon the photoelectric effect where high-energy photons (the x-rays) impinge upon a sample cross-section and eject deep core electrons (photoelectrons) with a kinetic energy (KE), and therefore binding energy (BE) corresponding the energy of the incident photons and unique to the particular surface species.[68,71]

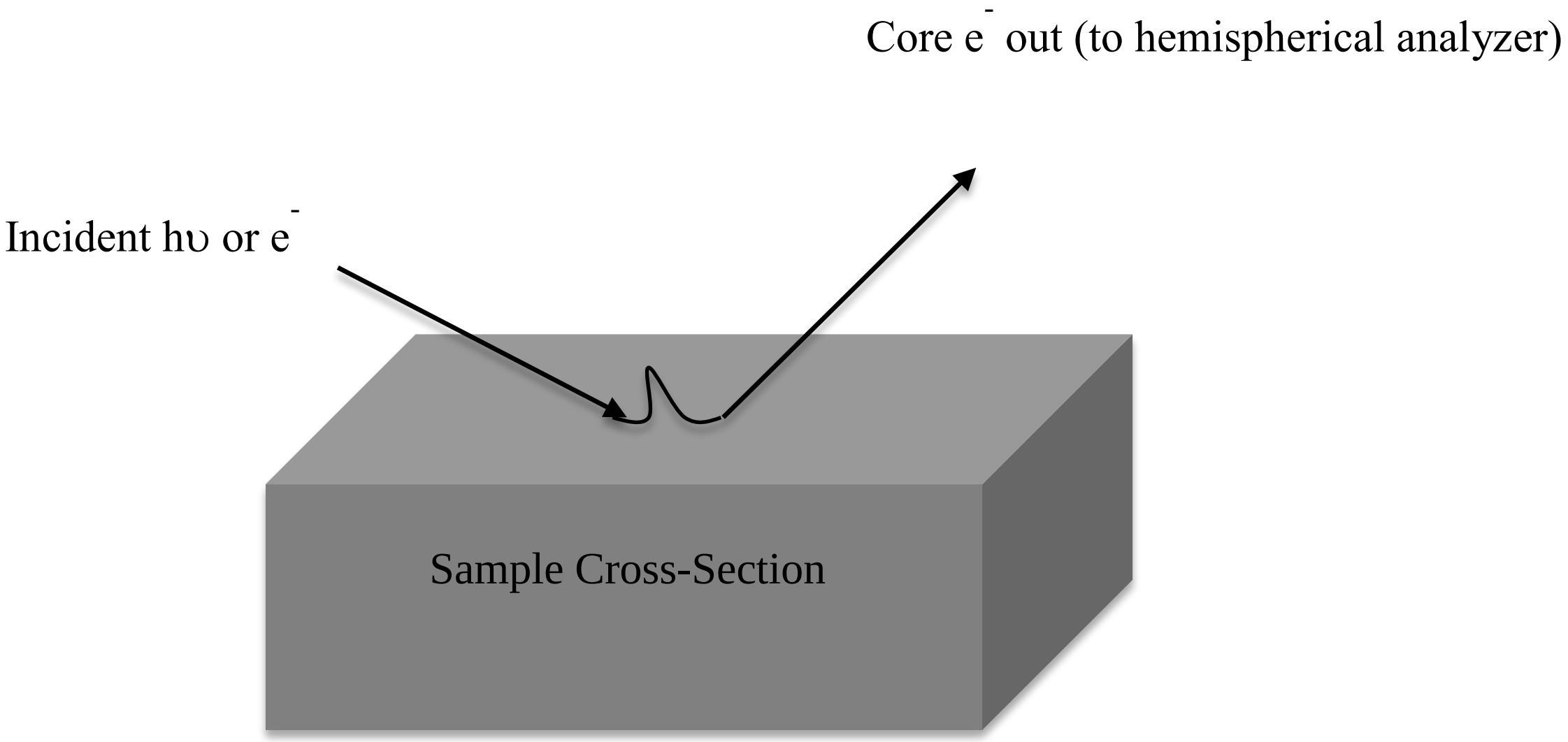

**Figure 1.4.1** Simple schematic of electron spectroscopy (XPS for the incident photon).

The basic photoemission process of Figure 1.4.1 is essentially the absorption of a photon with an energy in excess of its core electrons intrinsic binding energies and, since the core electrons do not participate in bonding, they are emitted with a kinetic energy unique to the atom from whence they came. These information-carrying, core-level electrons, after photoemission into vacuum, have kinetic energy, $E_k$, given by:

$$E_k = h\nu - E_b - \varphi \qquad (1.32)$$

where $E_k$, $h\upsilon$, $E_b$, and $\varphi$ are the electron kinetic energy, x-ray energy (typically ~14.26 keV), binding energy of emitted electrons as measured at analyzer, and the work function of the solid, respectively. Upon detection of the photoelectrons, information about the surface's binding energy can be inferred. Typically the XPS spectrum is a plot of intensity as a function of photoelectron kinetic energy correlated to its binding energy, which in turn corresponds to the chemical identity of elements present on the surface. Binding energy is specifically defined as the difference of energies between initial and final states of an atom such that the initial ground state has *n* electrons and resulting photoelectron or ion has *n-1* electrons.

Thus:
$$E_b = E_f(n-1) - E_i \qquad (1.33)$$

where $E_f$ and $E_i$ are the final and initial state electron energy functions. These energy functions serve to the exact correlation of kinetic energy of the electron to its binding energy where any non-initial or final state "event" is unphysical, that is, not a property of the particular element.

The quantification of virtually any chemical species is possible with XPS and this thesis involves elemental quantification of metals with their native oxide layer. Though it's not the principal surface-sensitive technique employed herein, XPS measures and identifies elements within 10 nm of the surface[68,71] and is useful in quantifying the chemical composition of the surface in an atom-by-atom basis. Calculation of atomic percent (At. %) is done by relating the object peak area to a standard concentration (often with respect to carbon 1s peak):

$$X_i = \frac{(At\,\%)_i}{\sum_{j=1}^{n}(At\,\%)_i}\, x\ 100\% \qquad (1.34)$$

where $X_i$, and *At %* indicate elemental composition (in % of atoms per cross-section) *At %* of element *i* relative to internal reference element *j*. In this work, when corroborated with particle-density calculations from NC-AFM data, quantification can often confirm the integrity of the topographic data, providing confirmation of the true chemical nature of both the support and the metal (oxide) nanostructures. After chemical shift identification and possible correction for charging (insulating supports often charge), quantification can be performed in the relevant software package.[72] Information regarding the electronic structure and identification of oxidation states for metal oxides –in this case for the supported nanostructures—can be ascertained from the corrected spectra and quantification with correctly inputted relative sensitivity factors (RSFs) for respective surface species from the Scofield library[6,71] Accuracy of the quantification and identity of the oxidation state for metal oxides is limited by the resolution efficiency of the analyzer and the monochromaticity of the X-ray source. Also, oxidation states with chemical shifts to a higher binding energy corresponds to higher (positive) oxidation states because of the enhanced Coulombic attraction between the photoelectron and its ion core.[6,71]In cases where the full width at half maximum (FWHM) of the major metal peak envelope is on the order of the peak separation energy of the different oxidation states, complex fitting procedures must be carried out (see following chapter).

# CHAPTER 2 EXPERIMENTAL DETAILS

## 2.1 EXPERIMENTAL SETUP

All NC-AFM data in this thesis was collected on the Molecular Imaging (Agilent) Pico Plus System and XPS/STM data was collected in both the Giorgi group's Specs/RHK Multi-Technique System and the Kratos Axis Ultra DLD system in the Center For Catalysis Research and Innovation (CCRI). AFM tips were purchased from Nanosensors™ (types: Point Probe Plus-Noncontact Long Cantilever (PPP-NCL)[73], Super Sharp Silicon-NCL (SSS-NCL)[73], assorted Diamond-Like Carbon (DLC)) and single-crystal samples were purchased from both MTI Corporation (YSZ(100)/(111), MgO(100)) and NanoScience® (HOPG). Finally, photochemical growth experiments on the cleaned single crystals were carried out in collarboration with Tse-Luen Wee of the Scaiano group. Experimental details of the NC-AFM and XPS setups as well as the protocol for the photochemical growth experiments are provided herein.

## 2.1.1 NONCONTACT ATOMIC FORCE MICROSCOPY SETUP

NC-AFM imaging was carried out in the SEM/AFM lab on 4$^{th}$ floor of the Biosciences (CCRI) building. The instrument itself is operated in intermittent contact mode or AC mode, in which the cantilever is driven by an alternating current. The microscope bundle includes a MAC controller that includes two operating modes or magnetic AC (MAC) and acoustic AC (AAC). The latter mode was the one employed throughout NC-AFM imaging and its working principle is essentially a voltage that oscillates at ~300 kHz and drives a piezo crystal located in the scanner module, particularly in the nose cone assembly (Figure 2.1.1). Typically, the drive frequency in AAC mode actually relies on the natural frequency or resonant frequency (RF) of the AFM probe or cantilever that is used. The tips used throughout the imaging cycles were principally Nanosensors™PPP-NCL (RF=190 kHz) and, particularly for post-growth images, SSS-NCL (RF=190 kHz).

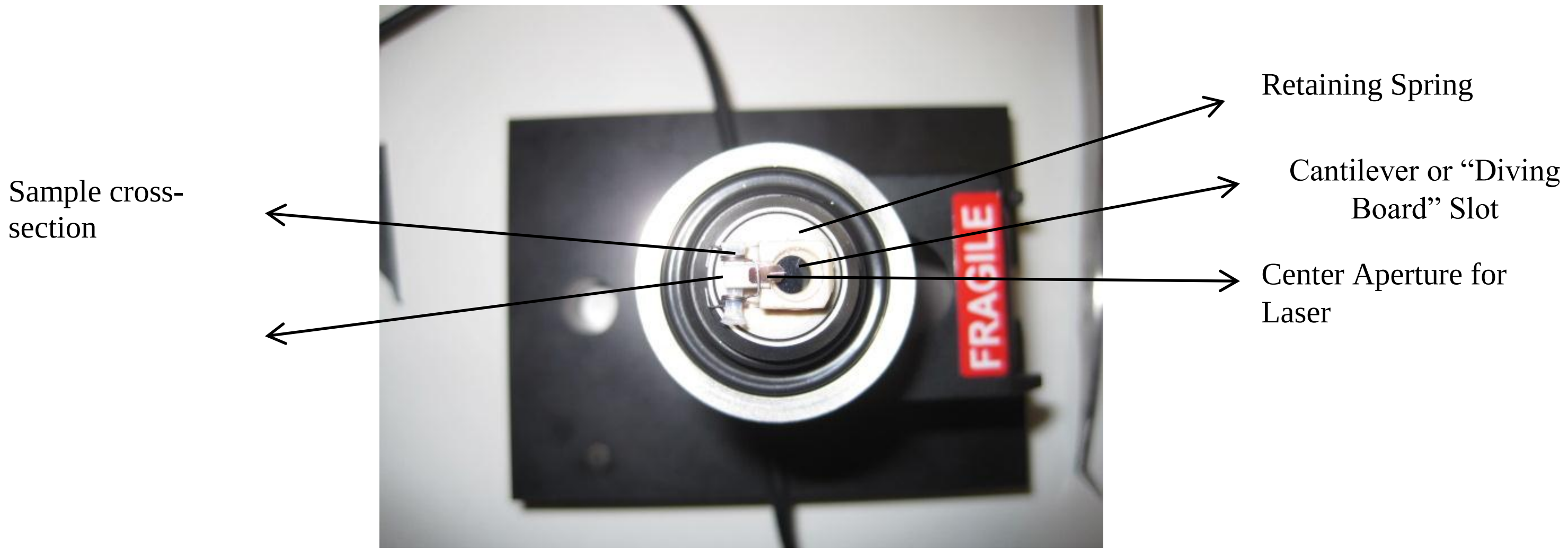

**Figure 2.1.1** AFM scanner module equipped with SSS-NCL probe

After proper tip mounting on the scanner module and proper sample mounting on the standard XYZ-sample stage (Figure 2.1.2) , the photodiode “quadrant” detector (Figure 2.1.3) assembly is fit by gently sliding it into the appropriate slot of the scanner module. A properly mounted tip is often held such that only 1/3 of the cantilever is held by the retaining spring. XY-translation of the detector is often necessary while centering the laser beam reflection from the cantilever into the center of the detector quadrant. Typically, the laser beam is aligned directly on the center of the cantilever until the highest amplitude is achieved on the Pico Scan 2500 Controller (see Figure 2.1.4) for that particular tip. The final aligned spot of the laser should appear such that its diffraction pattern forms an “X” pattern, visible optically through the Pico Plus Video System (Figure 2.1.4).

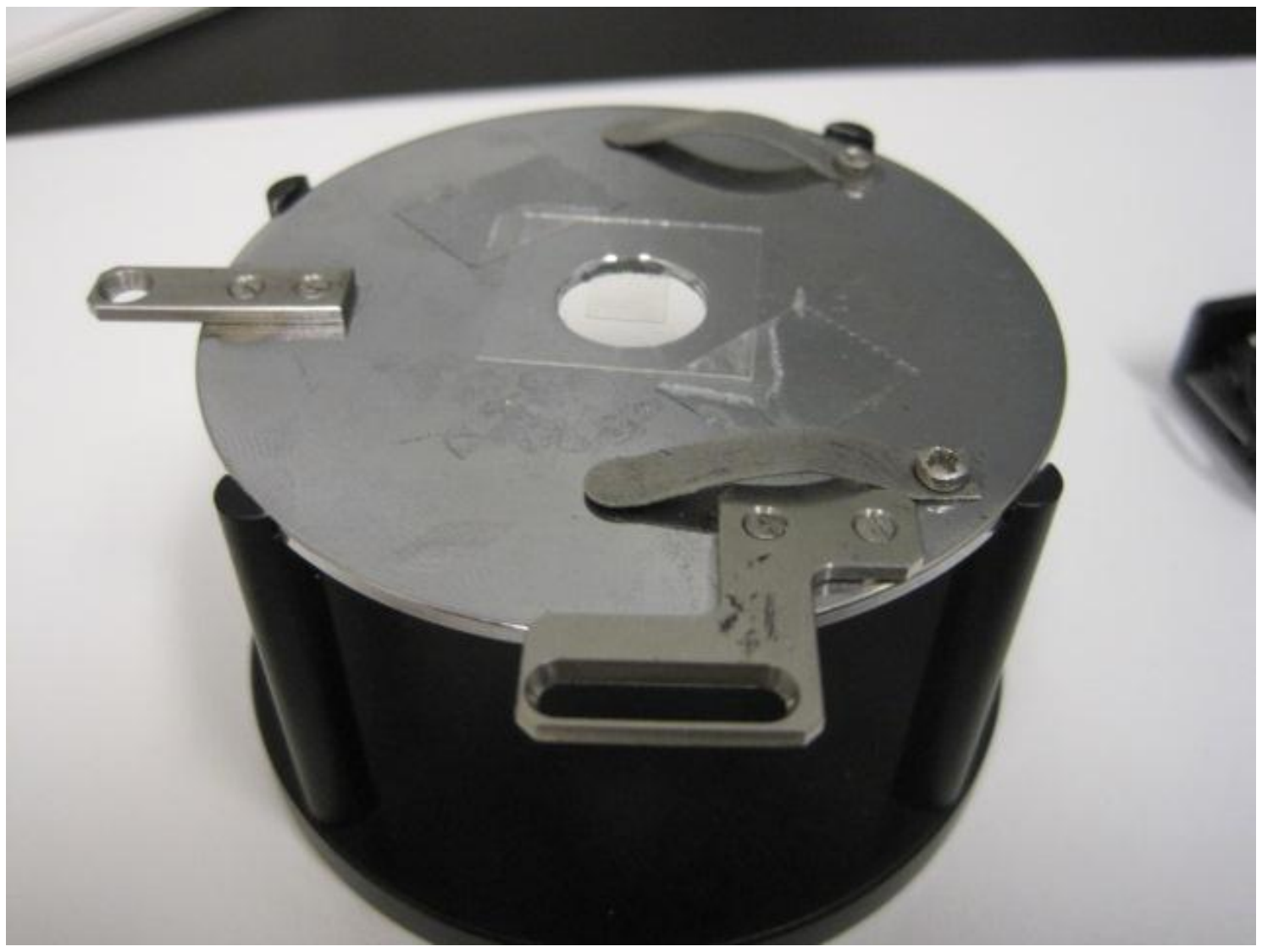

**Figure 2.1.2** Standard sample mounting stage (plate)

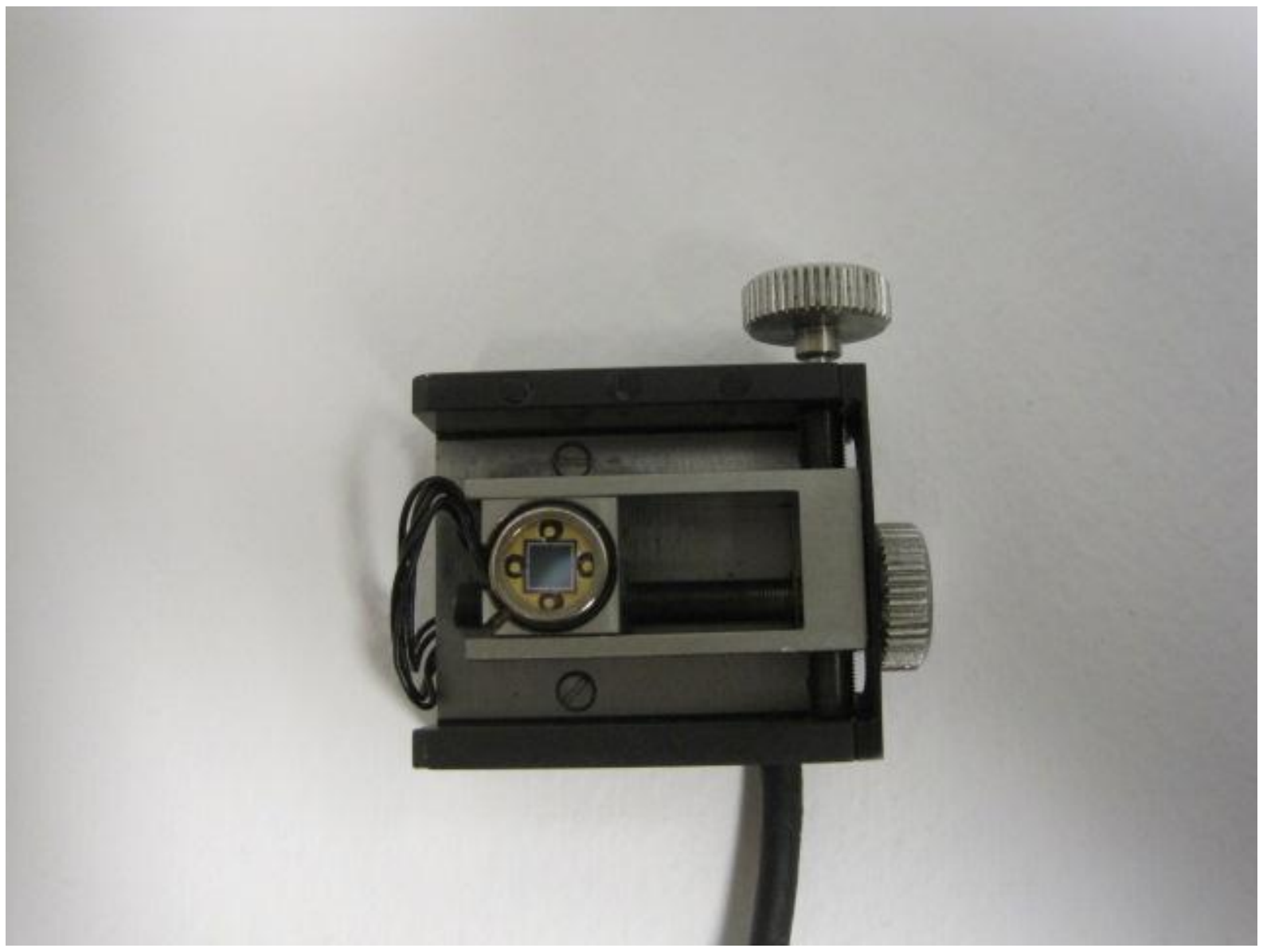

**Figure 2.1.3** Photodiode detector unit (bottom view) with photodiode chip “quadrant”

The connection diagram of the MAC mode system and for the system used in this work appears as the following:

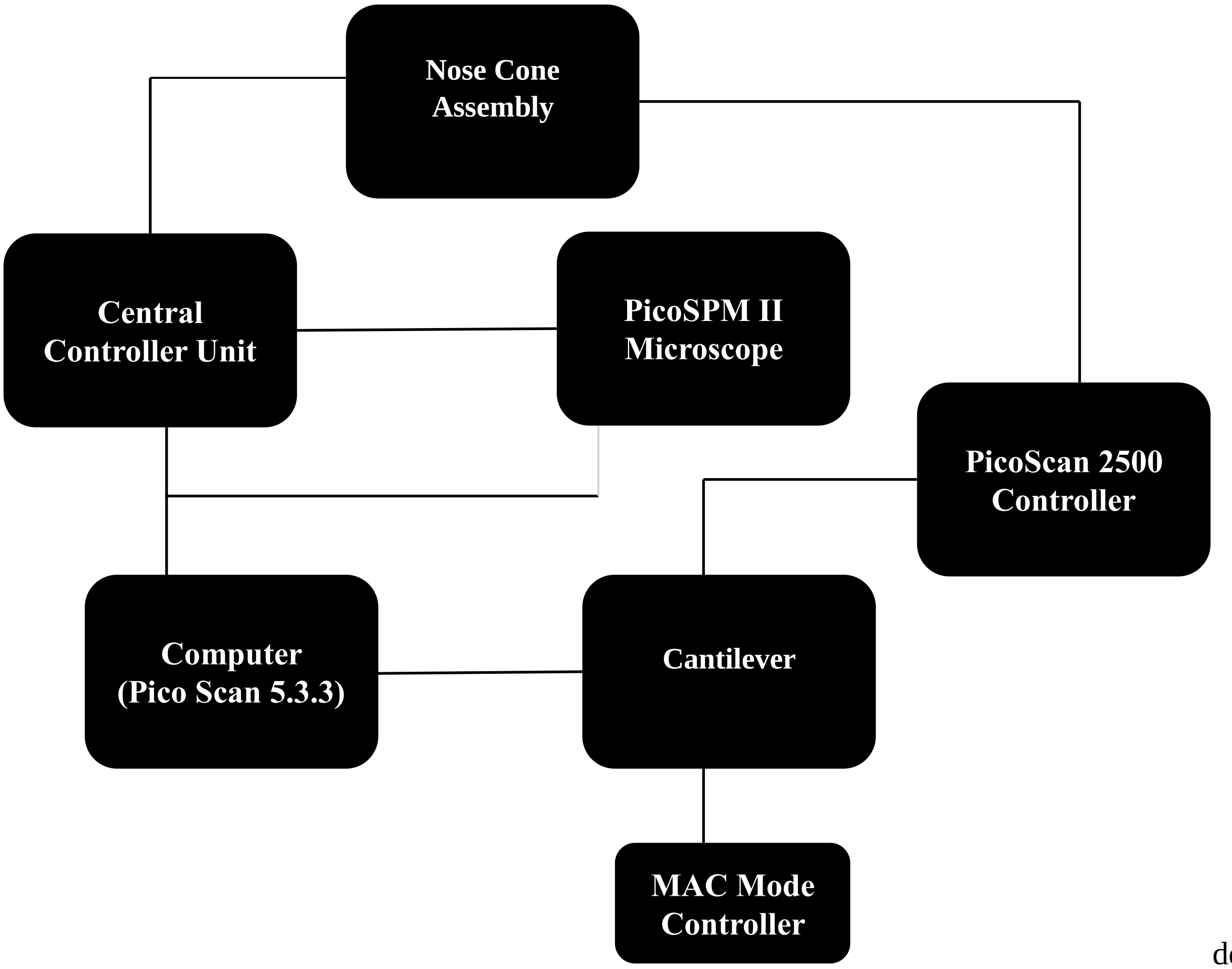

dd

**Figure 2.1.4** Connection diagram for the Pico Plus MAC Mode System

The Pico Plus SPM II microscope and video system are mounted together and housed in the Pico Isolation Chamber (Figure 2.1.5.) on a ~40 lb block suspended by bungee cords. The chamber is lined with foam to allow acoustic isolation as well as some isolation from electronic noise. When not in use, something like a phonebook is useful to keep the block propped up, thus preserving the spring constant of the bungee cords (resonance frequency ~1 Hz). The suspended block should be roughly 5-7” above the bottom of the isolation chamber. Since the microscope was not always entirely level on the block and due to the pendulum-style configuration of the AFM as well as the fact that it’s situated on the 4$^{th}$ floor of a building, polynomial flattening was often used during image acquisition to correct for the inherent drift and during image processing (next section).

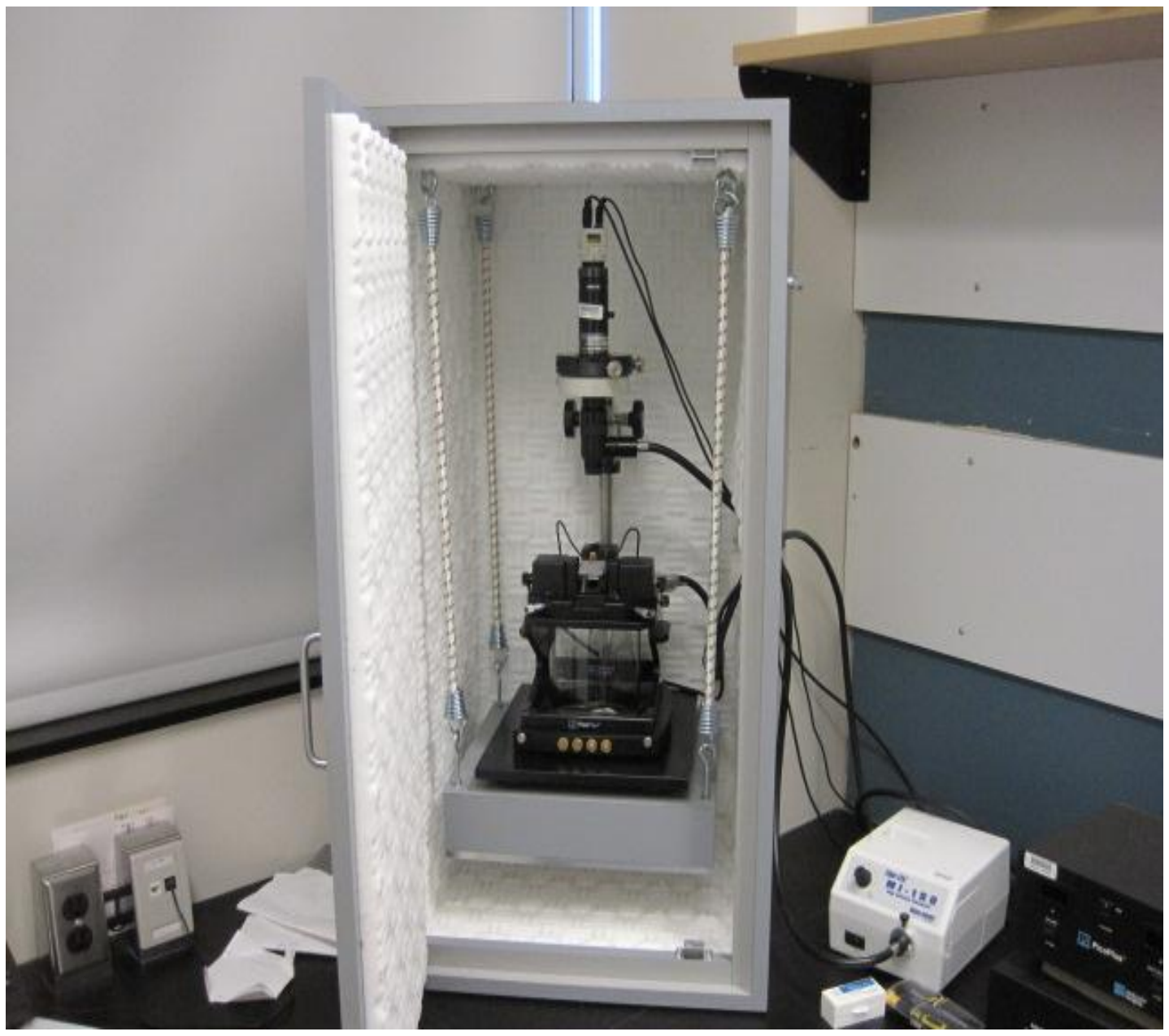

**Figure 2.1.5** Pico Plus Isolation Chamber used to house the SPM II microscope.

In addition to having the isolation chamber and ensuring proper daily maintenance of the microscope, the following conditions are ideal for obtaining high-resolution, *in air* AFM images: noise less than 60 dB (equivalent to human traffic and conversation within the room, noise from rush-hour traffic outside, etc); relative humidity 10-40% (it is ~20% in winter and ~35% in summer (according to building specifications); room temperature conditions (~23$^{o}$C); microscope should be isolated from direct sunlight and air pollution. Calibration of the piezo crystals is also necessary every so often to ensure that the open-loop system is providing reliable topographic data and this calibration is done by using a known standard sample provided by the manufacturer (annealed gold on Mica samples with steps of defined grading and pitch). Use of standard samples also serves to correct and/or reduce drift in the NC-AFM images that occurs due to lateral vibration of the building, improper tip and sample mounting, thermal and atmospheric effects, and noise. Since the system is open-loop, and since the MAC controller and microscope are grounded to their respective casings, grounding loops and coupling of electronic noise common in closed-loop systems was not very problematic throughout imaging.

### 2.1.2 NC-AFM IMAGE ACQUISITION AND TROUBLESHOOTING

NC-AFM images were collected in the Pico Scan 5.3.3 software originally provided with the instrument. At some point the software for the MAC controller was updated to another release (Pico View), which although a more user-friendly interface with a laser alignment plugin built right in, offered less control over the imaging parameters than Pico Scan 5.3.3. The following five tip parameters, as described previously (Chapter 1), were always important parameters worth noting before doing NC-AFM imaging in AC mode: Spring constant, *k*, of cantilever; RF, $f_0$, of cantilever; Q-factor, *Q*, of cantilever; oscillation amplitude, *A*, of cantilever; frequency shift (repulsive or attractive force profile), $\Delta f$. The first three parameters are defined by the tip (PPP-NCL: $k$=20 N/m, $f_0$=140-180 kHz, *Q* is very high but determined empirically with highest values found *in vacuo*). The latter two values could be adjusted as necessary during NC-AFM imaging (Figure 2.1.6). Other typical scanning parameters that can be manipulated during imaging are shown in Figure 2.1.6, with typical image buffers of topography (left and right trace), phase, and amplitude shown in Figure 2.1.7. Proper tip-mounting can ensure a smooth tuning process, which is the process of finding a cantilever's natural RF, that yields the proper RF with a resonance peak containing little to no baseline noise and a sharp, high-amplitude (5 V< A <10 V) response.

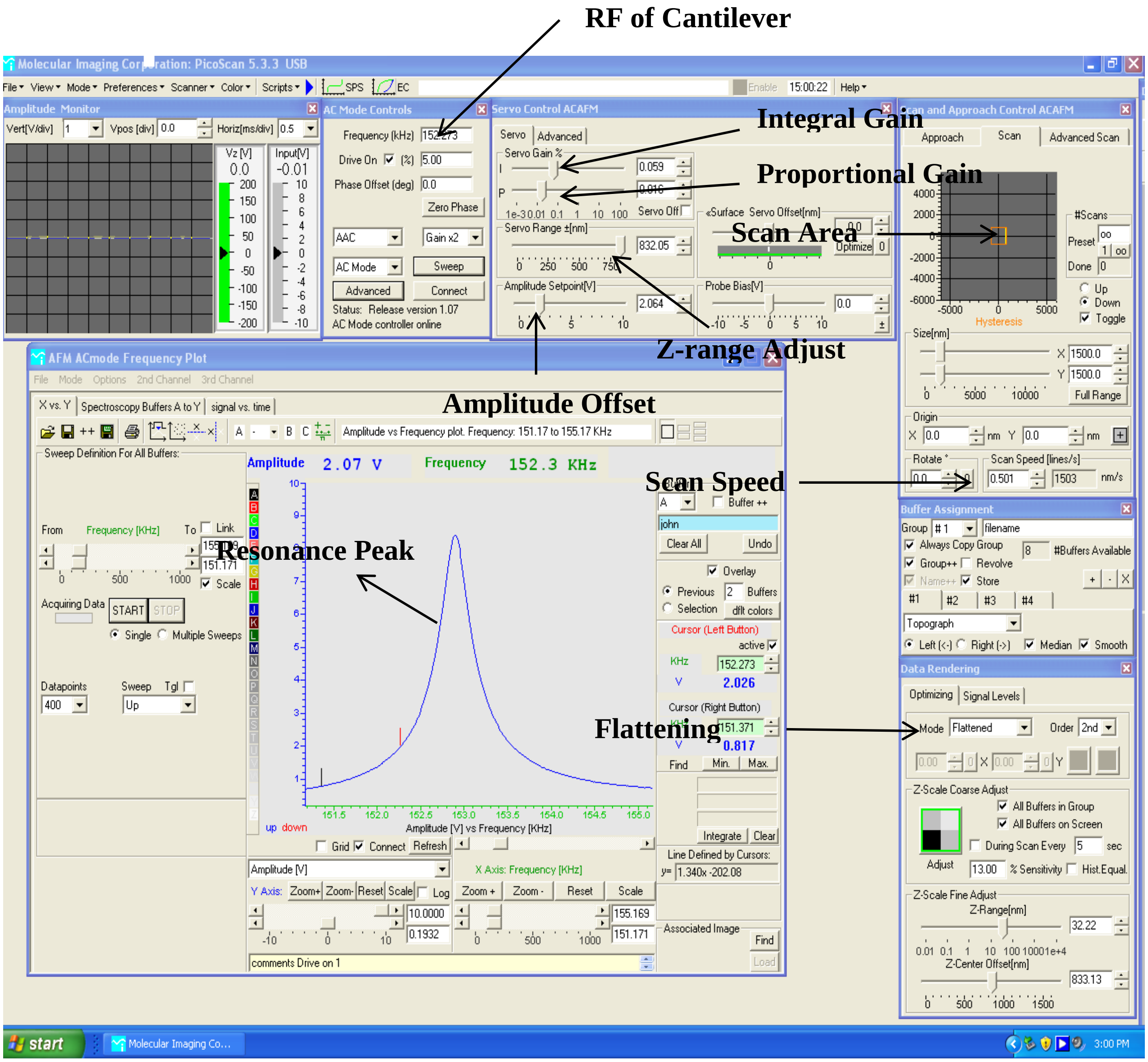

**Figure 2.1.6** Snapshot of the Pico Scan 5.3.3 NC-AFM imaging environment with labeled scanning parameters that often get calibrated in accordance to the type of tip and sample being imaged. In this case the SSS-NCL tip is being tuned (RF~152 kHz) before imaging. 5% drive is typically the default value for tuning cantilevers and is adjusted accordingly with the gain multiplier if the oscillation amplitude is above 10 V.

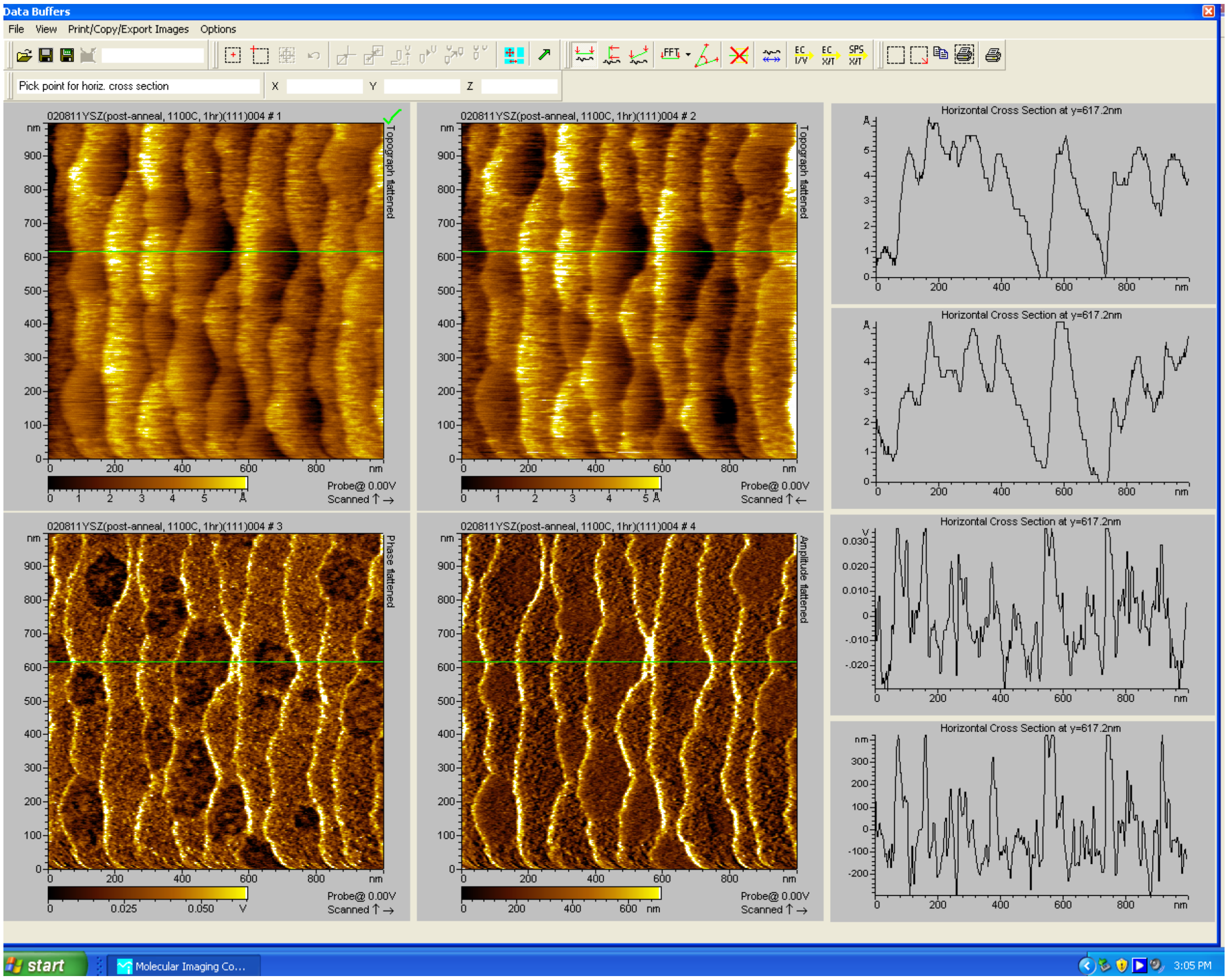

**Figure 2.1.7** Snapshot of the Pico Scan 5.3.3 NC-AFM imaging data buffers with top two panels for topographic data while the bottom two buffers provide phase (left) and amplitude (right) images. These images in particular were of a Type I YSZ(100) surface showing atomically flat pits after annealing to 1100$^{o}$C for 1 hr.

Often the tip-sample interaction is selected such that imaging is in the noncontact or intermittent contact regime, hence "NC"-AFM, and this is largely controlled by both the setpoint amplitude and the resonance offset (see red-line just left of resonance signal in Figure 2.1.6) of the cantilever. There are different schools of thought regarding true NC-AFM versus intermittent contact mode AFM, but for this thesis it is essentially repulsive, amplitude-modulated NC-AFM.

An *a priori* knowledge of the sample surface roughness parameters is always invaluable information when doing AFM imaging and since the samples in this study are all particularly "hard" single-crystal surfaces, the amplitude setpoint was often toggled between 80-90% of full extension where the tip is in contact. By slightly off-resonant imaging of the surface, that is, in the "repulsive" regime (just left of the resonance peak) relative to the free resonance of the tip, the tip-broadening effect (blunting) and tip-induced damage of the surfaces was avoided. Of course, the parameters are unique for every sample, but typical scanning parameters for high-resolution AFM images are given in the following table:

**Table 2.1** Typical NC-AFM scanning parameters for an annealed YSZ(100) surface with PPP-NCL probe (RF=169.42 kHZ, $A_{osc}$=4.632[a])

| Scan | Scan Area | Scan Speed (line/s) | I-Gain (%) | P-Gain (%) | Drive (%) | Setpoint Amplitude (V) | Z-range[b] (nm) |
|---|---|---|---|---|---|---|---|
| **1** | 2 µm x 2 µm | 0.350 | 0.012 | 0.022 | 1.000 | 4.289 | 501.73 |
| **2** | 1 µm x 1 µm | 0.650 | 0.001 | 0.022 | 0.840 | 4.977 | 450.02 |
| **3** | 500 nm x 500 nm | 0.800 | 0.010 | 0.022 | 0.340 | 2.941 | <400 |

[a] This value is related intrinsically to the setpoint amplitude (actual voltage based on tip-sample distance) by the physical setpoint established (in nm) (button not shown in Figure 2.1.6) during the approach, which is typically to stop cantilever at 90% extension.

[b] Typically varies sample to sample and depending on tip quality, but this value is always <832.05 nm (full z-range motion of the small scanner module) and if it appears >832.05 in servo window, the wrong scanner module is probably selected and scan areas <1 $\mu m^2$ will be very noisy.

Despite the expertise of the user, one is always faced with many issues of varying degrees of complexity concerning instrument calibration and image quality when doing AFM, and it would be an olympic task to capture each and every problem, so, instead, after extensive time on the system, a comprehensive list of errors was enough to generate a troubleshooting flowchart (Figure 2.1.8). The flowchart is split into two frames (red for hardware and blue for operational

`

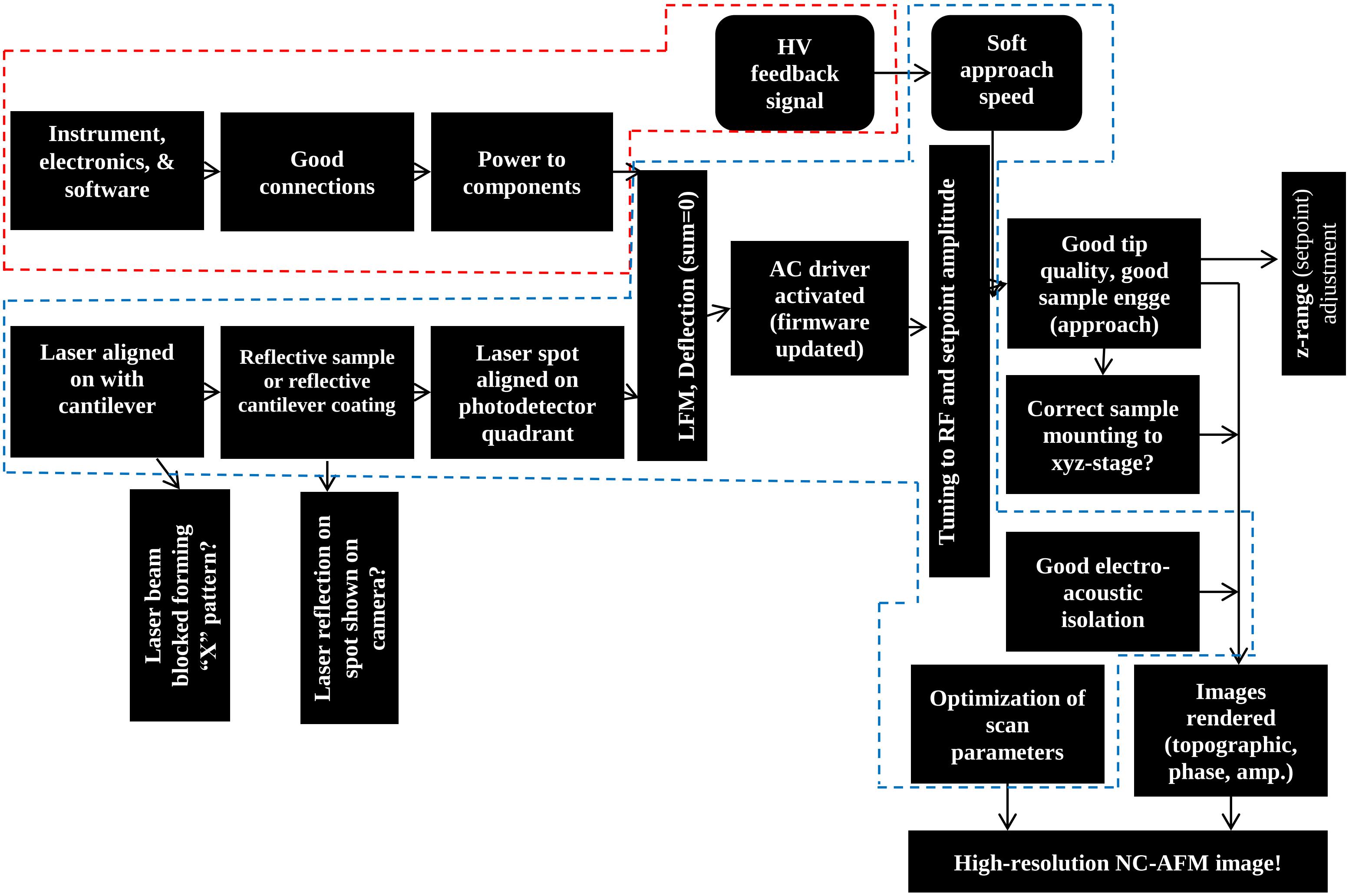

**Figure 2.1.8** Troubleshooting flowchart for Pico Plus MAC System operated in AC mode. Red frame is the hardware workflow and blue frame describes the more involved operational workflow.

parameters) and assumes that the ideal environmental conditions described earlier as well as correction configuration of the system (Figure 2.1.4) are maintained. Though time troubleshooting is mostly spent in the blue frame of the flowchart, the most common cause of poor image resolution was often preventable at the tip quality or tip/sample mounting steps. Also, improper control over the z-range, coupled with poor tip-sample geometry, often lead to false engages, blunting or contamination of tip, and overall noisy topographic images. A fine-tuning of the optimization parameters that are sample-dependent as well as a working knowledge of the expected surface features are therefore the ingredients to achieving optimal resolution for ambient NC-AFM images.

### 2.1.3 NC-AFM IMAGE ANALYSIS AND PROCESSING

NC-AFM image analysis including the following processes: grain analysis; levelling; FFT noise filtering; flood analysis; profiling; roughness calculations; streak removal, Gaussian smoothing; and matrix deconvolution were all performed using WSxM 5.0 Develop 4.1 software created by Ignacio Horcas, *et al*.[74] The images (*.stp file extension), all of which were collected in the Pico Scan 5.3.3 software, had compatibility issues with other SPM freeware such as Gwyddion and SPIP, so for the purpose of consistency, WSxM was used throughout the image analysis. It should also be noted that any grain analysis, roughness calculations, particle density calculations, or flood analysis (for HOPG-$Co_2O_3$ system in particular) were performed on the untreated, raw image files. As such, the data presented herein are the truest values without applying nontrivial corrections for tip-sample convolution and other calculations that require the absolute dimensions of the cantilever, and so a full tip characterization, in every case.

Keeping in mind the lateral (xy) resolution limit of ~20nm and axial (z) resolution of down to 0.1Å, the grain analysis was performed on the raw topographic NC-AFM images. In cases where the local contrast was sharp enough such that edge features of nanoparticles or nanoclusters was obvious, though, the phase channel images could be used for more precise lateral size calculations. The following topography image demonstrates the cluster height and lateral size measurement using the FWHM of a line scan. In terms of nomenclature:

“nanostructures” and “nanoparticles (NPs)” are used interchangeably throughout with “nanoclusters (NCs)” being 3D nanostructures or agglomerates of NPs.

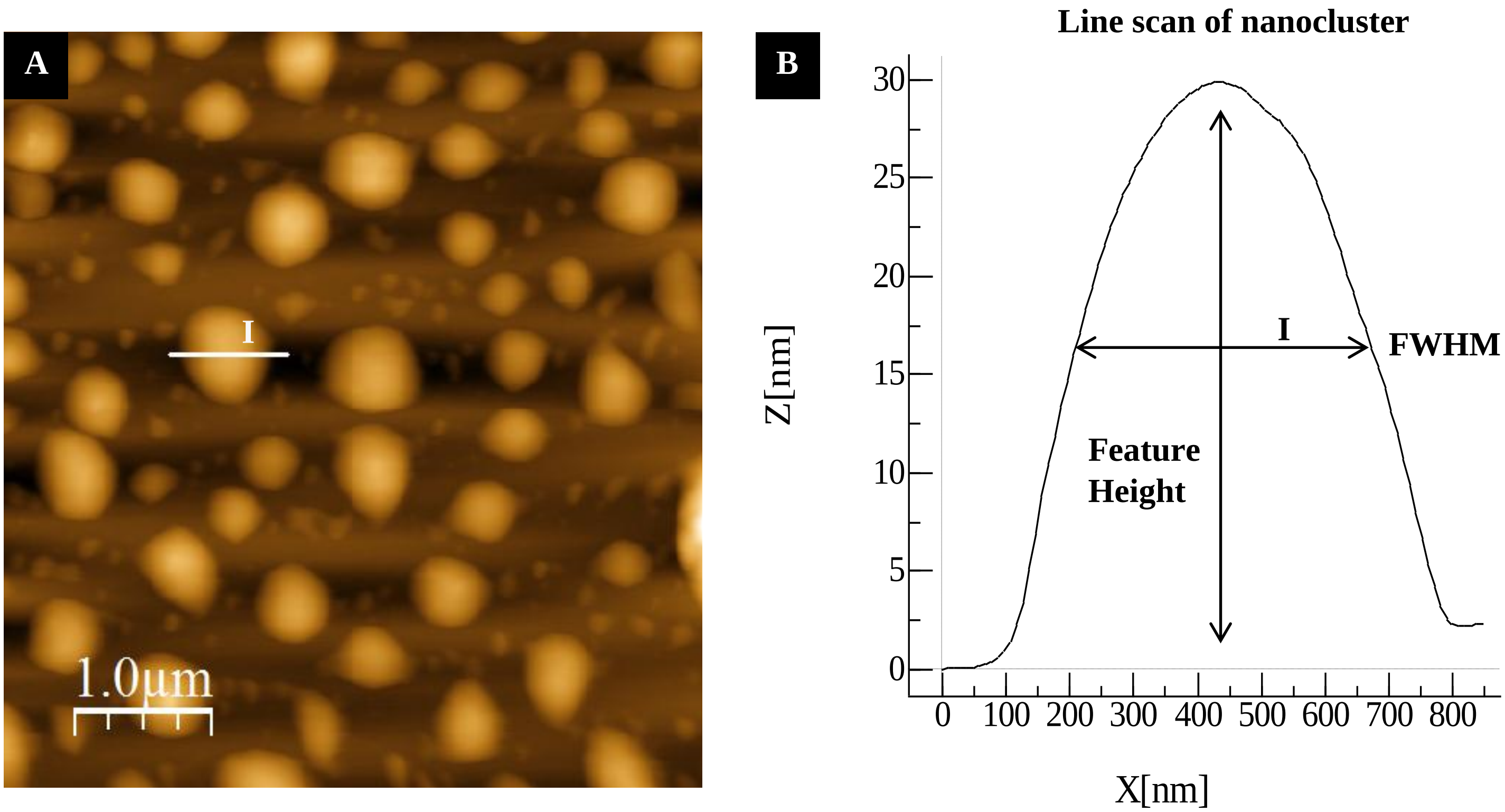

**Figure 2.1.9** (A) Topographic NC-AFM image of $Co_2O_3$-nanocluster on the YSZ(100) surface system. (B) Line profile corresponding to the trace “I” in A, demonstrating an example of the typical lateral particle/cluster size calculation during grain analysis of images. Note that for smaller nanostructures (<20 nm), the tip-broadening

Prior to analysis and often during NC-AFM image cycles, flattening of the image is required because of the “bowing” effect of many pendulum-style AFM scanners. Using the flattening algorithms in Pico Scan 5.3.3. was often sufficient for ensuring consistency between the trace and retrace (i.e. the raster pattern of the AFM tip in the left and right directions) data and, depending on the sample properties, often accounted for any drift effects. However, sample features can often cause tip-broadening or convolution effects as well as shadowing, which is evident by the darkened zones of topographic images (Figure 2.1.10). These shadow zones are perceived as changes in the local topography around large features, but they are often just artifacts resulting from the recalibration of the tip's setpoint amplitude and *z*-range after scanning the feature. Typically this can be calibrated and adjusted during the image cycle, but when it is not, secondary flattening of specific regions for a background correction can be performed in WSxM.

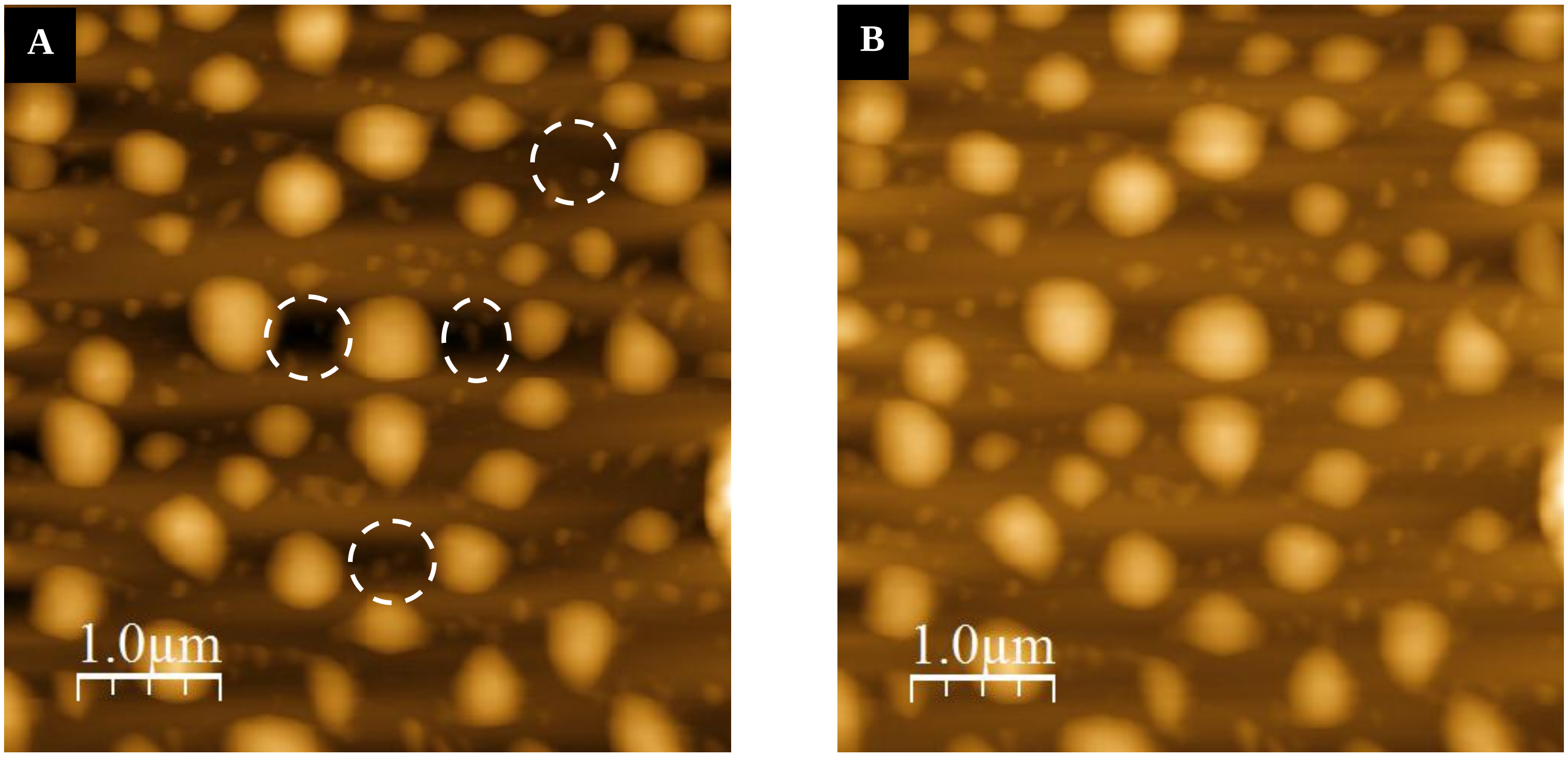

**Figure 2.1.10** Topographic NC-AFM images with A) Shadow artifacts present in AFM image ( enclosed by dashed white circles) and then B) flattened with respect to the highest image feature. Region-specific flattening in WSxM works by creating a filter that contains all regions not included in the flattening process and then applying a simple flatten to the regions not selected (i.e. the shadow artifacts). This flattening process allows for better local contrast of the features that end up in the shadow zones, which were caused by the tip-cantilever overshooting.

Two more common image treatments were for the removal of electronic noise artifacts in NC-AFM images by using Fast Fourier Transform (FFT) filtering and for removal of horizontal streaks or lines where the cantilever has drifted temporarily out of range during the scan. Figure 2.1.11 shows that electronic noise yields a well-defined peak and this can be effectively subtracted from the image by applying any filter (in this case a Hamming window was used) to filter out exclusively the higher frequencies. Mechanical noise often compromises image resolution so smoothing functions were used, albeit very infrequently due to the tradeoff between overall image resolution and the reduction of resolution in local features.[64]

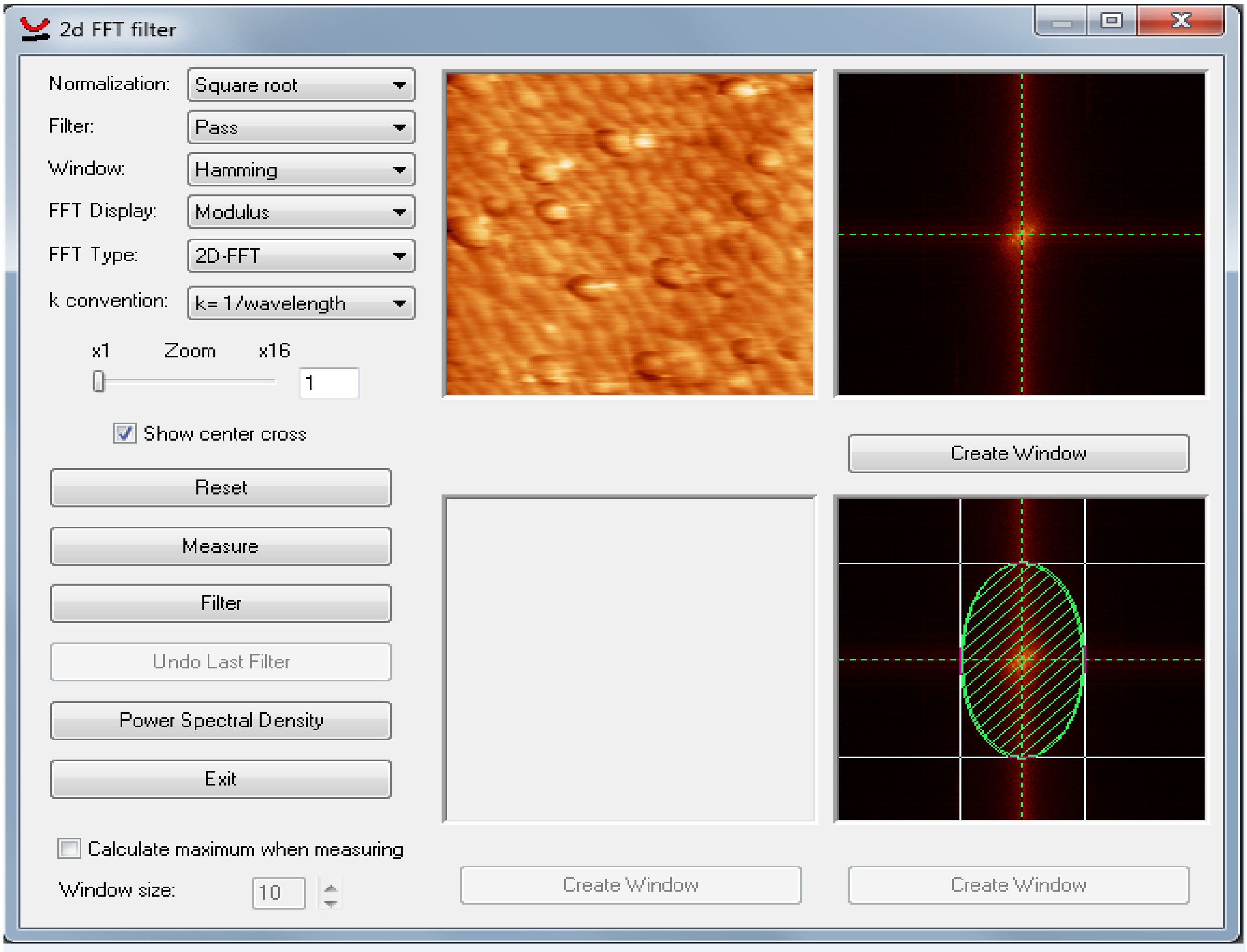

**Figure 2.1.11** Snapshot of the 2D FFT filter function in WSxM 5.0.

The FFT filter is the most computationally robust filter in terms of eliminating the type and frequency of noise that is observed. Typical point-spread mappings (Figure 2.1.11) often show low frequencies in the origin, but virtually any degree of noise can be reduced using low or high-pass filters, again, taking note of the aforementioned tradeoff with image smoothing/filtering.

Roughness parameters (Chapter 1) and calculations were performed in the WSxM software and are summarized comprehensively in the chapters to follow. Flooding analysis that was performed on the HOPG-$Co_2O_3$ system was used to corroborate the XPS quantification of $Co_2O_3$NPs and is essentially a function that allows flooding of the specific regions of an AFM image (flooding itself defined by minimum and maximum height parameters) and producing the % coverage of these flooded zones (i.e. $Co_2O_3$ feature) while omitting the background.

## 2.2.1 SAMPLE PREPARATION PROTOCOL

Single-crystal samples of YSZ(100), YSZ(111), MgO(100) were purchased from MTI Corporation, all with a nominal RMS roughness guarantee of < 5Å.[3] Single-crystal samples are cut from the boule using a diamond-wire saw with their defined orientations only having ±0.5$^o$ error. The chemical and mechanical polishing method is largely proprietary information but common for all oxide single crystals manufactured by MTI. Typical methods include using slurries containing inert nanoparticles such as silica to clean the surface, followed by a thermal anneal to 1000$^o$C to remove adventitious carbon over-layers. HOPG samples used in this study were purchased from NanoScience and mechanically cleaved for the purpose of surface cleaning (these samples were used as-received and not annealed) prior to AFM imaging. More particular sample preparation details are provided within results chapter for the clean single-crystal surface of YSZ(111) (Chapter 3.1.1). The dimensions of the MgO(100) samples used were the same as YSZ(111) while the YSZ(100) sample thickness and diameter was either 0.5 mm or 1 mm and 8 mm or 10 mm, respectively.

Ambient (in air) heating in order to obtain clean single-crystal oxide samples was performed in Lindberg/MPH tube furnace fitted with a modified Barber-Colman 7SD temperature controller.[75] Samples were put in alumina boats and covered in a specially made quartz tube during thermal treatments. The modification of the controller was done with the help of the University of Ottawa Electronics Department to allow a temperature ramp program to T > 800$^{o}$C. The controller itself works off a solid-state relay signal that essentially provides step-wise heating program (ramp rate~3$^{o}$C/minute). A current is instantaneously triggered between on/off states throughout the ramp cycle until the target temperature, as sensed by a K-type thermocouple, is reached and then this current is held for an amount of time proportional to maintaining the target temperature. The program is halted and triggered many times at this temperature if the target temperature is over- or undershot, respectively. Ideally a temperature program with more linear ramp-rate behavior is preferred, but for the simple thermal treatments done in this work, this configuration was sufficient.

### 2.2.2 EXPERIMENTAL: GROWTH OF $Co_2O_3$ ON SINGLE-CRYSTAL SUBSTRATES

Liquid-phase photochemical growth by nucleation on our single-crystal substrates (YSZ(100)/(111), MgO(100), and HOPG) was done in collaboration the Scaiano group at the University of Ottawa with Dr. Tse-Luen(Erika) Wee. The photochemical growth of $Co_2O_3$NPs on the single-crystal surfaces employs the initial photoreduction of $CoCl_2$ precursor salt in dry solvent (acetonitrile purged with Ar) in a clean glass vial containing the clean single-crystal substrate and the photoinitiator, Irgacure-907™ (I-907).[16,76] I-907 is known to be a potent reducing agent and ideal in producing the necessary α-amino radical to reduce the $Co^{2+}$ to $Co^{+}$ and then to its metallic $Co^{0}$ state (scheme in chapter 4). First, however, the specific α-amino radical must be generated by a Norrish I photocleavage, which is cleavage of the carbon-carbon bond between the α-carbon and carbonyl of the I-907 upon exposure to UV-A light (315 nm < λ < 415 nm)[77], producing two acyl radicals, one of which, viz., the α-amino radical is important in the photoreduction process.[16,78] The mechanism for this is depicted below:

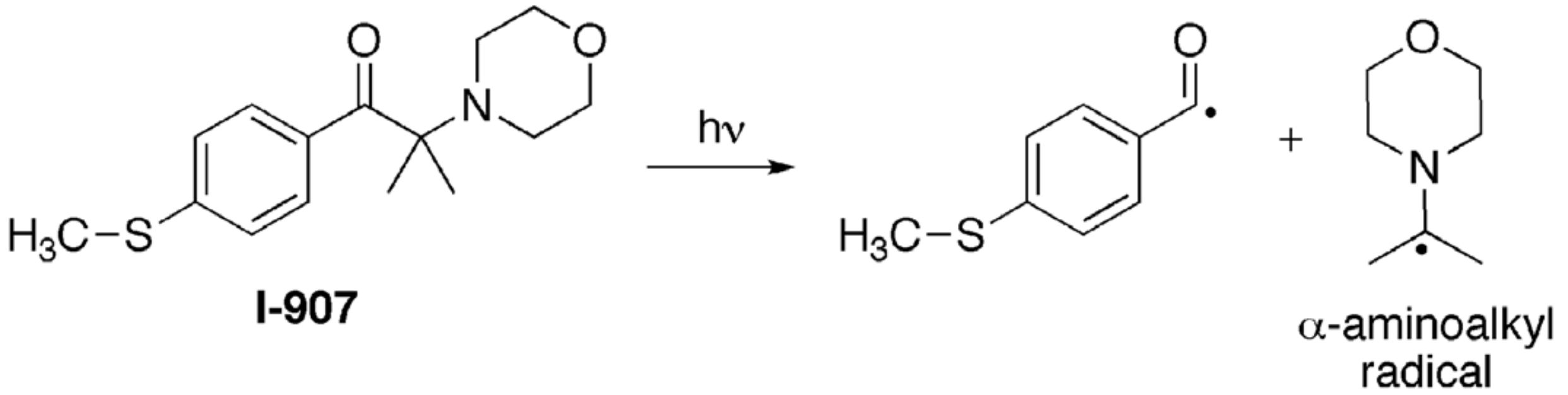

**Figure 2.2.1** Photocleavage of the I-907 photoinitiator by a Norrish I reaction, producing the α-amino radical necessary for reduction of the $CoCl_2$ to CoNPs on the single crystal.

The photoreduction of the transparent light blue $CoCl_2$ solution is evident by a colour change to dark green and then to a sharp blue-ish black colour as the CoNPs start to precipitate. [16] The CoNPs have been known to cluster in the Scaiano group's initial experiments on nanocrystalline diamond (NCD) due to their high paramagnetism,[76] but this clustering behavior is a function of the surface features and number of nucleation points, thus, different behavior on the single-crystal substrates is expected. Because of their high instability towards oxygen, the CoNPs or CoNCs that have grown on the surface undergo spontaneous oxidation to $Co_2O_3$ after they are taken out of inert (Ar) atmosphere and into ambient conditions. This was actually evident on initial samples with the YSZ(100) support that had excess $Co_2O_3$ coverage (with residual $CoCl_2$ salt solution) where the $Co_2O_3$ species appeared as a translucent white film. To correct this, loading concentrations of the $CoCl_2$ and amount of $Co_2O_3$ on surface were adjusted to 0.05 mM and 0.18 µg (1.05 nmol), respectively. Confirmation of the $Co_2O_3$ surface species is found via

XPS with grain size and morphology characterization done as per the NC-AFM method (AAC mode) previously discussed.

## 2.3.1 EXPERIMENTAL SETUP: XPS

XPS work on the post-growth samples in this study were characterized with primarily the Kratos Axis Ultra DLD XPS setup[79] in the CCRI and also on the custom designed system built by Specs Gmbh (Germany) for the Giorgi lab.[80] The latter system is part of a custom designed ultra-high vacuum setup (Figure 2.3.1) that houses XPS and ultraviolet photoelectron spectroscopy (UPS), low energy electron diffraction (LEED), quadrupole mass spectrometer (RGA mode), ion bombardment gun (particularly $Ar^+$) and metal evaporator setups. The other half of the chamber (RHK) houses the SPM setup for AFM/STM. Isolation of mechanical/acoustic noise is achieved via a set of vibration isolation air legs. A preparation chamber serving to be the load-lock as well as a homebuilt multiport cube (not shown) allow for sample loading and transfer via manipulator arms to the main chamber. The Specs and RHK chambers are separated by a gate valve system and the entire system is pumped down using a combination of turbo, roughing (backing), and ion pumps. Typical pressures of the main chamber are held at ~ 6.0 x $10^{-10}$ mbar with base pressures sometimes reaching down to 2 x $10^{-10}$ mbar after bake-out. It takes roughly 2 hours for samples to be transferred from the load-lock chamber to the Specs chamber for XPS analysis. Additionally, the manipulator arm also has the capacity to allow for resistive heating of samples, so for single-crystal samples that require vacuum annealing for cleaning or for studies *in situ* it can actually take up to a few days to properly mount the sample.

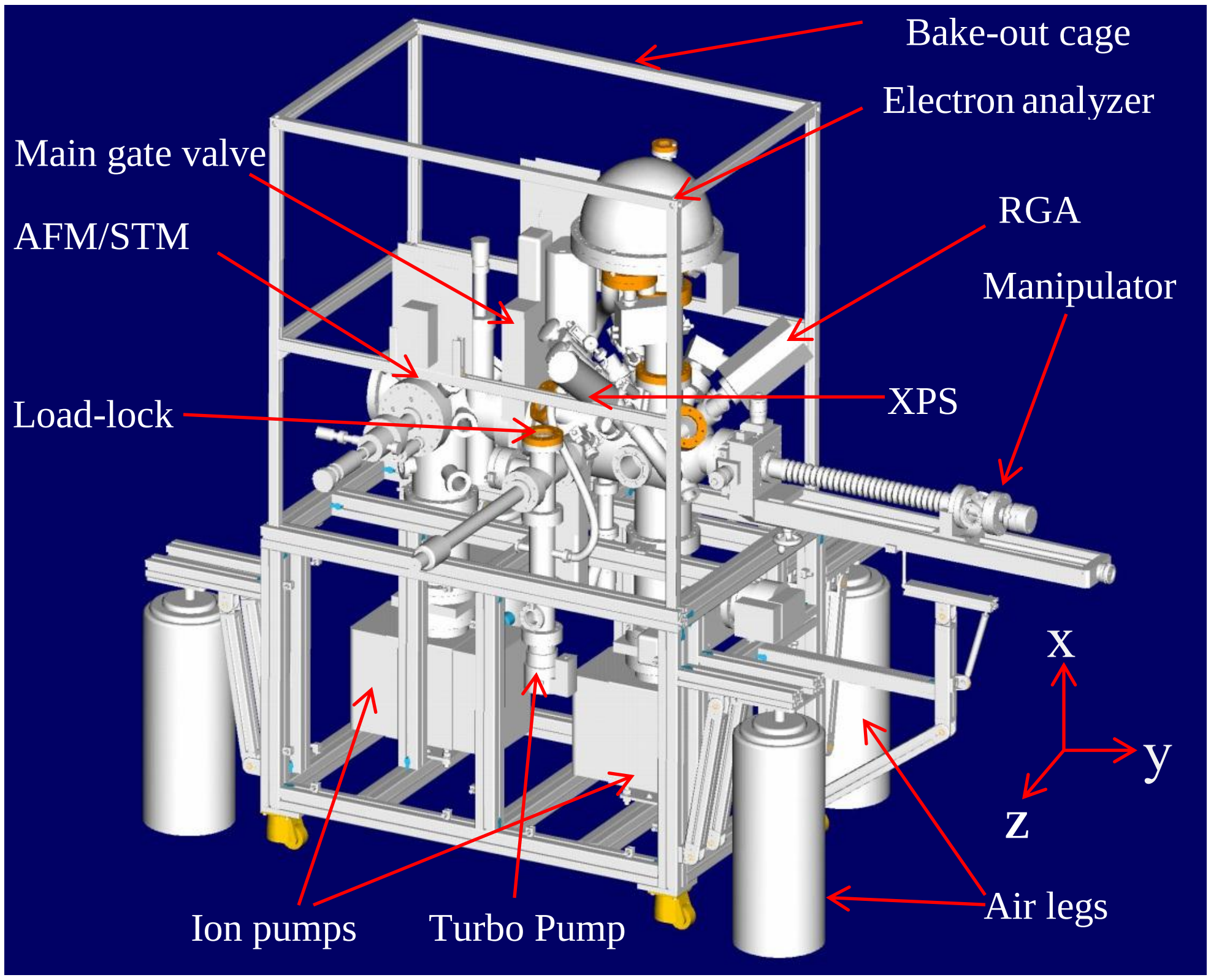

**Figure 2.3.1** Render of the Specs/RHK multi-technique UHV system

The commercial XPS system used has the following components:

i) X-ray source (XR-50) containing both Mg(300 W) and Al (400 W) anodes for $K_\alpha$ radiation of 12.53 keV and 14.26 keV, respectively. For this study we used the Al $K_\alpha$ source operated at ~14 keV (26 mA).

ii) Water cooling control unit for the XR-50 (CCX 50) which cools the X-ray source and has required flow rate of at least 2.7 L/min, which is interlocked with the "operate" function of the X-ray power supply. The chiller unit uses the building's recycled water supply and supplies the proper flow rate of water such that the control unit can operate under its 3-5 bar pressure requirement.

iii) The XPS system is powered by a Phoibos 100 SCD power supply and photoelectron detection was achieved with a hemispherical analyzer and detector unit (HSA, Phoibos100, SPECS). Detection of electrons that are kinetically excited within materials is achieved with a dynamic range for this analyzer being 0-3500 eV. The unit also contains an electron multiplier component that allows for enhancement of the electron count as well as overall sensitivity. Sample alignment is typically normal to the analyzer. Alignment optics are listed in the table below (Table 2.2) that also summarizes common XPS scanning parameters for this system and for the Kratos Axis Ultra DLD system that were used for the purposes of this particular study. The table is for survey and higher resolution component scans of specific regions.

**Table 2.2** Typical XPS acquisition parameters and lens functions for the HSA system. These parameters were used for Al $K_\alpha$ (14.26 keV) source with a nominal sample position (for RT) of {x=69.500 cm, y=11.245 cm, 11.625 cm, $\theta_{sampleholder}$=25$^o$, $d_{gun-sample}$=12.5 mm}

| **HSA Parameters** | | **Scan Mode (Specs Lab)** | | |
|---|---|---|---|---|
| **Entrance Slit Aperture** | 7 mm x 20 mm | **Scan Parameter** | **Survey** | **Component** |
| **Exit Slit** | Open | **Energy Step (eV)** | 0.2-0.4 | 0.1-0.2 |
| **Iris Aperture** | 32 mm | **Dwell Time (s)** | 0.3 | 0.4-3.0 |
| **Lens Mode** | Medium | **Number of Scans** | 1 | 2-10 |
| **Analysis Mode** | Fixed Analyzer Transmission | **Pass Energy (eV)** | 10 | 30 |

## 2.3.2 XPS SPECTRA ANALYSIS

Fitting procedures for the XPS spectra were all performed in the Casa XPS software package developed by Neal Fairley.[72] No one particular motif for analysis or peak fitting can be applied because of the varying nature of samples, but there is a general logistic structure that can be adhered to when attempting an accurate interpretation of spectra. The Kratos Axis Ultra DLD (CCRI) spectrometer was used for 3 of the 4 samples analyzed (all except YSZ(100)-$Co_2O_3$ system). Previous work using both spectrometers on a YSZ(100) support confirmed the resolution to be marginally better in the Kratos system (<0.08 eV).[12] Ultimately, for this study, quantification of the cobalt oxide species was necessary to confirm both the nature of the oxide as $Co_2O_3$ and to corroborate the post-growth particle densities calculated from the NC-AFM data. Quantification and fitting of the XPS data begins with the identification of regions (i.e. Co 2p, O 1s, C 1s, Zr 3d, Y 3d, etc). Upon selecting the appropriate peak envelopes from either the survey or component scan, a background subtraction is applied, which often invokes the Shirley type background, a weighted-averaging function that removes unphysical peak asymmetries.[81] Quantification of the carefully selected region is done after background subtraction using least-squares fitting of the selected peak envelope. Different functions are employed for the fitting and quantification of species in XPS and the line-shapes typically used are Gaussian-Lorentzian (GL) profiles. Usually a GL(30) is applied, means a mostly Gaussian distribution with n=30% Lorentzian peak shape. This value, though, is transient and is a question of the relative sensitivity factor (RSF) culled from the Scofield cross-section library in Casa as well as the FWHM area for that particular region.[72] Baseline calibration is often necessary, especially in the event of insulating surfaces (i.e. YSZ(100)/(111)), thus, before quantification the charge-compensated spectrum must be generated (in the case of YSZ, the baseline is calibrated with reference to the Zr $3d_{5/2}$ 4+ peak at 182.6 eV) or else all synthetic line-shapes and fitting parameters would be intrinsically incorrect for the sample. Ultimately, quantification is partly a function of the researcher's chemical intuition and also a working knowledge of the instrument's calibration to ensure removal of unphysical peak broadening. In ensuring the latter, it is well-documented[12] that the Ag $3_{5/2}$ peak envelope of sputter-cleaned silver yields an average FWHM of 1.14 eV, so any deviations from this would imply calibration errors, improper acquisition parameters, or damage to the intrinsic properties of the material. Any internal reference in lieu of silver would also be a plausible internal reference for calibration of spectrometer resolution, so long as there is consistency in the preparation and mounting of the sample as well as with the acquisition parameters used.

# CHAPTER 3 CHARACTERIZATION OF THE SINGLE-CRYSTAL SUPPORTS BY NC-AFM

The following two chapters comprises the NC-AFM work done on the cleaned (annealed) single-crystal supports (surfaces are prior to growth or "clean" samples). Imaging was performed in acoustic AC, amplitude modulated NC-AFM typically within hours after (*ex-situ*) thermal treatments. The series of insulating supports chosen have varying surface polarity/ ionicity from relatively nonpolar in the YSZ(111) to polar in the YSZ(100) to nonpolar-ionic in MgO(100). Finally, a conductive and relatively inert support, HOPG, was chosen as the last single-crystal surface used in this work for the photochemically induced growth of cobalt oxide nanostructures. Background information on the support surfaces as well as the extensive NC-AFM characterization and XPS/NC-AFM characterization of the single-crystal supports and nanostructure-support systems, respectively, follows this page.

*Please note that all NC-AFM image sets are presented at the beginning of the sections with all other analysis presented thereafter in the body of the text.

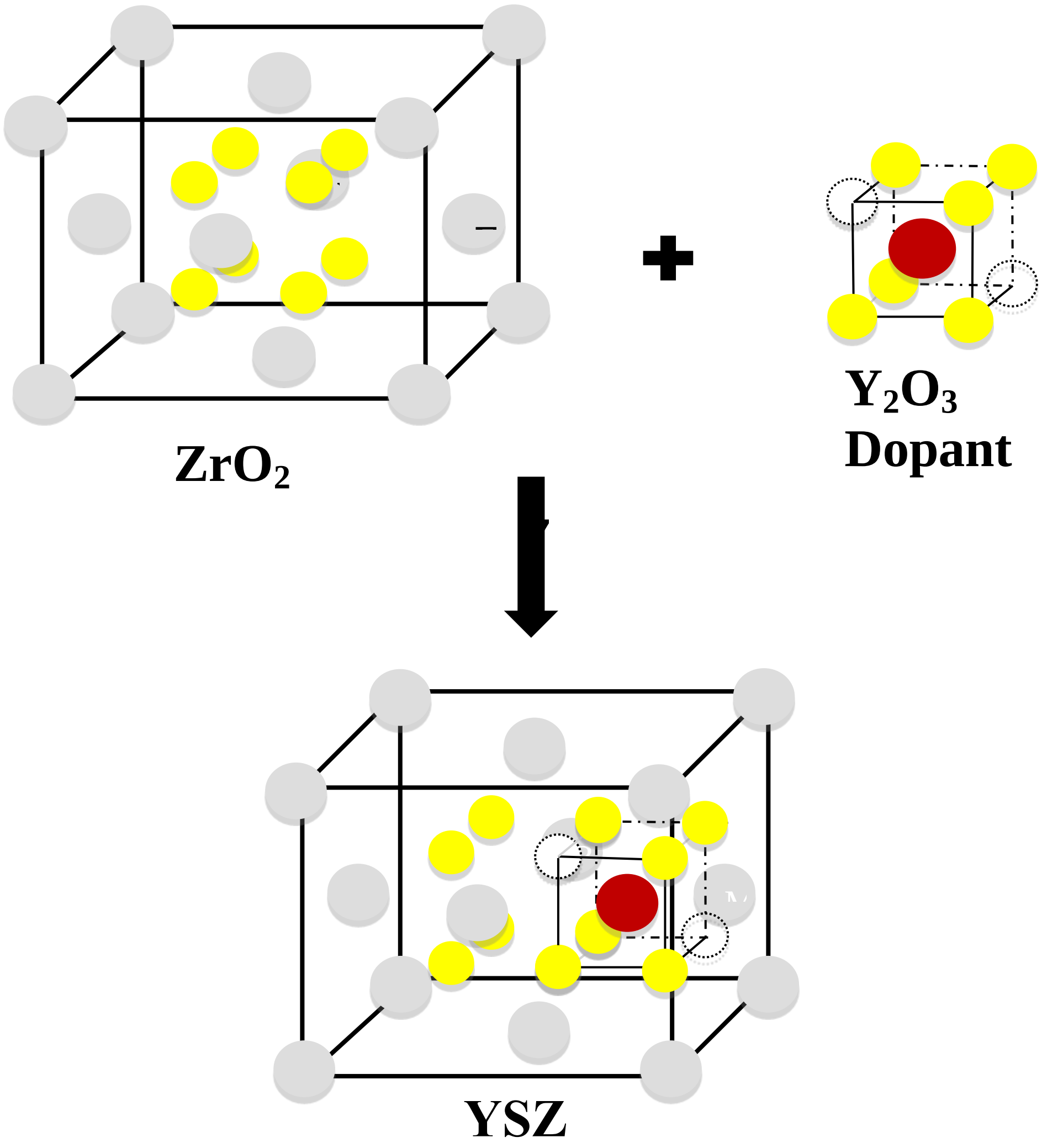

**Figure 3.1.1** Ball-and-stick sketch of stoichiometric doping of the fluorite crystal structure of FCC Zirconia by Yttria wherein one oxygen vacancy is generated for every Yttria molecule introduced, sometimes rendering the surface of YSZ(100) oxygen-terminated. The light greyish white balls represent $Zr^{4+}$, yellow balls represent $O^{2-}$, red ball represents $Y^{3+}$, and dashed disks represent $O^{2-}$ vacancies. This is the stoichiometric (ideal) substitution of $Y_2O_3$ into the fluorite crystal structure ($CaF_2$) of $ZrO_2$

Yttria-stabilized Zirconia (YSZ) is an insulating material used most ubiquitously as in solid oxide fuel cells (SOFCs) and oxygen sensors due to its high ionic conductivity properties. Upon doping a Zirconia ($ZrO_2$) lattice with ~10-12% Yttria ($Y_2O_3$), vacancies at the oxygen sites of the $ZrO_2$ lattice (Figure 3.1.1) are introduced and these vacancies allow for ion conduction channels to form in the material, giving it relatively higher conductivity under certain doping, temperature and pressure conditions.[12,82–93] This means a band-gap for the YSZ material has various lower and upper limits between 4-6 eV, but is still classifiably a high band gap, insulating material.[94,95] Though an exciting, complex material, its insulating nature makes it an ideal candidate for NC-AFM surface topographic studies as a lot of its reconstructive and unpredictable surface properties can be traced right back to its basic structure. In YSZ, the lattice of $ZrO_2$ is typically doped by the acceptor oxide, $Y_2O_3$. However, a multitude of others have been employed in other reports, particularly in the application of improving ionic conductivity in SOFCs and some examples include ScSZ[96], CaO[92], and MgO[97]. Virtually all reports for YSZ indicate that the highest ionic conductivity is in fact achieved at 8-9 mol% of $Y_2O_3$, which is effectively also the concentration that fully stabilizes the Zirconia.[82–92,98,99] The face-centered cubic (FCC) lattice (unit cell parameter, a=5.124 Å [100]) of $Zr^{4+}$ with 4-13 mol% of $Y_2O_3$ invokes distribution of oxygen vacancies into the lattice, but the location of these vacancies remains a point of contention. Actually, the sample-to-sample variability of single-crystal YSZ surfaces often raises more questions than it answers, which is why the body of experimental work performed on it often deviates from the computational/theoretical studies.[92,101–105] Much of the stability of any metal oxide surface, especially YSZ, is predicated on the local environmental conditions like dopant concentration, temperature, and defect structures, so it makes the task of predicting stability of the surfaces quite complex. This complexity is also effectively a motivation for this study as local surface topography work using NC-AFM still remains critical and largely difficult from both the point of view of NC-AFM and of spectroscopy since the measurement probe in either case may or may not induce changes to local surface features. Nevertheless, NC-AFM is a nonaggressive surface science tool that should provide the most accurate characterization of local surface features on the highly reactive surface supports used in this thesis. This particular work is with two different crystal cuts of YSZ as single-crystal supports for growth of nanostructures and because the two cuts of (100) and (111) surface plane orientations are polar and nonpolar, respectively, growth behaviour of nanostructures should vary according to the spatial and chemical information generated from the NC-AFM topographic and phase channel data, respectively.

### 3.1.1 THE NONPOLAR YSZ(111) SURFACE

The first single crystal surface under investigation was that of the YSZ(111) surface. As indicated by the ball-and-stick model of YSZ(111) in Figure 3.1.1, the pure stoichiometric YSZ(111) surface has a fcc(111) surface plane with a surface plane parameter a=3.62 Å (note that this is not the bulk unit cell parameter, a=5.124 Å of YSZ) for hexagonal closest packing (hcp), which, experimentally, is the most stable surface configuration for YSZ.[1-4] The surface of this plane, as visualized along the c-plane(“basal” plane) shows hexagonal closest packing arrangement with respect to the $Zr^{4+}$ and $O^{2-}$ (Figure 3.1.1) As mentioned, all YSZ single-crystal surfaces are inherently reactive and also quite defective due to the random distribution of the $Y_2O_3$-induced vacancies, however, the (111) orientation is still the most densely packed among all surface types in accordance to clean, ideal crystal surfaces (see chapter 1); this would inevitably have consequences during the growth of $CoO_x$-nanostructures.[106,107]. As mentioned, it is speculated that YSZ(111) surface is more stable[108], with less low-coordinated sites that give rise to surface defects, as compared to other associated surfaces like the YSZ(100) or MgO(100).[109] The nonpolar YSZ surface, especially after annealing treatments, exposes the low energy (111) face readily and, though cubic fluorite-like structures of (111) orientations are almost always polar, the oxygen diffusion through YSZ during annealing processes induces relatively "nonpolar" surface stability.[28,110,111]Of course, this is easier to prove computationally by inducing the reconstruction of the most stable surface in the presence of O-vacancy-induced defects, but a challenge is, however, associated with characterizing this experimentally and whether the YSZ(111) surface itself can be easily made atomically flat. The characteristics of both the atomic scale periodicity and the frequency of defect sites, both of which require the atomic resolution of UHV NC-AFM/STM.[109,112] Local surface topography data in the NC-AFM, particularly RMS roughness, can provide the geometric corrugation and, if optimal resolution is obtained, atomic corrugation in some cases. NC-AFM images of the stepped, hill-and-valley surface structure of clean YSZ(111) should provide insight how this surface could be of use as a model support for photochemical  growth of nanostructures on surface defects.

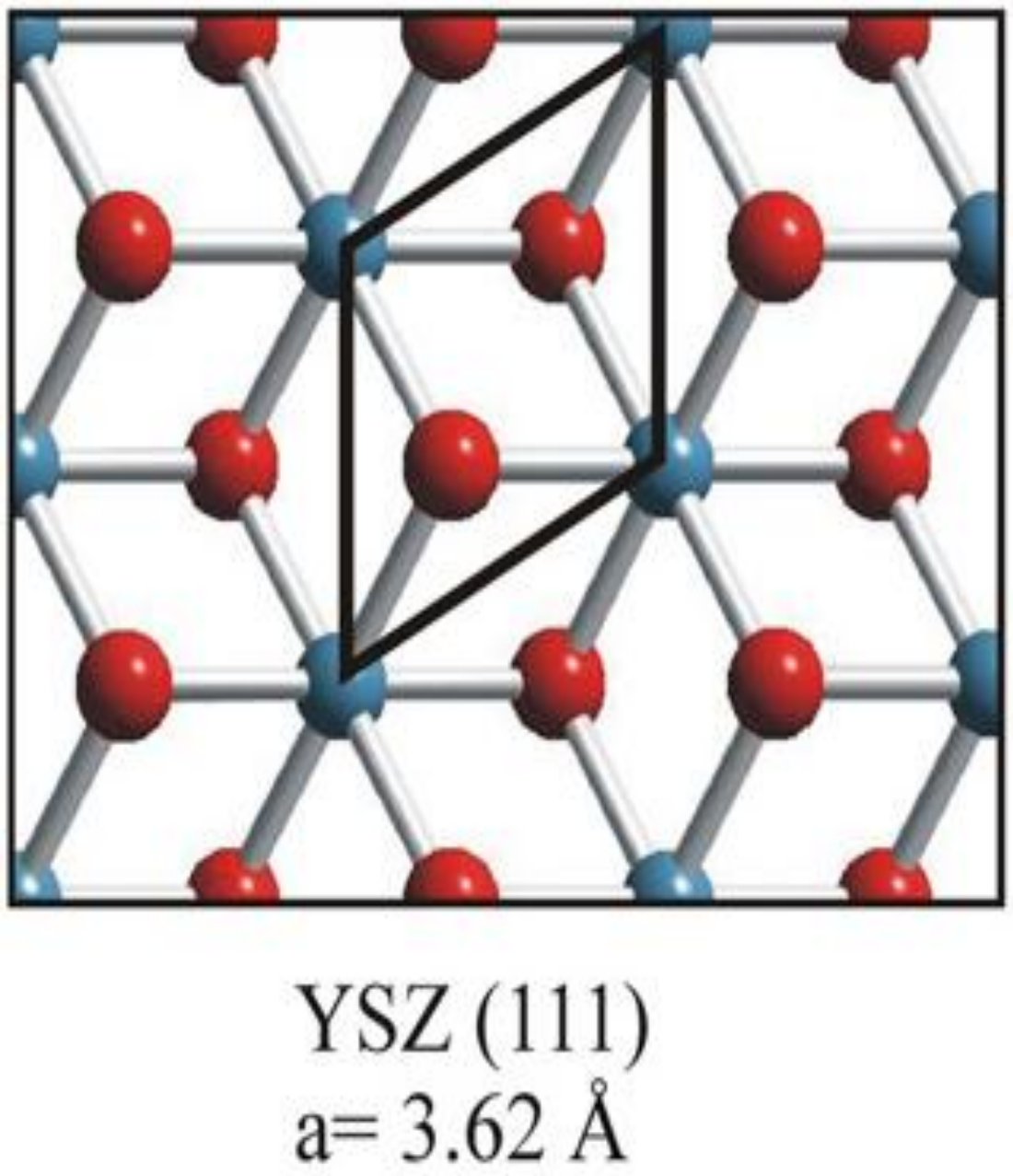

**Figure 3.1.2:** Ball-and-stick model of the YSZ (111) surface with associated hexagonal unit cell parameter of a=3.62 Å.[113]

The following figure depicts an analogous atomically flat surface for (111)-Samarium-doped Ceria (SDC):

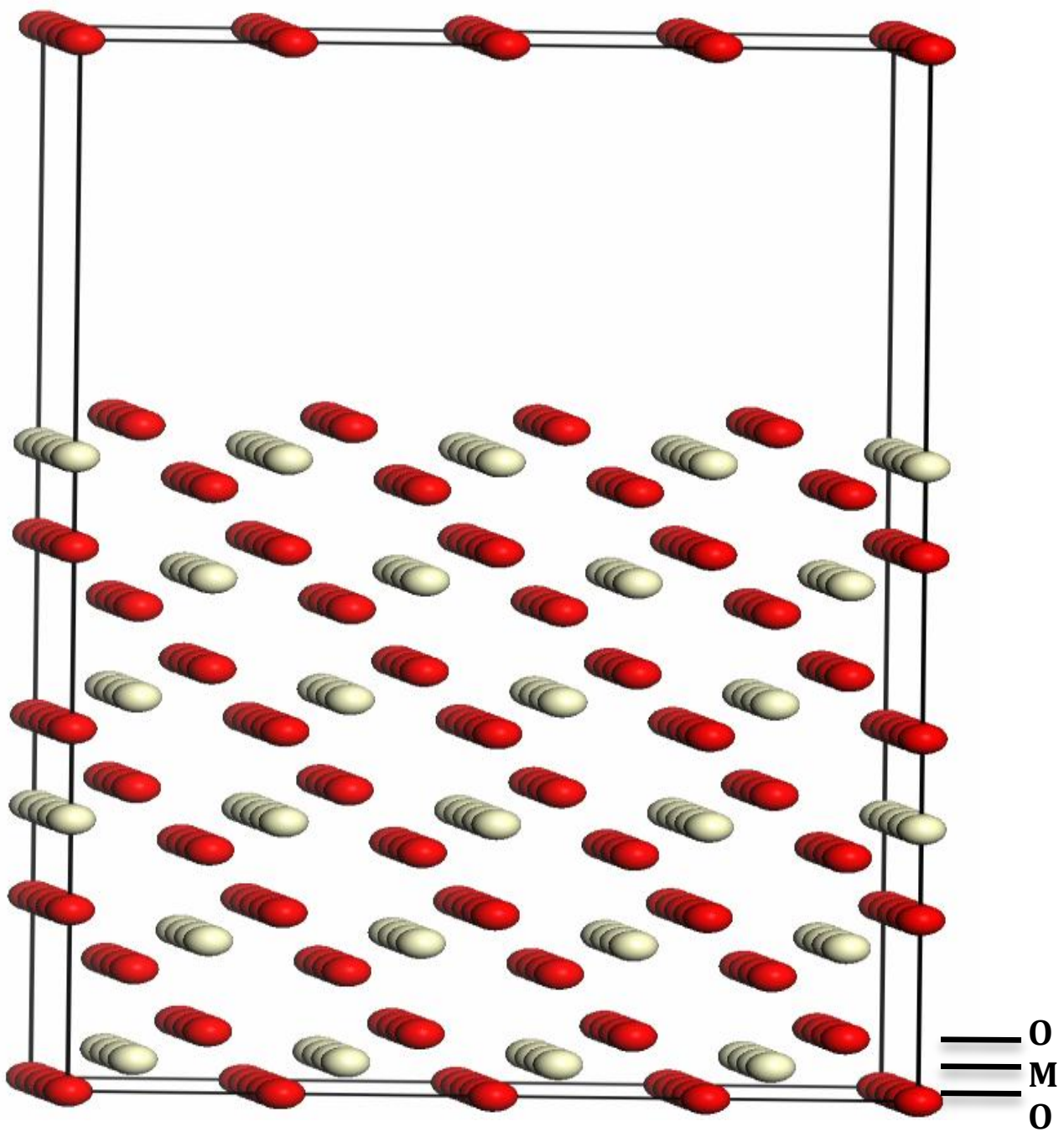

**Figure 3.1.3** Side view (slightly perspective) of (111) surface plane (for Samarium-doped Cerium surface in this case). This is an analogous crystal surface plane for $ZrO_2$(111) (not yet doped with $Y_2O_3$). Red and white balls are for oxygen and any metal, respectively. The Oxide-Metal-Oxide (OMO) layers make up the bulk material and often reconstruct to stabilize the surface energy.

Figure 3.1.3 offers a glimpse into the nature of the (111) surface as it is seen without defects, however, experimentally, this is often not the case and renders the surface unstable and also more reactive. The geometric corrugation of relatively defect-free surfaces such as SDC(111) or $CaF_2$(111)[109,114] can sometimes differ drastically than the atomic scale corrugation of the some material due to the presence of atomic defects. In the case of low-index cuts of YSZ single-crystals the story is very much the same (especially true for (100)-orientation, as we will see), and since atomic corrugation is beyond the scope of ambient NC-AFC performed in this study, the relative number of defects resolvable on the surface as a function of thermal treatment is helpful in predicting defect-induced growth behaviour of nanostructures and, eventually, could have implications for development of kinetic growth models.

### 3.1.2 SAMPLE PREPARATION AND THERMAL TREATMENT: CLEAN YSZ (111)

As is the case for any single crystal surface, the surface's features and morphology depend strongly on the sample history. Both square (1 SP) and round (2 SP) YSZ(111) samples (13% $Y_2O_3$) of 8 mm diameter and 0.5 mm thickness were purchased from MTI Corporation. In their well-established cleaning and mechanical polishing procedure (CMP), the samples are cut from a boule ((111) ± 30$^o$ according to manufacturer) using a precise diamond wire saw and subsequently polished in a number of $Al_2O_3$ and $SiO_2$ colloids. Finally, an initial heat treatment to 1000$^o$C is performed to remove the adventitious carbon over-layer that is created from the CMP process. This was done by putting samples in alumina boats and putting in a quartz tube furnace open on both ends to air.

Other cleaning procedures for these kind of single crystals involve $Ar^+$ sputtering or sonication in acetone processes, but their efficacy in leaving reproducibly clean and periodic surfaces is ill-defined.[115] To some degree, surface variability between samples can be corrected with extensive thermal annealing cycles at temperatures between 900-1200 $^oC$ and serve to "prepare" the clean surfaces prior to metal oxide growth. Because the manufacturers (MTI Corp.) initially annealed the samples to 1000 $^oC$ to meet the 0.5 Å < RMS < 10 Å guarantee, a number of the as-received samples (~1/4) showed parallel terraces and reduced noise in the ambient NC-AFM imaging comparable to well annealed surfaces. Being a highly reproducible surface with a well-documented annealing behavior, preliminary imaging of only a 3-4 samples was required before selecting the adequate sample for growth experiments.

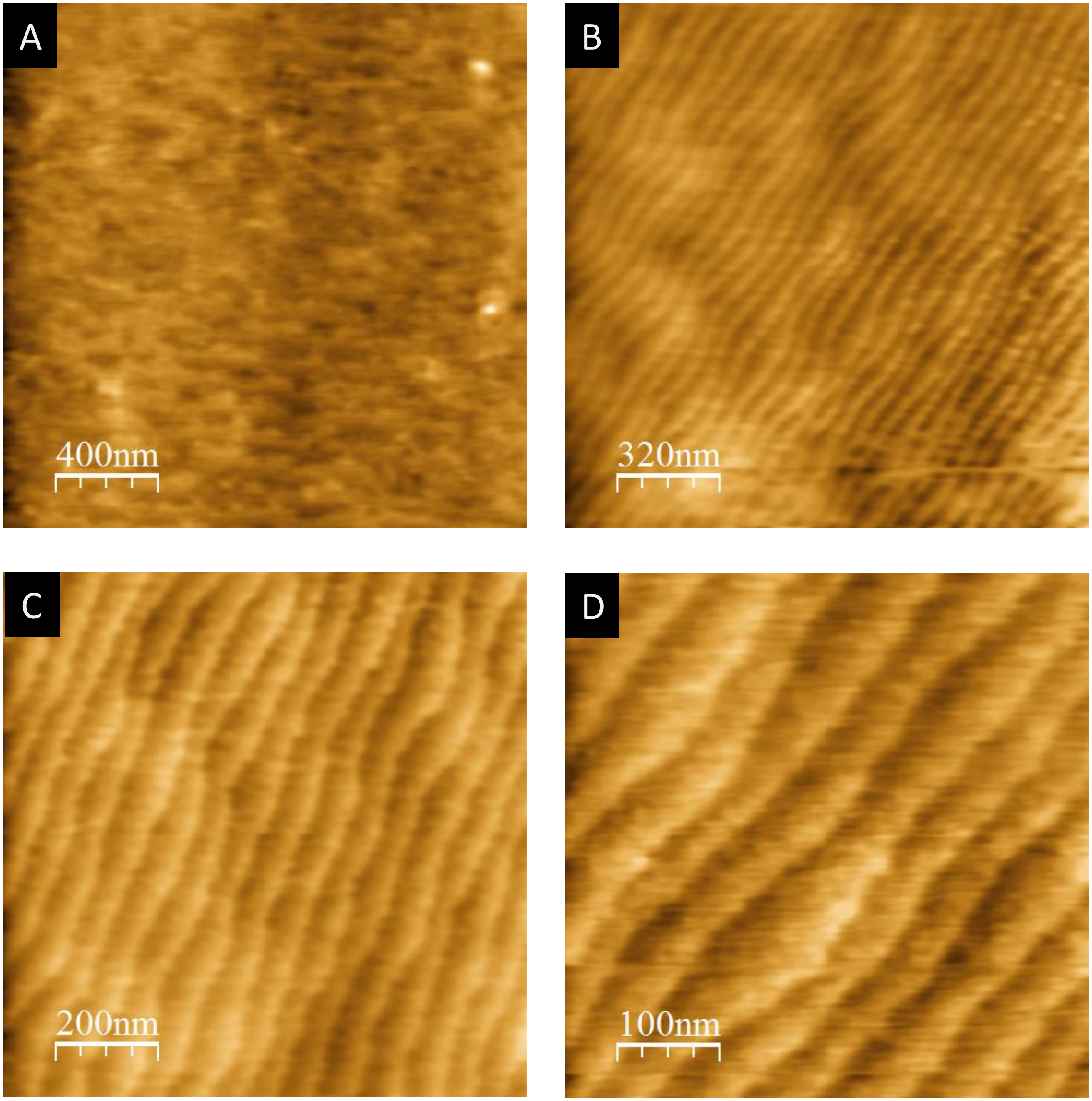

**Figure 3.1.4** High resolution NC-AFM images of the clean, pre-growth YSZ (111) single-crystal surface **A)** 2.0 µm x 2.0 µm micrograph of as-received single crystal (nominal z-range=5.221 Å). Micrographs **B)-D)** depict sample after ambient thermal cycle in air at 1000 $^{o}$C for 1 hr. **B)** 1.6 µm x 1.6 µm micrograph of post-annealed sample (nominal $z_{range}$=4.485 Å). **C)** 1.0 µm x 1.0 µm micrograph of post-annealed sample (nominal $z_{range}$=4.565 Å). **D)** 500 nm x 500 nm micrograph of post-annealed sample (nominal $z_{range}$= 3.727 Å)

NC-AFM imaging of 5-6 as-received single crystals of YSZ(111) was performed before 1-2 of them were selected for further surface preparation. Despite the roughness guarantee of <10 Å (often < 5 Å) established by the manufacturer, the as-received sample typically contains surface contaminants and extrinsic defects that impose a challenge on NC-AFM imaging, especially if there are interesting surface features. Naturally, it was important to select the appropriate scan parameters when imaging and filters during image processing (see previous chapter to details proper). "Soft" approaches were used for imaging of any as-received samples because of the high RMS of the surfaces. This was done by ensuring that the approach ceased at ~80% of the set-point value for that particular scan and that the scan was taken in the repulsion (or "deflective") regime of the tip-sample force profile. This effectively makes the cantilever tip less susceptible to the following: crashing; "false approaches"; unwanted interactions with adsorbate over-layers; tip contamination prior to imaging of the cleaned samples. Because amplitude parameters are the principal parameters in any AFM study, they require delicate control, thus it was important throughout the imaging cycles to maintain an understanding of how to adapt these parameters to the type of sample being imaged. The nominal z-range for the as-received sample was 5.221 Å with an associated RMS (and geometric corrugation) of 1.006 Å, well within the RMS limit reported by MTI Corp. The effect of the surface adsorbates and contamination on surface cleanliness is reported comprehensively elsewhere,[11] but horizontal streaks and other effects due to mechanical noise in the image were eliminated using low-pass and Gaussian filters.15

Being strictly an ambient NC-AFM study, thermal treatment of the as-received system was performed to effectively remove surface contaminants and defects while introducing some reconstruction to the surface to allow it to be both thermodynamically stable and reproducible.

Since the ambient annealing cycles were done in an oxygen-rich environment, the delicate harmony of limitations imparted by the instrument and by the surface variability itself would make it impossible to study the reconstruction or annealing kinetics explicitly. A study of this nature would be restricted to a UHV-study that employs techniques that would warrant control over partial pressures of oxygen and the ability to study the concentration of oxidation-induced defects *in situ* (ie. UHV-STM).[93] The task of studying surface reconstructions of highly reactive, acidic metal oxides like YSZ as a function of annealing temperature and time by ambient NC-AFM requires atomic resolution, which is next to impossible.[63]

Topographic images (Figures 3.1.4b-d) of the post-annealed sample reveal that after heating the samples for extended periods, the clean surface develops these homogenously distributed terrace features. As mentioned, this was seen in numerous samples ($n \geq 6$) and is a direct result of the annealing treatment. It was observed that these terraces would not form at temperatures < 700 $^{o}$C and typically at least 1 hr was required above this threshold to notice evolution of any surface features at all. Referring to Figure 3.1.5, the average terrace width for the YSZ(111) surface was found to be 42.8 ± 7.4 nm with average step heights of 1.9 ± 0.8 Å, which is equal to just over $2/3d_{(111)}$ ($d_{111}$=2.98 Å, which, by equation 1.29, is interplanar spacing parameter for cubic $ZrO_2$ and corresponds to one OMO-layer). Though nothing can be said about "atomic" flatness, the post-annealed sample had a geometric corrugation vis-à-vis RMS roughness calculations of 0.832 Å, comparably less than that found for the as-received sample of 1.006 Å. As it is quite evident that they are not completely linear, the terraces adopt this undulating surface pattern and this was ascribed to the crystal manufacturing itself and step-edge effects, but the true {111} fcc-plane is still exposed. Ultimately, reconstructive processes and diffusion paths for oxygen to the surface during annealing did not play an obvious role in terms of defect formation and this was evidenced by the appreciably small decrease in surface roughness.

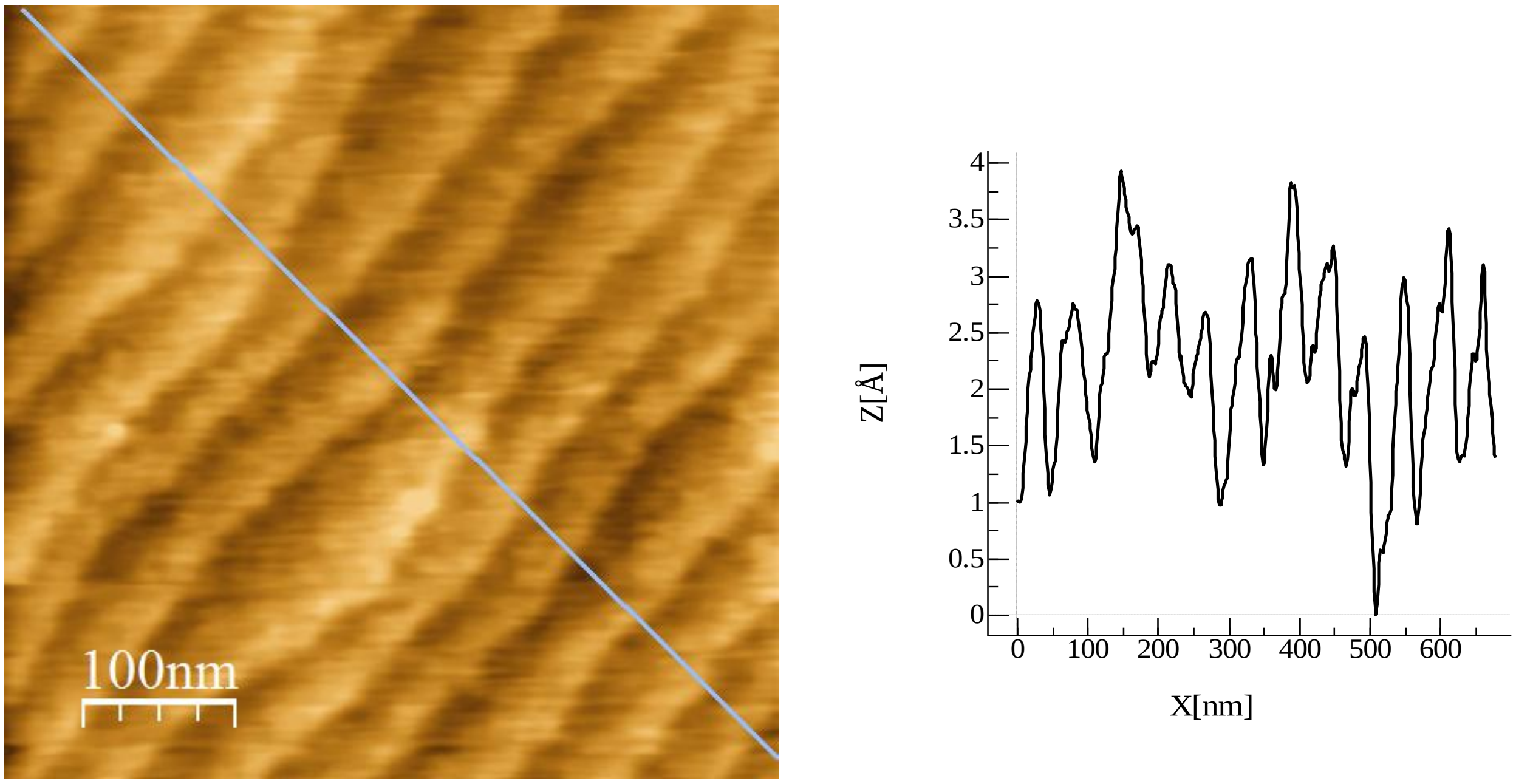

**Figure 3.1.5** NC-AFM height profile for YSZ(111) surface (also in Figure 3.1.4d), revealing a high density of terraces with minimal edge defects. This step-edge morphology is indicative of a surface flattening or "smoothing" process that occurs on the single-crystal surface after a thermal annealing treatment.

This waviness is also obvious from the baseline in the profile of Figure 3.1.5 where it varies between 0-1.25 Å, keeping in mind, too, that this profile was traced out after real-time flattening of the raw data and choosing the flattest surface plane. The undulations, again, did not affect the overall stability of the sample and this is in accordance to other findings indicating that the nonpolar YSZ(111) surface is the most stable.[108] The reported surface kurtosis, $S_{ku}$, parameter for the as-received and cleaned YSZ(111) surface was on the order of 2.8-3.8 (Table 3.0), indicative of a Gaussian-like surface[116] with a feature height distribution that does not deviate significantly from the mean height defined by the RMS. Kurtosis values of the raw images were vital in gauging the necessity for flattening filters during image processing since most of the post-annealed single-crystal surfaces in this study had relatively similar "peakedness" of ~3.0. In summary, the roughness values from the ambient NC-AFM characterization of the "cleaned" YSZ(111) surface are as follows:

**Table 3.0:** Ambient NC-AFM surface data for pre- and post-annealed YSZ(111) sample

| Image Size/Sample | [a]RMS (Å) | Roughness (Å) | [b]Maximum Height (Å) | Kurtosis, $S_{ku}$ |
|---|---|---|---|---|
| **2 µm x 2 µm/YSZ(111) as-received** | 1.006 | 0.797 | 5.221 | 3.843 |
| **1.6 µm x 1.6 µm/YSZ(111) annealed** | 0.832 | 0.661 | 4.485 | 3.221 |
| **1 µm x 1 µm/YSZ(111) annealed** | 0.857 | 0.683 | 4.565 | 3.222 |
| **500 nm x 500 nm/ YSZ(111) annealed** | 0.849 | 0.686 | 3.727 | 2.818 |

[a] Also defined as the geometric corrugation of the surface in this work.
[b] Maximum peak-to-valley height fofalculated by excluding extrinsic topographical defects due to contaminants or tip effects

Reduction of the adventitious carbon over-layer and other surface contaminants was evidenced from the 17.4% decrease in RMS roughness, a value in agreement with previous reports involving the YSZ(100) surface.[11] Even with the increase in stepped features and arrangement into a hill-and-valley surface structure after annealing, the maximum peak-to-valley height (Table 3.0) still decreases by 14.1%. It can be concluded that this thermal treatment process was both adequate and reproducible in yielding a "clean" single-crystal substrate conducive as a support template for further experiments involving the photochemically induced growth of mixed cobalt oxide nanostructures.

## 3.2.1 THE POLAR YSZ (100) SURFACE

It follows from the previous section that YSZ is a very non-trivial surface to study and is susceptible to reconstruction even under stable, ambient conditions;[11] YSZ(100) proves to be no exception. The surface of YSZ(100) is polar, especially when bulk terminated, as discussed, and it as well as other low-index terminations including (111) just are known to undergo major reconstruction. Particularly for the case of YSZ(100), this propensity for reconstruction and sample variability (see following discussion) lends itself to a material responsible for various surface configurations. Actually, the distribution and mobility of the O-vacancies in YSZ(100) often make the surface of YSZ(100) more unpredictable experimentally and unstable than what is reported for the (111) and (101) surfaces.[117] In YSZ(100) there is a dipole moment perpendicular to its {100}-repeating plane direction (i.e. normal to the (010) and (011) direction) this feature is effectively what makes it a "polar" surface. Further, it is classifiably polar on the basis of surface energies and Tasker's rule[10] for Type 3 surfaces (which are defined as "polar surfaces" and often include most $CaF_2$-like cubic structures) as well. Local charging effects often occur in the sample if $O^{2-}$ vacancy termination occurs near a defect or impurity meaning that the bulk and surface are infrequently charge-neutral. These effects, if not mitigated, prove to complicate NC-AFM imaging of local surface features and also affect nucleation of nanostructures even on clean, thermodynamically stable surfaces. When thermally activated, the diffusion of intrinsic defects and rearrangement of atoms towards a stable surface in YSZ(100) can give rise to a variety of complex surface nanostructures as we will see.

Extensive characterization work both *in vacuo* and in ambient conditions has been performed on the YSZ(100) surface and, despite a wellspring of data, control over the surface morphology is still somewhat elusive. Though similar to the YSZ(111) system, the cleaned

YSZ(100) system typically yields 3 distinct surface types: I, II, and III, which were established by previous experiments done in this group.[12] YSZ(100) samples used in this study ideally fall under the description of at least one of the following qualitative surface types:[11,12]

Type **(I):** Surface contains parallel step and terrace features, often containing defects such as kinks, ad-atoms, or (very infrequently) atomically flat pits. Usually this is the smoothest surface, reporting the lowest RMS roughness values.

Type **(II):** Surface contains many defect structures, typically in the form of deep pits, and in such high frequency that the stepped surface geometry seen in Type 1 surfaces is completely distorted.

Type (**III**): Surface contains distinct rectangular holes distributed throughout with atomically flat bottoms. Also observed are ridges between the rectangular features that measure about ~1/2**a** (where unit cell parameter, a, is 5.124 Å [3,11,118,119]) and the surface itself, though within the <5 Å RMS roughness guarantee, often has the highest RMS roughness values across the three types.

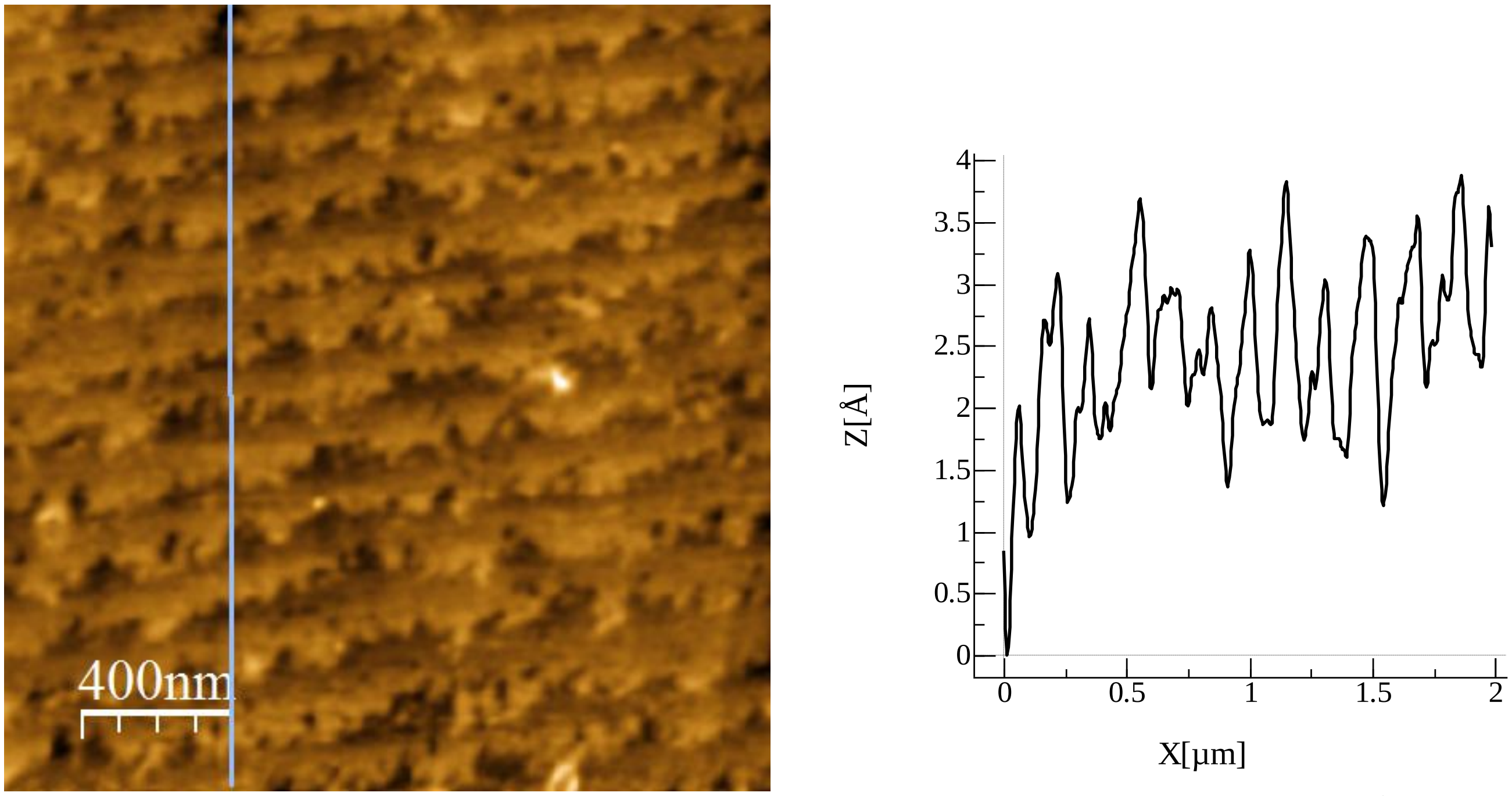

**Figure 3.2.1** 2.0 μm x 2.0 μm NC-AFM micrograph of a Type I surface (post-anneal, 1000°C, 1 hr) with associated height profile on the right, corresponding to the grey trace line (note that the profile begins at the top of the trace). Parallel steps and terrace features dominate this and frequently contain defects in the form of small pits or kinks. RMS roughness was 0.7380Å.

Most frequently encountered (~75% of the time) was the Type I surface pictured in Figure 3.2.1. Parallel steps with minimal bunching after one thermal treatment to 1000 °C were usually accompanied by defect structures accumulating along the step edges that separate the terraces. Pits or kinks were often observed with depths of ~0.5-0.6**a** and the main contribution to the still very minimal RMS roughness values was usually from these defect features.

The prominent parallel step features evident in the Type I surface disappear or are at least vaguely present in the Type II surface and do not even register in the Type III surface. Reasons abound, albeit speculative, to account for this surface variability, many of which speak to the interaction of defects and the kinetic activity on the underlying (100) substrate geometry during annealing. In the latter case it is often found on other analogue surfaces that non-uniform defect pits or clusters of ad-atoms can form in weaker areas of the surface during heating and form large clusters of these defect pits. [120,121]The abrasive CMP process can often induce impurity-based defects or random step defects in perpendicular directions to the stable (100) direction; these can often be resolved in NC-AFM imaging.[122] There is reason to believe that some of the cleaned YSZ(100) samples exhibit the aforementioned ripening of pit defects as a function of temperature. Figure 3.2.2 depicts a quasi-Type II surface wherein extrinsic defects are not that abundant but if they do occur, they tend to preferentially collect as small clusters in the individual step edges and as larger clusters at junctions bridging multiple step edges. Unfortunately, a convolution of AFM probe and defect morphology limits a completely rigorous characterization of the defect structures but based on a library of previous images, especially of the Type III surface, it can be safely concluded that the clusters are spherical or truncated spheres appearing as rectangles in the XY direction.[12] Again, because of the lower defect concentration that is quantified from the surface (~20 defects $\mu m^{-2}$), this particular surface is in between a Type II and Type III. Distortion of the parallel step geometry of the surface in Figure 3.2.2 is attributed formation of defects, dislocations, and lattice mismatches that become mobile during thermal agitation, which in turn perturb the atomic periodicity of the surface along its (100) cleavage plane.[114,122,123] At higher temperatures and heating for longer periods, the Type III surfaces are shown to form rectangular islands or pits with atomically flat holes[11,12] but in this study the surfaces are cleaned via ambient thermal annealing to generate the most common, stoichiometric, step-and-terrace Type I surfaces. A mechanism for nanostructure organization on the substrate surface may also be realized.

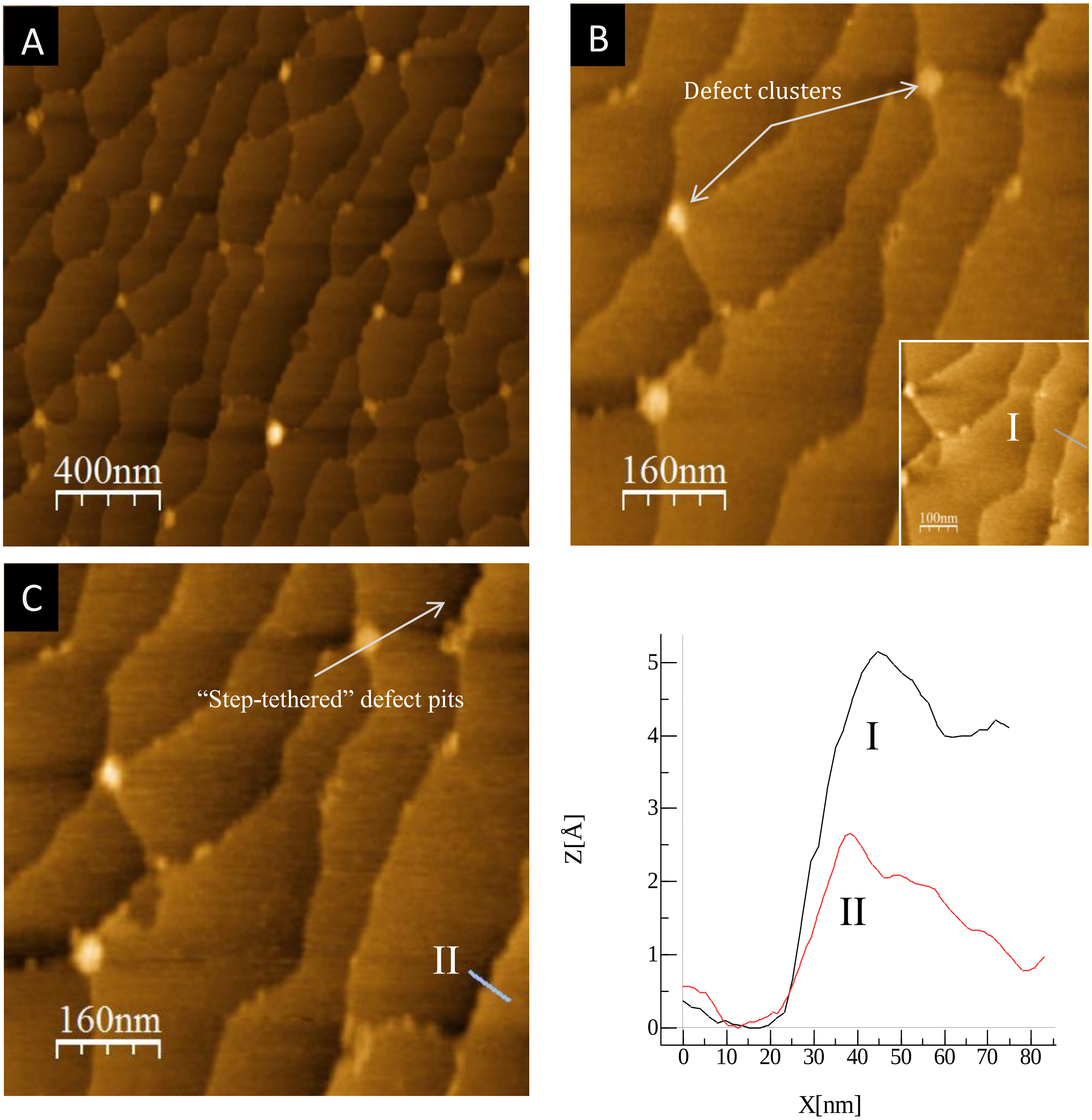

**Figure 3.2.2** Ambient NC-AFM images of clean YSZ(100) (post-anneal, 1000 °C, 1 hr) surface depicting a roughly Type II surface orientation with defect structures (pits or clusters) forming preferentially at the step edges. **A)** 2.0 μm x 2.0 μm scan (z-range=4.697Å, RMS=1.681 Å, scan speed=0.2 line/s) **B)** 819 nm x 819 nm scan (z-range=3.754 Å, RMS=1.265 Å, scan speed=0.5 line/s) revealing defect pits of ~1**a** (recall a=5.124Å) length with nearly flat bottoms in the line profile for I in **C)**. 500 nm x 500 nm scan (z-range=5.372Å, RMS=1.003 Å, scan speed=~1.0 line/s) showing terrace step length of ~1/2**a**, as shown in height profile for II in **C)**, and spherical defect clusters formed at the step features.

### 3.2.2 SAMPLE PREPARATION AND THERMAL TREATMENT: CLEAN YSZ(100)

The YSZ(100) single-crystal samples were, in large part, prepared identically to the YSZ(111) samples (details in Chapter 3.1.1). Again, the samples were produced from cutting the crystal sample from the boule at <100> ± $30^o$ orientation, rendering a sample with < 5 Å RMS.[3] Due to increased variability between samples for this particular surface, ~7-10 samples were annealed and characterized before an adequate candidate was selected for the further experiments involving nucleation of nanostructures. There is a well-documented study reported elsewhere that involves thermal annealing with $O_2$ cycles (back pressures ~5 x $10^{-6}$ mbar), producing a clean vicinal YSZ(100) surface.[11] NC-AFM image analysis of the clean Type I YSZ(100) is reported in the following section.

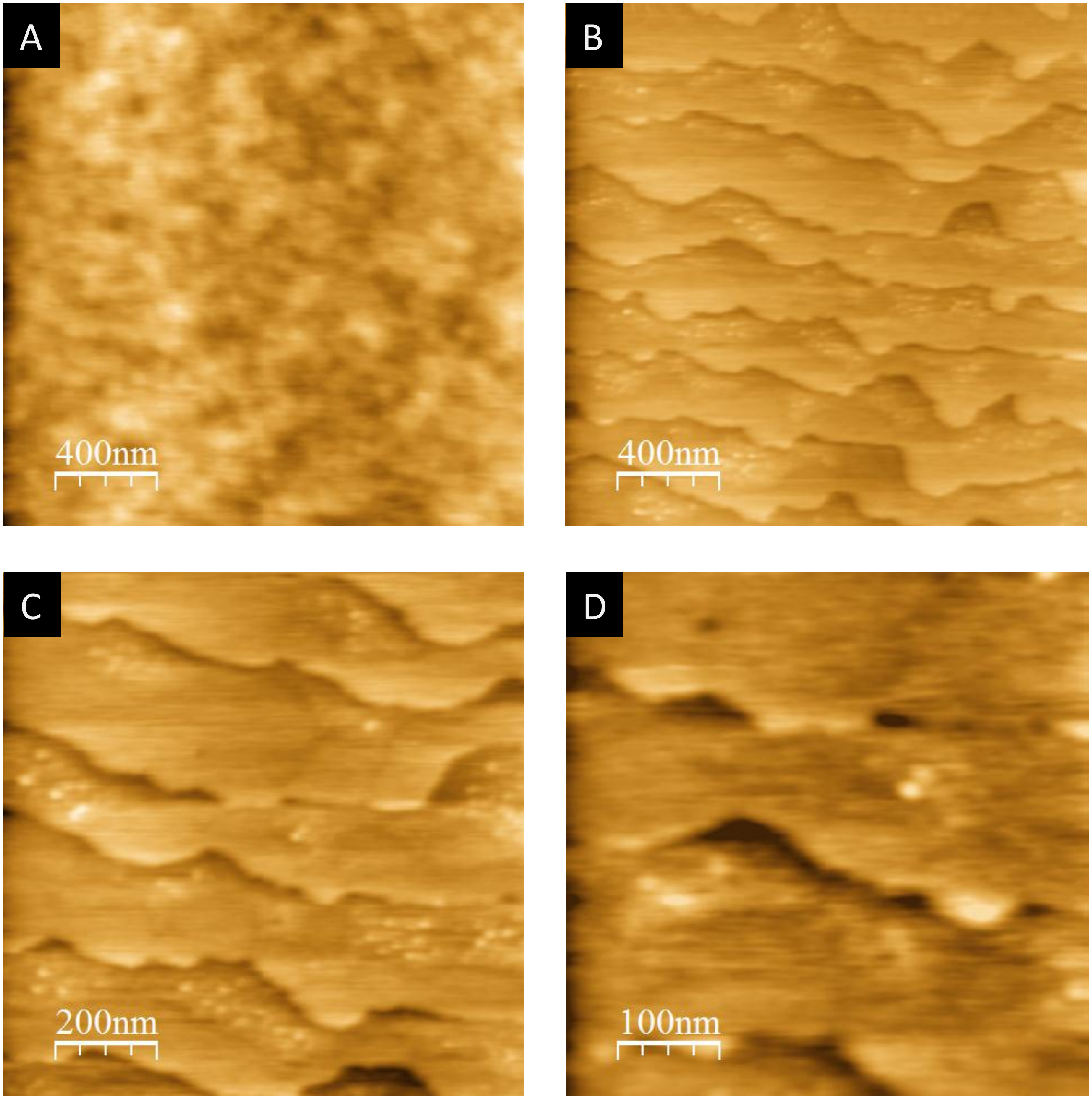

**Figure 3.2.3** High resolution NC-AFM images of the clean, pre-growth YSZ (100) sample. **A)** 2.0 µm x 2.0 µm micrograph of as-received single crystal (nominal z-range=9.762 Å). Micrographs **B)-D)** depict sample after ambient thermal cycle in air at 1000 $^{o}$C for 1 hr. **B)** 2 µm x 2 µm micrograph of single crystal, post-anneal (nominal $z_{range}$=7.673 Å). **C)** 1.0 µm x 1.0 µm micrograph of single crystal, post-anneal (nominal $z_{range}$=5.993 Å). **D)** 500 nm x 500 nm micrograph of single crystal, post-anneal (nominal $z_{range}$= 4.759 Å). Images required 2$^{nd}$-order polynomial flattening during acquisition.

Similar to Figure 3.2.1, the post-annealed clean YSZ(100) sample obtained after thermal annealing yielded a minimally defective surface with a homogenous distribution of parallel steps and terraces. Image analysis also revealed small defect clusters (shown in Figure 3.2.33c) atop the terraces, with associated terrace-specific RMS roughness of 0.6294 Å (calculated from the 1 x 1 $\mu m^2$ topographic image in Figure 3.2.3c above). Shown below was the associated phase image, which yielded excellent local contrast around the terrace defects, which appear to show Volmer Weber-like island growth out of the of the surface.

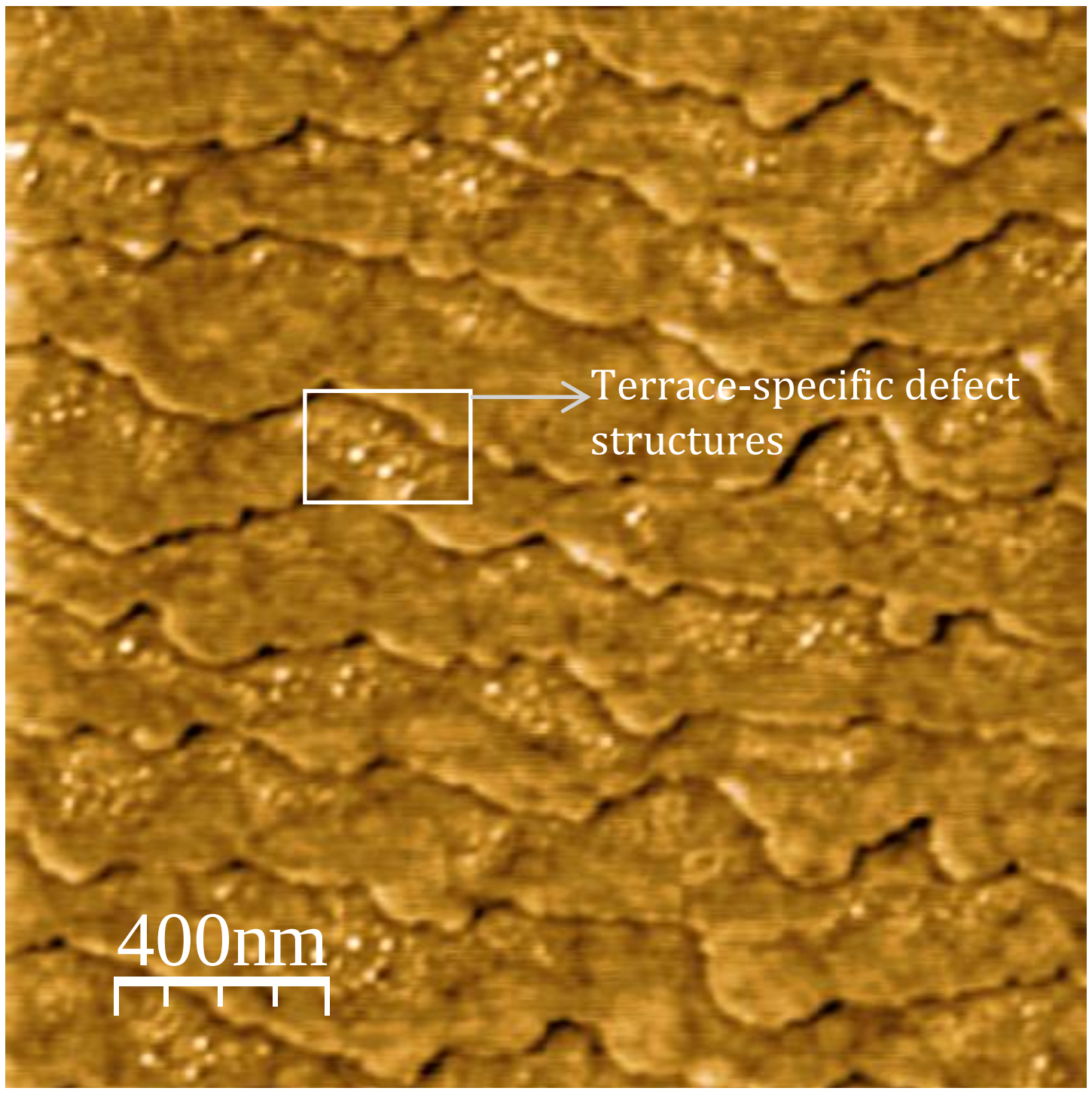

**Figure 3.2.4** 2.0 µm x 2.0 µm NC-AFM phase image of clean YSZ(100) surface (corresponding topography micrograph in Figure 3.2.3b) containing clusters of terrace-specific defects. Saturation of the signal and local contrast in the phase image suggest that the defects are non-stoichiometric with respect to the surface.

This value was calculated by an area-weighted mean, which excludes outlier values due to feedback parameters and cantilever-induced artifacts. The phase images prove to be a useful aid in calculating roughness parameters, particularly that of the terrace-specific RMS roughness because the defect sites on the terraces have a different elastic modulus component from just the terraces themselves. As Figure 3.2.4 depicts, brighter or saturated spots indicate presence of a dewetting-like over-layer in the form of clusters. The stability of imaging does not suggest the likelihood of surface modification of the AFM probe, but this was a consideration at the time. Unlike the YSZ(111) sample, the terraces were separated with very nonlinear step-edge sites, often in the form of kinks that sometimes occupy up to ¾ the width of the terraces. These defect kinks could actually mean the onset of larger reconstructive processes that are observed in the more defective Type III surfaces, but only further annealing sessions can confirm this detail. In fact, it is found that after the 1 hr of initial annealing, the lowest values of RMS roughness, terrace-specific RMS roughness, maximum height, and roughness are typically observed. [12]Step heights for the selected surface (Figure 3.2.3) were on the order of ~1/2**a** at 2.4± 0.4 Å with average terrace widths of 199 ± 52Å. The surface was relatively more defective than that of the YSZ(111) surfaces after annealing and also posted higher roughness parameters. Qualitatively, this would be evident by observing the pristine, virtually defect-free terraces of the YSZ(111) surface. Table 3.1 outlines the roughness parameters that were calculated from the NC-AFM images obtained for the chosen Type I surface (Figure 3.2.3-4). Note that the clean, post-annealed surface (Figure 3.2.3) shows a decrease of RMS roughness from 2.001 Å to 1.180 Å. Surface kurtosis value deviated more than compared to the YSZ(111) surface largely because of the higher defect concentration within the terraces. This caused the peakedness to vary with respect to a higher max peak-to-valley (9.762 Å for <100> vs. 5.221 Å for <111> orientation) feature height and step height than what was found in the YSZ(111) surface. Ideally, because of the peak-and-valley nature of the surface and relatively low defect concentration, the Type I surface would serve to be the best template in terms of the "controlled" or substrate-directed growth of sub-monolayer growth of cobalt oxide nanostructures. Since the cleaned surface offers little predictability sample to sample, this part of the work proved to be longest since many single-crystal samples were studied before proceeding to further experiments. An empirical probability of Type I surface after thermal treatment (1000 $^{o}$C, 1 hr) was found to be ~$(3/5)^n$ (n=number of samples). The chemical nature of the defect clusters found on this particular sample offers many possibilities including Yttria segregation or a dewetting-like process enhanced by surface diffusion of defects during heating, but these are the subject of extensive experimentation reported elsewhere[12] and are thus outside the scope of this work. Also, cantilever-tip effects due to electrostatic interactions with a charged or polar surface, especially that of YSZ, during imaging cannot be ruled out as a cause for some of the defect structures.[114,124]

**Table 3.1** Roughness parameters calculated from the NC-AFM data for the clean YSZ(100) Type I surface

| Image Size/Sample | [a]RMS (Å) | Roughness (Å) | [b]Maximum Height (Å) | [c]Kurtosis, $S_{ku}$ |
|---|---|---|---|---|
| **2 x 2 μm$^2$/YSZ(100) as-received** | 2.001 | 1.622 | 9.762 | 3.100 |
| **2 x 2 μm$^2$/YSZ(100) annealed** | 1.180 | 0.887 | 7.673 | 5.299 |
| **1 x 1 μm$^2$/YSZ(100) annealed** | 0.909 | 0.697 | 5.993 | 3.975 |
| **500 x 500 nm/YSZ(100) annealed** | 0.711 | 0.527 | 4.759 | 4.882 |

[a] Also defined as the geometric corrugation of the surface in this work.
[b] Maximum peak-to-valley height for surface. Calculated by excluding extrinsic topographical defects due to contaminants or tip effects
[c] Calculated on the basis that raw images needed to be planar tilted with respect to the highest feature since values >> 3 were computed before this correction.

## 3.3.1 THE IONIC MgO(100) SURFACE

Due to the surface orientation <100> for the YSZ, coupled with the inhomogenous distribution of charge through the single crystal itself upon doping with Yttria, we established that its surface is quite polar.[12,89,125,126] This result is actually in sharp contrast with the surface of MgO(100) despite sharing the <100> orientation because this surface is in fact a well-known ionic "checkerboard" crystal surface. [127–130] Before elaborating on the single-crystals annealing behavious, it is important to gain a fundamental understanding of the surface. Similarly to the other alkaline-earth oxides, MgO is a rocksalt oxide that has close packing layers of anions and cations such that the surface appears as alternating positive and negative charges. [131] The inter-penetrating layers of cations and anions actually make the ideal surface ionically neutral, with the <100> orientation being the most stable, nonpolar surface for that particular oxide.[9]Of course, no surface is pristine or ideal, especially in ambient conditions, so the stability and *relative* nonpolarity of the MgO(100) surface are always points of contention The surface of MgO(100) looks largely similar to other (100) single-crystal samples since the crystal ledges often run in the very stable <100> direction, yielding that parallel-stepped surface morphology already seen with both YSZ surfaces. Often the only features that vary would be roughness, defect structures, step heights due to varying lattice parameters, and reactivity with local environment. In fact, the latter characteristic proves to be the most interesting from a characterization point of view. This is best captured in the following remark: " Polarity is not only a question of orientation of the surface but also of the actual termination".[9] Figure 3.3.1 shrewdly depicts the atomic arrangement of the MgO(100) surface, which is found to be relatively nonpolar by the Tasker Type I classification of rocksalt surfaces with <100> orientation. [9,10]

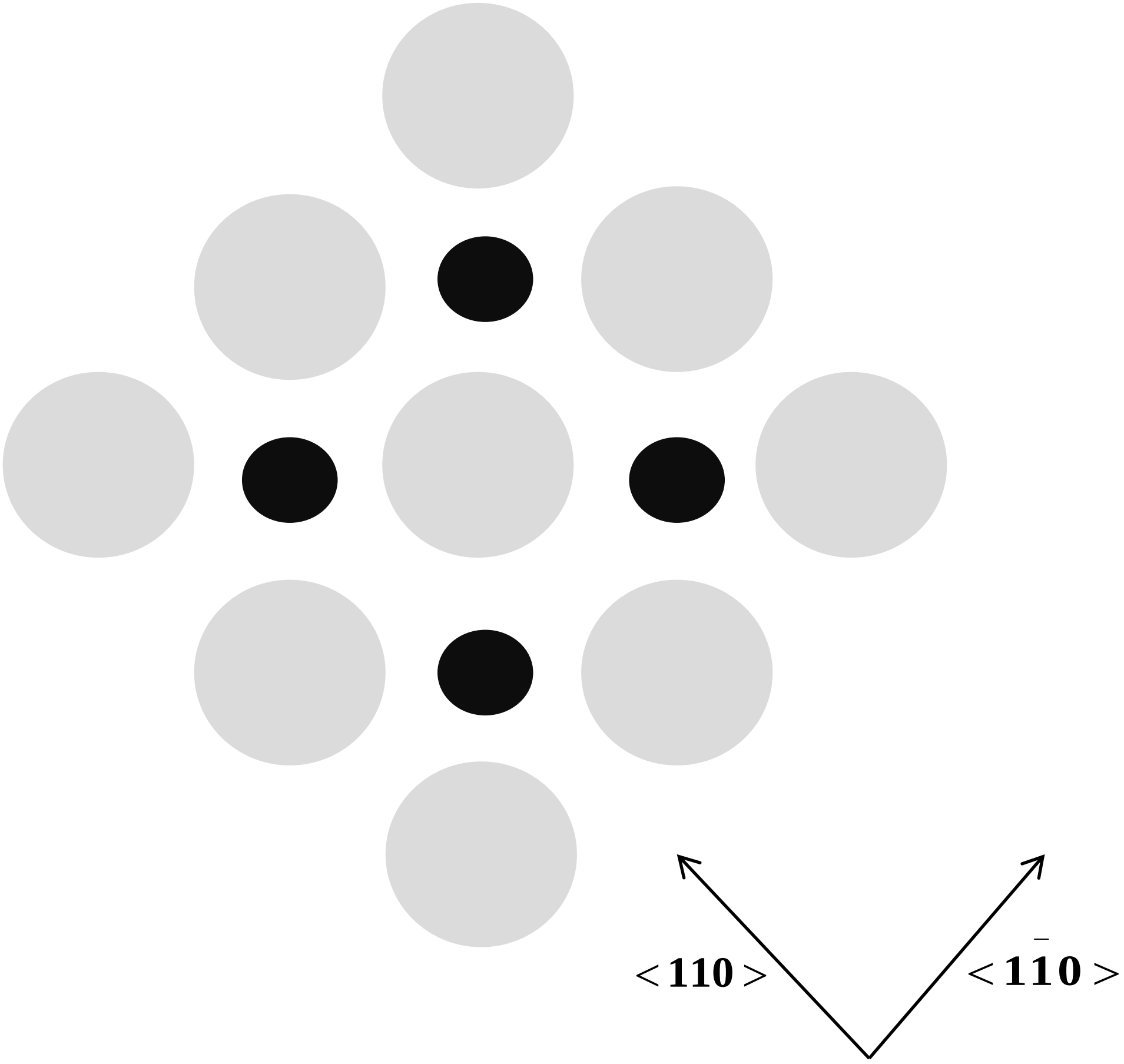

**Figure 3.3.1** Atomic "checker-board" surface structure of MgO(100) with $O^{2-}$ ions represented by large grey circles and $Mg^{2+}$ ions by the small black circles. The ideal <100> orientation allows the MgO(100) surface to be ionic neutral (nonpolar) without requirements for charge stabilization.

The necessary and sufficient condition of the relatively nonpolar nature of the surface is derived from frozen bulk termination models wherein atomic layers at the surface are assigned the same composition as the bulk, and these layers are stacked such that one dipole-free unit cell leaves the surface region empty; by these reports MgO(100) is ideally nonpolar with polarity compensation, meaning that defective MgO(100) has weak polar-termination sometimes. [9] The instability of even the clean MgO(100) surface in ambient conditions gives rise to the surface's rich chemistry. Speaking to this point, the reactivity of MgO surfaces, particularly the (100) surface, towards atmospheric water and carbon dioxide has actually hindered the atomic-scale resolution of the MgO(100) surface in many past studies.[9,127–129] It would therefore be egregious to proceed without reporting on the instability towards ambient conditions and how they may hinder NC-AFM characterization. Recent reports actually indicate that surface roughening in constant 15% relative humidity (RH) of the as-received MgO(100) increases from 4.2Å up to 2.82 nm in some cases.[132] This makes an annealing treatment for the clean MgO(100) vital since the surface can roughen by up to an order of magnitude in just two weeks. Even after annealing, the surface is still sensitive to the enhanced roughening from ambient water so much so that, in one case, at %RH values <1%, the surface is fully hydroxylated (~1ML).[133] The MgO(100) surface is found to undergo the following reactions:[132]

$$MgO + H_2O_{(g)} \rightarrow Mg(OH)_2 \quad (3.1)$$

$$MgO + CO_{2(g)} \rightarrow MgCO_3 \quad (3.2)$$

Presence of surface-tethered hydroxyl peaks in the Fourier-Transform Infrared (FT-IR) spectrum around 3700 $cm^{-1}$ are often proof of this energetically favourable surface hydroxylation. Since it is the case that MgO surfaces are often fully hydroxylated in ambient conditions

(15% < %RH < 45%), one should consider the reaction of $Mg(OH)_2$ with $CO_{2(g)}$ as well:[130]

$$Mg(OH)_2 + CO_{2(g)} \rightarrow MgCO_3 + H_2O \quad (3.3)$$

Qualitatively, NC-AFM imaging shows surface structures of $Mg(OH)_2$, referred to as "hillocks", in the form of saturated white spots occurring at terrace steps.[132] This is important to note, especially for the imaging of the MgO(100) surfaces (as-received and clean), and should be reproducible.

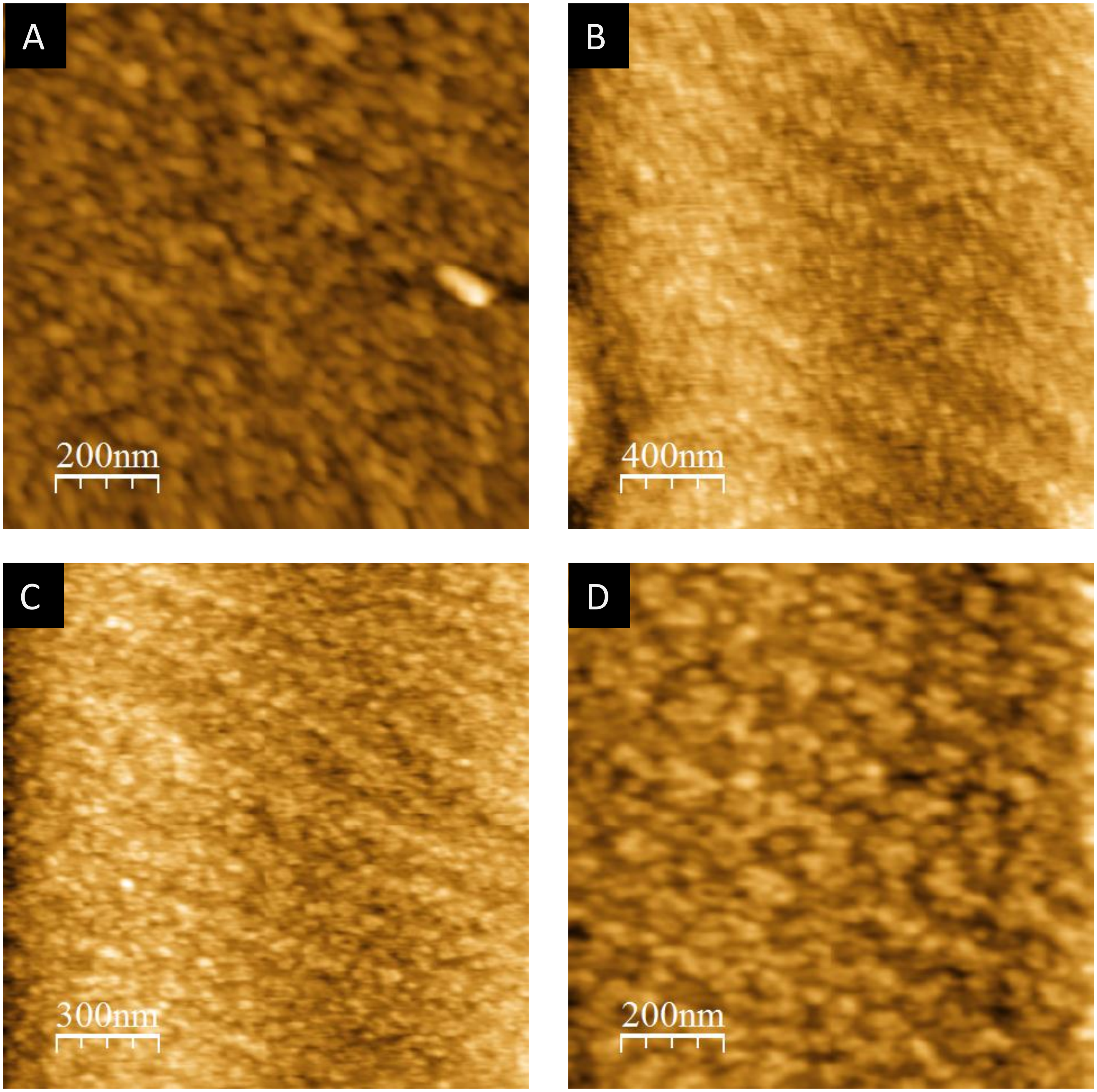

**Figure 3.3.2** NC-AFM images of clean, pre-growth MgO(100) surface. A) 1.0 µm x 1.0 µm micrograph of as-received single crystal, B)-D) clean single crystal (annealed 1hr, 1000$^{o}$C) 2.0 µm x 2.0 µm, 1.5 µm x 1.5 µm, and 1.0 µm x 1.0 µm micrographs, respectively. White spots (saturated features) indicate both $MgCO_3$ and $Mg(OH)_2$ groups where ambient carbon dioxide and water have chemisorbed. Scan speed: 0.4-0.8 line/s.

Freshly cleaved MgO(100) single crystals were thermally treated exactly the same as the YSZ(100)/(111) samples with following exceptions: i) sample substrates were only removed from the vacuum sealed blister-pack if they were to be annealed and imaged within a 24 hr window; ii) single-crystal size was 10 x 10 x 0.5 $mm^3$ (1SP) with RMS roughness < 10Å (larger due to adsorbate-induced surface roughening); iii) only 4 samples were imaged prior to selection for of the growth substrate; iv) Diamond-like carbon (DLC) cantilever-tips were used intermittently for NC-AFM imaging due to increased hardness/roughness and tip-induced charging of the surface.; v) samples were stored under partial vacuum (~0.08 mbar) when not in use to avoid exposure to high %RH. (see previous chapter for all experimental details). Figure 3.3.2 shows four panels of the surface of MgO(100) single-crystal substrate in ambient conditions chosen for the growth experiment. In panel A is the 1.0 x 1.0 $\mu m^2$ NC-AFM image of the as-received sample within a few hours complete exposure to air (%RH ~20 ± 5%). The surface already appeared to be quite roughened, RMS roughness of 6.968 Å, with virtually indistinct step features throughout the surface. Measurement of the terrace features was not possible through image analysis but it is suspected that the adsorbate features, mainly mixed $H_2O$-OH binding to MgO sites[134], are preferentially binding to step/corner/kink sites on the MgO(100) surface. The kinetic model or mechanism of this binding remains a rich debate, but it is well-documented that the hydroxylation of the surface at %RH>15% means full, self-limiting termination of the surface (>1 ML).[127,134–136] As per Figure 3.3.3, there is enhancement of roughness at the step-edge sites due to clustering of hydroxyl groups and possibly due to increased mobility during the annealing cycle. This formation of hillocks at step-edge sites is in agreement with other reports showing that hopping and mobility of the surface hydroxyls decreases by nearly a 50-fold factor[2,135,136] at room-temperature/ambient conditions, meaning the aggregation of water monomers is favourable and the surface assumes thermodynamic stability.[134,137] NC-AFM imaging (Figure 3.3.3) confirms that the saturation of the hillocks increases where the step edges form. Thus, the $Mg(OH)_2$ hillocks are basically immobilized at room temperature, of course barring any tip-induced mobility or interactions. That being said, of all single-crystal substrates characterized in this study, the MgO(100) posed the greatest difficulty in terms of imaging due to the tip-adsorbate interactions, often causing immediate blunting of the cantilever-tip. To elucidate this issue, DLC tips (nominal RF=300 kHz) and super-sharp Silicon (SSS-NCL, nominal RF=~160 kHz) (tip apex <3 nm radius), among many other chemically and physically harder tips, were employed but all exhibited the same blunting and deflection behavior half-way through the scan. Panels B-D show the post-annealed (clean) surface of MgO(100) with a RMS roughness down to 1.453 Å (for the 1.0 x 1.0 $\mu m^2$ image). The 79.2% decrease in RMS roughness is attributed to desorption of many of the hydroxyl and carbonate residues due the thermal treatment, bringing the MgO(100) closer to its stoichiometric surface than in the as-received sample. In contrast to the YSZ samples, MgO(100) shows an enhanced roughening that is much larger than what is induced by the adventitious carbon over-layer found on the YSZ samples.

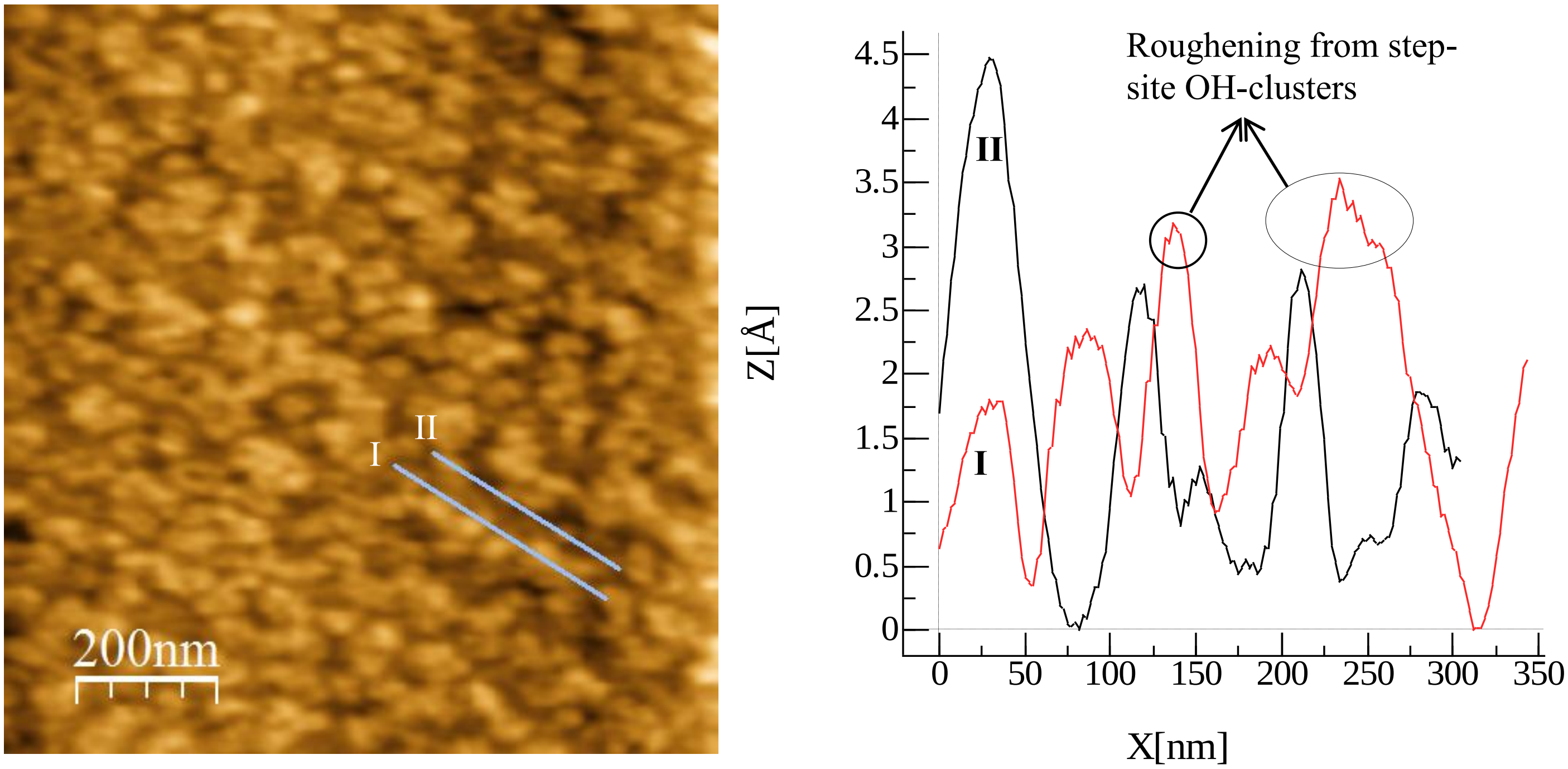

**Figure 3.3.3** NC-AFM line profiles corresponding the 1.0 µm x 1.0 µm image (Figure 3.3.2d) of clean MgO(100). Trace I (red profile) shows clusters, presumably of $Mg(OH)_2$ character, bunching within step edge, suggesting a higher average roughness and therefore higher probability of OH-binding at those sites rather than atop the terraces themselves. Hillock heights were on the order of 1/2**a** (where a=4.213 Å for MgO[2,3] and a=4.766 Å for $Mg(OH)_2$)

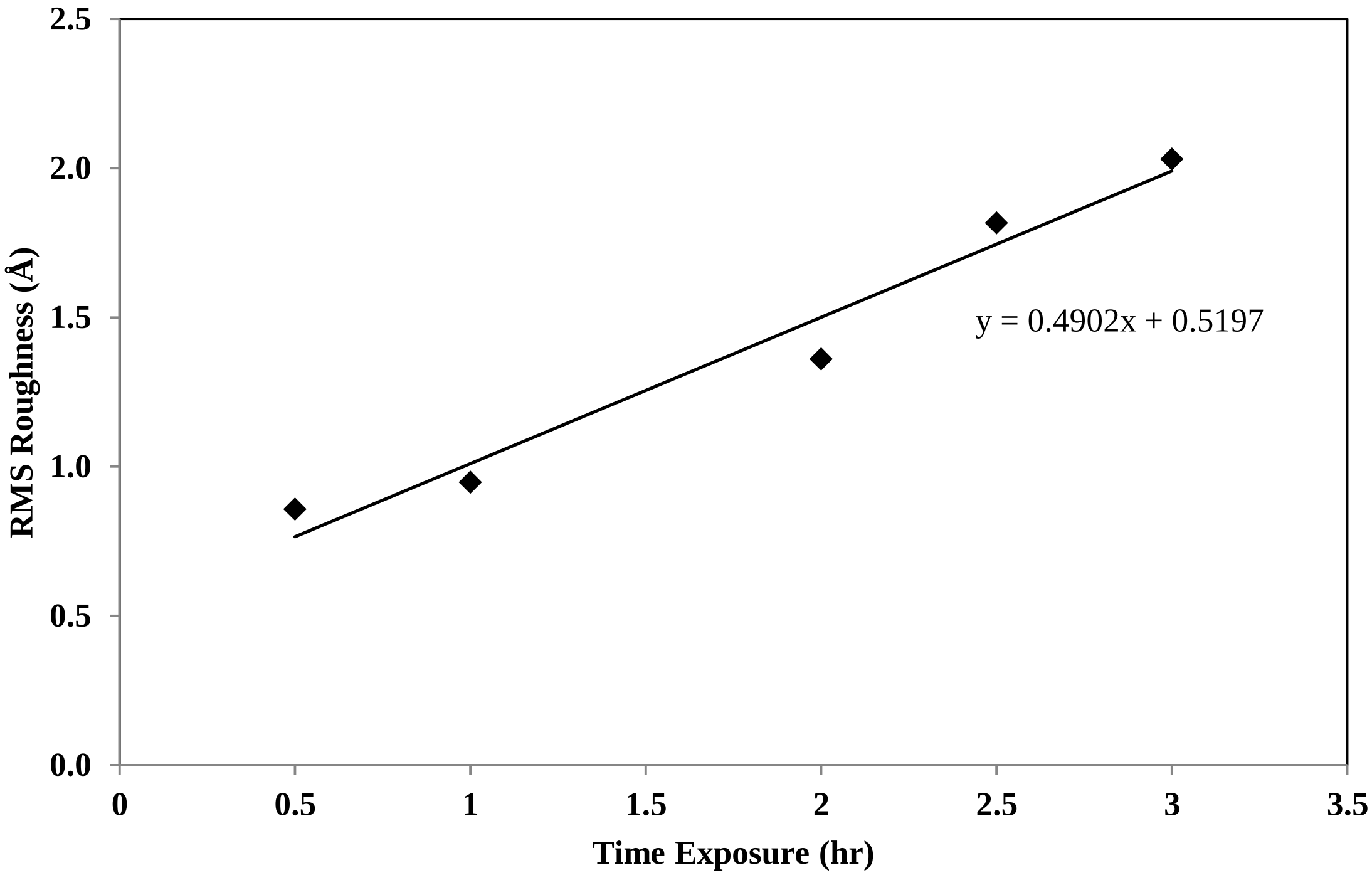

**Figure 3.3.4** Surface roughening of as-received MgO(100) single-crystal surface as a function of exposure time to ambient conditions (22 $^{o}$C, %RH 25 ± 5 %) during NC-AFM Imaging.

Chemisorption events are energetically favourable ($\Delta G_r^o$=-35.5 kJ/mol[127] for $Mg(OH)_2$, no data available for $MgCO_3$) on MgO(100) and, in fact, it was found that surface roughening shows a nearly linear response with respect to surface exposure time to air during NC-AFM imaging (Figure 3.3.4). The implications of this finding suggest that a kinetic model for the adsorption of $H_2O$ and subsequent hydroxylation of the surface in ambient conditions can be realized, but as mentioned, the process of hydroxylation happens so fast at %RH > 0.01% that it would be difficult, even on the basis of roughness parameters. The RMS roughness shows a nearly proportional increase of roughly 0.4902 Å/hr where the %RH threshold for MgO(100) is 10% for 1 monolayer (ML) of $Mg(OH)_2$.[135] The MgO(100) is best described as having almost exclusively $Mg(OH)_2$ character in ambient conditions, which effectively means the stoichiometric surface refers to the adsorbate-induced termination of MgO(100) to form $Mg(OH)_2$ moieties with 100% coverage (> 1 ML, where a ML is an OH-OH group chemisorbed per $Mg^{2+}$ site[134]).

**Table 3.2** Ambient NC-AFM surface data for pre- and post-annealed (clean) MgO(100)

| Image Size/Sample | RMS (Å) | Roughness (Å) | Maximum Height (Å) | Kurtosis, $S_{ku}$ |
|---|---|---|---|---|
| **2 µm x 2 µm/MgO(100) as-received** | 6.968 | 5.303 | 26.236 | 8.108 |
| **2 µm x 2 µm/MgO(100) annealed** | 1.809 | 1.452 | 8.589 | 3.080 |
| **1.5 µm x 1.5 µm/ MgO(100) annealed** | 1.729 | 1.376 | 10.939 | 3.454 |
| **1 µm x 1 µm/MgO(100) annealed** | 1.453 | 1.170 | 6.637 | 2.912 |

Though the roughness data (Table 3.2) calculated for the surface suggests relatively lower contamination in the annealed samples than for the as-received, the stoichiometry of the surface drastically changes when MgO(100) is exposed to ambient environment. Instabilities of the AFM tip-cantilever and an obvious distortion to the images of the topography suggest that there are strong interactions between the droplet-like defect structures (“hillocks”) formed on the surface and the tip. These were unavoidable circumstances and were present under various scanning parameters and tip configurations, so similar limitations would also beset NC-AFM image resolution of post-growth images (next chapter). This serves to further reinforce the inherent difficulty that belies SPM work on ionic surfaces like MgO(100) in ambient conditions.[128,129,131,133] Luckily, the residual parallel step-and-terrace morphology of the surface may allow some control over morphology and orientation of the nanostructures.

## 3.4 HOPG SURFACE

Perhaps the most ubiquitous and highly-cited nanomaterial in recent history, graphene is the essential building block of a plethora of carbon allotropes including carbon nanotubes[138] and $C_{60}$ Buckminster Fullerene[139] to name a few. Another one of those allotropes is highly-ordered pyrolytic graphite (HOPG). Knowledge of the honeycomb-like graphene material dates back to roughly 1947 when Wallace postulated that a single sheet of $sp^2$-hybridized carbon would have an internally uniform dispersion of energy and form what are called Brillouin zones (K-points) about which the graphene sheet adopts its thermodynamic stability.[140] Keeping with the ambiguous nature of its history, the first synthesis of a tiered graphene system was supposedly based on the method of Boehm who worked on the reduction of graphene oxide, effectively laying the groundwork for the most popular, bottom-up synthetic technique for making graphene to date.[141] Many experimental studies with the $sp^2$-carbon material would eventually lead to the production, on very large scales, of crystals like HOPG. Decades later the tantalizingly large number of applications and properties of graphene were exploited by Nobel Laureates, Novoselov and Geim. In particular, their method of mechanical exfoliation became a method of isolating single crystal layers of clean graphene (sheets) from its graphitic, intercalated form.[4,141,142] In this work, preparation of the single-crystal HOPG surface was done simply by this mechanical exfoliation method (Scotch-tape pilling method) from an HOPG single crystal and no annealing process was needed to prepare the “clean” surface. This single-crystal surface, although pristine, often contains many defect sites that are often in the form of cleavage steps.

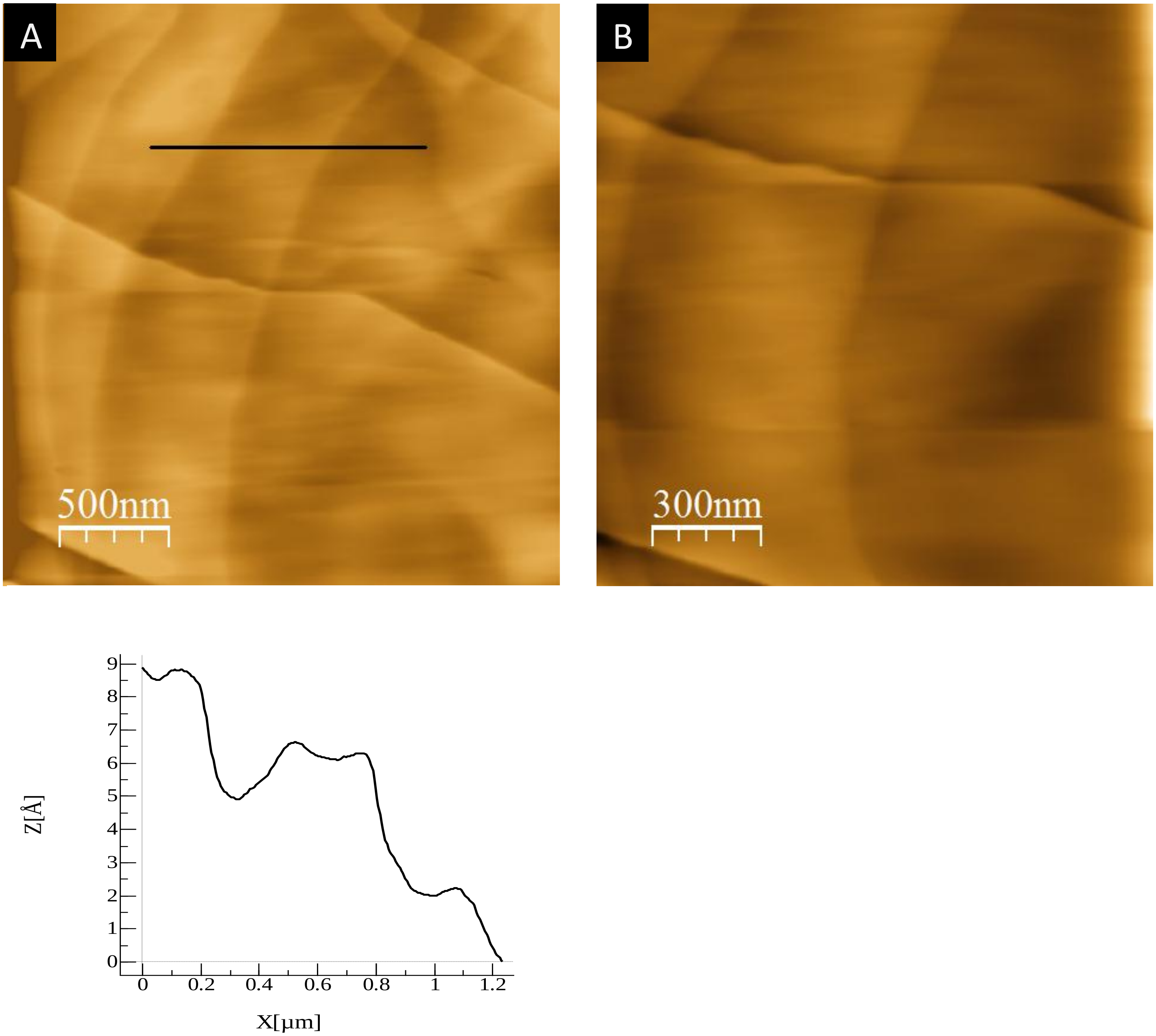

**Figure 3.4.1** A) 2.5 µm x 2.5 µm NC-AFM image of clean, freshly cleaved single-crystal HOPG Step-size of ~3.687 Å; close to interlayer spacing diameter, a=3.41Å[3,4] B) 1.5 µm x 1.5 µm NC-AFM image of clean, freshly cleaved HOPG sample (Z-range=1.637 nm, scan speed ~0.8 line/s)

Figure 3.4.1 depicts the freshly cleaved surface of HOPG used for the nucleation of cobalt oxide nanostructures. After mechanical cleavage the HOPG surface is rendered quite defective, consisting of mostly cleavage planes and step edges where multiple layers of graphene are exposed. This and other representative samples had appreciably high max peak-to-valley heights (z-ranges, Figure 3.4.1) and anomalously high kurtosis factors of ~ 6.776, meaning the surface required extensive flattening. The HOPG surface was not annealed and the calculated RMS was 3.421 nm (Figure 3.4.1a). The HOPG surface largely consists of flawless slabs of carbon atoms presumably, albeit not resolved by our ambient NC-AFM imaging, arranged in honeycomb structures and separated by linear cleavage steps. The cleavage steps comprise the bulk of the single-crystal surface's defect sites and undoubtedly play an important role in the structure and orientation features photochemically grown cobalt oxide nanostructures. Relatively inert DLC AFM (at 75% approach position) cantilever tips were required for stable imaging of the pristine HOPG single-crystal surfaces, but the PPP- or SSS-NCL (Si, uncoated) tips would be used for all subsequent imaging of the post-growth samples (Chapter 4) for the sake of consistency.

# CHAPTER 4 NC-AFM/XPS CHARACTERIZATION OF $CO_2O_3$NP-SINGLE-CRYSTAL SYSTEMS

## 4.1. THE PHOTOCHEMICAL SYNTHESIS OF COBALT (III) OXIDE NANOSTRUCTURES

The last portion of this work involved photochemical synthesis of Cobalt (III) Oxide nanostructures on the surfaces of the clean surfaces extensively described in the preceding chapter. Since the process is substrate-dependent, a detailed characterization of each surface and its respective morphology was critical for a complete picture of the metal-oxide supported Cobalt (III) Oxide systems. The basis of the work done by Tse-Luen Wee, J.C. Scaiano et al is to photochemically synthesize nanostructures whose morphology can be controlled by factors such as spectral properties, time-dependence, and intensity of the exciting light.[16] This work, by extension, is to ascertain structure-property relationships between the clean surface and the morphological properties of the photochemically grown metal nanostructures. Various surfaces, from highly pristine in the YSZ(111) to highly defective in the MgO(100), provide templates for this substrate defect-site-induced growth to occur. The nature of the surface should serve to not only govern the morphology of formed nanostructures, but also their distribution about the surface. Our particular interest is to explore the photochemically induced growth of these cobalt oxide nanostructures, which is both a completely novel liquid-phase reaction and a well-controlled growth process, on other supports and subsequently characterize their morphology and electronic structure via NC-AFM and XPS, respectively. We expect that the range of surface features and defects reported for the clean surfaces (Chapter 3) will cause growth and morphology behavior that largely contrasts what is reported for the NCD support. Though the nature of the cobalt oxide species should be mostly $Co_2O_3$, it is also expected that some contribution from other cobalt oxide species, especially the thermodynamically stable $Co_3O_4$, will be found in the XP-spectra collected.

It should be noted that while NC-AFM/XPS analyses are performed on the $Co_2O_3$NP-support systems, experiments exploring their viability as model catalyst systems were not done in this work.

## 4.1.1 REACTION MECHANISM FOR PHOTOCHEMICAL SYNTHESIS OF COBALT (III) OXIDE NANOSTRUCTURES

The novel, aromatic ketone-based photochemical synthesis of the $Co_2O_3$ nanoparticles ($Co_2O_3$NPs) is described in the following scheme:

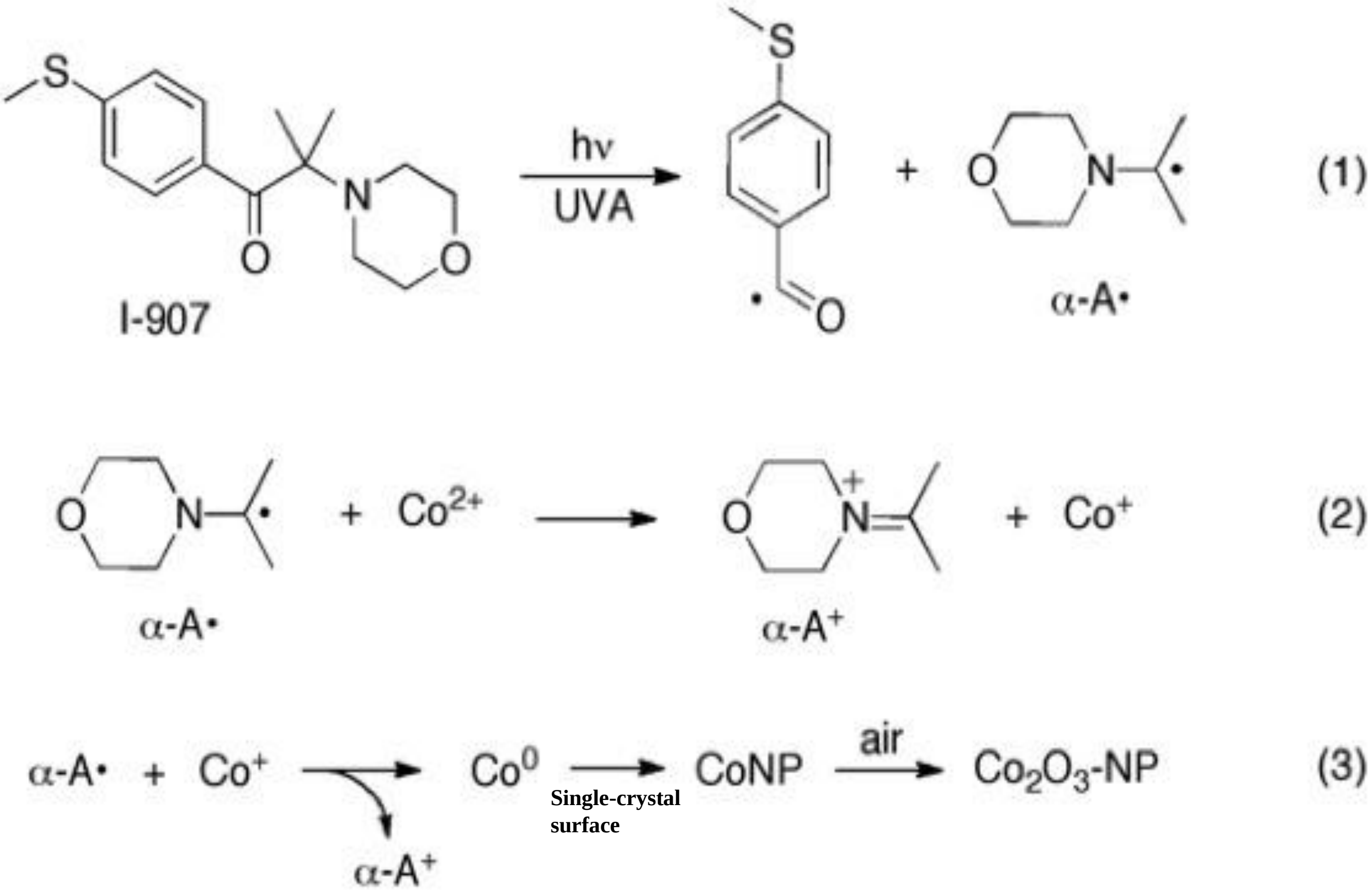

**Figure 4.1.1** Photochemical growth mechanism of CoNPs with their immediate exposure to air and subsequent oxidation to the $Co_2O_3$ nanostructures [16]

Initially, the photoinitiator (Irgacure-907™) is irradiated with UV-A light to generate the α-aminoalkyl radical (α-A$^{\cdot}$) via the Norish Type 1 reaction. The $CoCl_2$ precursor salt solution is then reduced in solution (solvent: acetonitrile) to the $Co^+$ ion in which subsequently gets further reduced to metallic Co. During normal procedures in the Scaiano group, the Co atoms nucleate in solution forming suspended nanoparticles of well-defined size. Part and parcel with this, nucleation also occurs at the vessel walls and on other nucleation points of nearby surfaces. Similarly, the metallic Cobalt species can form nanostructures on the single-crystal surface as CoNPs at nucleation sites formed by the defect structures and other surface features on the substrate. Because the CoNPs are highly paramagnetic[76] and, depending on the surface, they may cluster quite easily at room temperature. Once exposed long enough to air, the CoNPs oxidize to form into the $Co_2O_3$NPs, effectively losing their paramagnetism. [16,76] Immobilization of the $Co_2O_3$NPs (and other metastable oxides) to defective surface sites on the single-crystal supports likely occurs concurrently to this process. The $Co_2O_3$NPs grown on nanocrystalline diamond (NCD) supports have a reported average size of ~20 nm (lateral size, defined by the diameter determined from TEM).[16] These are typically the catalyst particles used in the development of new, cobalt-based water-oxidation nanocomposite materials.

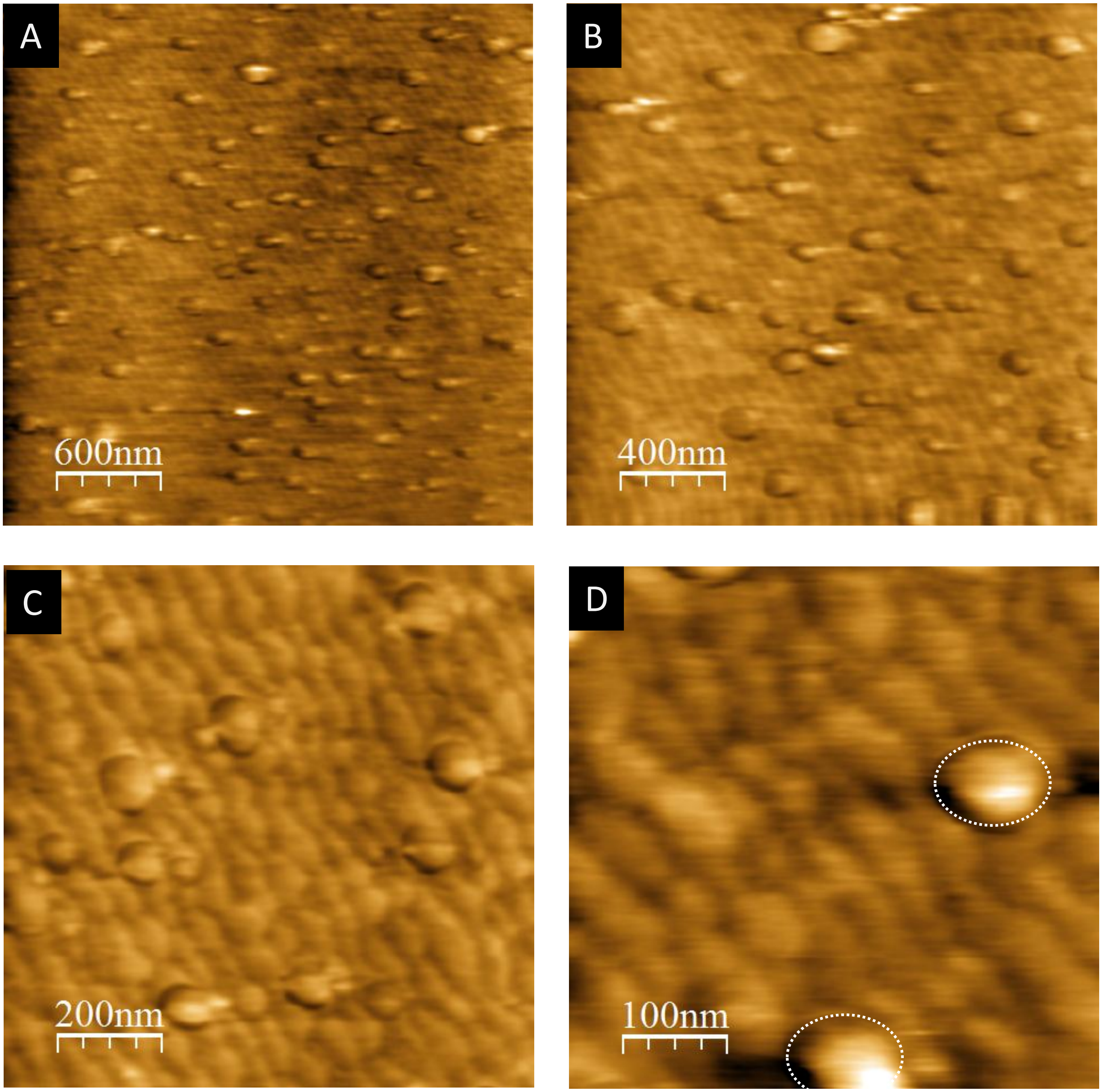

**Figure 4.2.1** High resolution NC-AFM image set of the $Co_2O_3$ nanostructures grown onto the clean YSZ(111) single-crystal surface. A) Large-scale 3.0 µm x 3.0 µm image of post-growth $Co_2O_3$-YSZ(111) system B) 2.0 µm x 2.0 µm image of the YSZ(111) surface a few hours post-growth. C) 1.0 µm x 1.0 µm image showing pancake-like growth of $Co_2O_3$ nanoclusters across 2-4 terrace widths. D) 500 nm x 500 nm image with the spherical $Co_2O_3$ nanoclusters, as circumscribed by the dashed traces. Scan speed=0.2-0.5 line/s

Immediately after deposition, the YSZ(111)-$Co_2O_3$NP system was studied via NC-AFM (Figure 4.2.1). The nanostructures show lateral growth, often straddling up to 3 or 4 terraces (recalling $W_{terrace}$ ~42 nm) and with a cluster or grain size distribution that was very narrow, with average particle size and height of 110.9 ± 37.7 nm and 3.24 ± 1.90 nm, respectively. Confirmation of the crystallite size was attempted by using the Scherrer equation[143] correlated to x-ray diffractometry (XRD) data, but it is inconsistent for non-nano-crystalline sizes (i.e. >0.1 µm) and there were issues with $Co_2O_3$ peak resolution for all systems due to extremely low cobalt loading (see experimental chapter). However, the expected $Co_2O_3$ phase was observed with a few contributions from CoO/$Co_3O_4$, but signal strength was diminished due to dilute sample loadings. The pancake-like morphology of the nanostructures appears to be mostly circular with no evidence of sharp faceting (Figure 4.2.1d), which is in agreement with initial reports of spherical $Co_2O_3$NPs on NCD.[16,76] Because the particles appear to be so flat, the high density of terraces that comprise the YSZ(111) support were also resolvable by AFM. While the parallel steps and terraces of the clean YSZ(111) single crystal were both linear and pristine, they appear to have a number of defects after the photochemical growth process. The effects of the $CoCl_2$ photoreduction and the local environment likely played a role in inducing defects within the parallel step edges and atop the terraces of the support, in turn affecting the $Co_2O_3$NP morphology, density, and mobility during nucleation. It is speculated that the YSZ(111) was pristine enough after thermal treatments that mobility of the $Co^0$ seeds on the inert surface was not limited by terrace or step features and the minimal $Co_2O_3$NP growth is simply proportional to a lack of defect/nucleation points. NC-AFM image and *ex situ* XPS analyses revealed that the $Co_2O_3$NPs had a density of ~8.75 NPs/µm$^2$ and 1.65 atom % Co (calculated based on major Co 2p peak fits at RSF=19.2) on the surface. The clean YSZ(111) surface posted the lowest defect concentration and corrugation of all 4 supports in this study, with the former being the principal reason for relatively low $Co_2O_3$NP surface coverage. Since nucleation of the nanostructures is intrinsically a function of nucleation points (defect structures), their ability to aggregate into clusters in solution would depend on a kinetic model of defect structure formation during the photochemical growth. Although the nanostructures self-organize quickly on a pristine surface, they also face a lowered mobility as the surface generates kink sites and other defect structures while in solution; this is one working explanation for the lateral growth mode of the nanostructures across multiple terraces. Further, the relatively small particle heights could be related to this lateral growth process since the terrace widths are small enough and nominally flat (as mentioned, this support yielded lowest RMS roughness values) enough to promote growth across terrace features. Though the nanostructures are not entirely commensurate with respect to the underlying surface, the support surface proves to be an excellent template for their photochemical growth and self-organization. Figure 4.2.2 depicts the high-resolution XP-spectrum of the Co 2p region and confirmation of the $Co_2O_3$ species on the YSZ(111) support. Metallic Co has a binding energy of 777.9 eV[6] for Co 2p with the higher oxides like $Co_2O_3$ shifting this to ~779.4 eV[14],~779.6 eV in other reports,[144] in good agreement with the value obtained herein (779.7 eV). Spin-orbit coupling of the 2p region between $2p_{3/2}$ and $2p_{1/2}$ (found further downfield but not shown) was resolved by XPS, as well as the shake-up satellites appearing as shoulders at the higher binding energies for the 2p peaks. Though they have lower intensity, the satellite peaks are shown in the GL peak-fitting curves in Figure 4.2.2 to confirm that the satellite peaks often contribute to unphysical peak broadening to oxidized cobalt samples.[144,145] Other oxidized cobalt species like CoO and $Co_3O_4$ appear in the 780.2 eV[14] and 780.7 eV[6] regions, respectively, so confirmation of which species was actually obtained after nucleation is ultimately contingent upon the resolution of the XPS instrument used. In any case, the photochemical growth of $Co_2O_3$ nanostructures, as confirmed by XPS, occurred on the YSZ(111) single-crystal support rather sparsely and showed a lateral growth mode across multiple step features, as confirmed by NC-AFM.

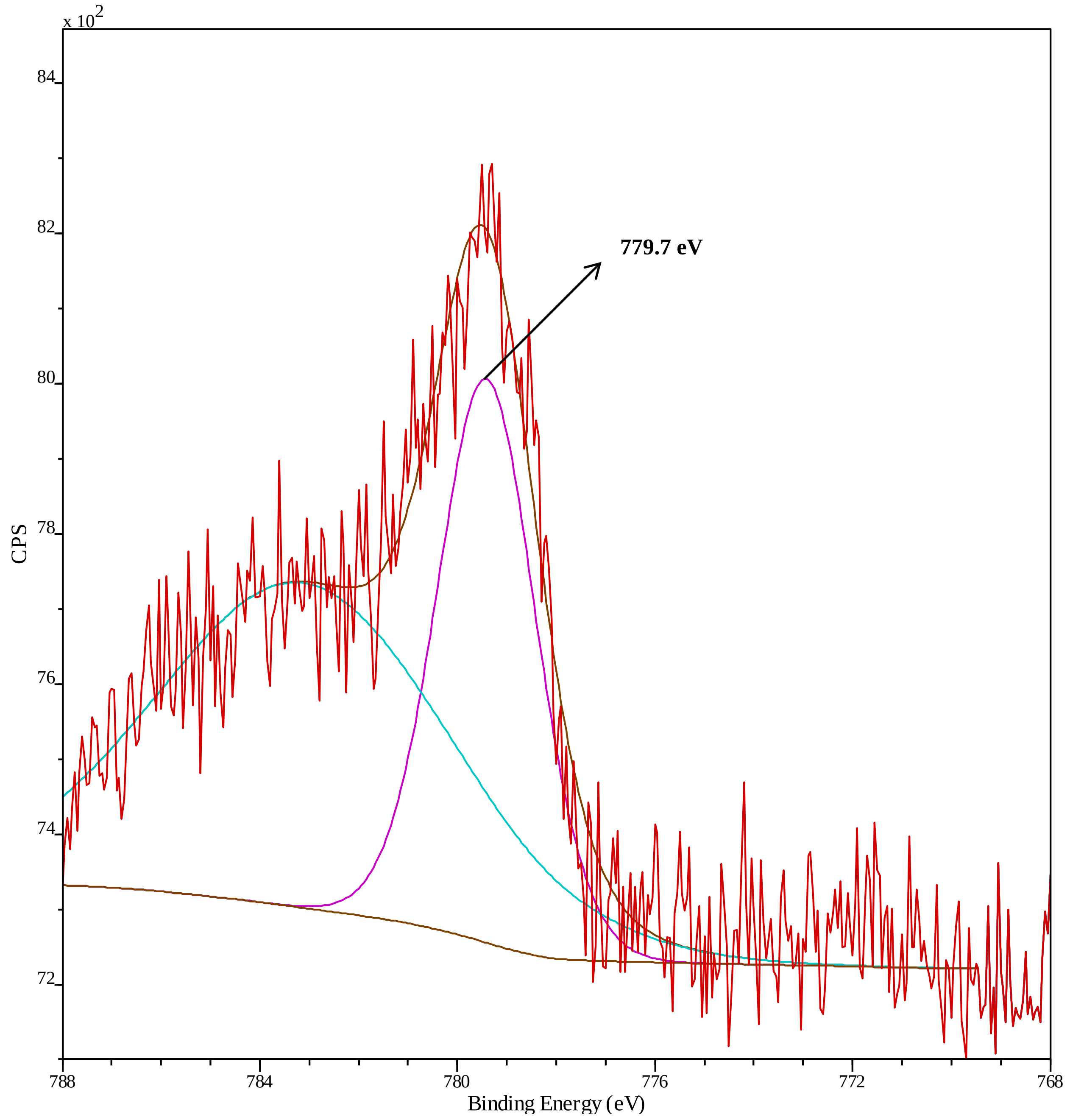

**Figure 4.2.2** XP-spectrum corresponding to the high-resolution scan of the Co region of the YSZ(111)-$Co_2O_3$ surface. The Co $2p_{3/2}$ envelope was found at 779.7 eV, corresponding to both (major, purple peak) $Co_2O_3$ and $CoO_x$ species (minor, light blue peak). [5,6] The Shirley background is fit clearly below the whole Co 2p envelope (red line). Prior to fitting procedure, the whole spectrum was charge-corrected with respect to the $Zr^{4+}$ $3d_{5/2}$ peak envelope which occurs at 182.6 eV. Obtained on the KRATOS Axis Ultra DLD spectrometer (CCRI).

## 4.3.1 NC-AFM/XPS ANALYSIS OF THE POST-GROWTH YSZ(100) SYSTEM

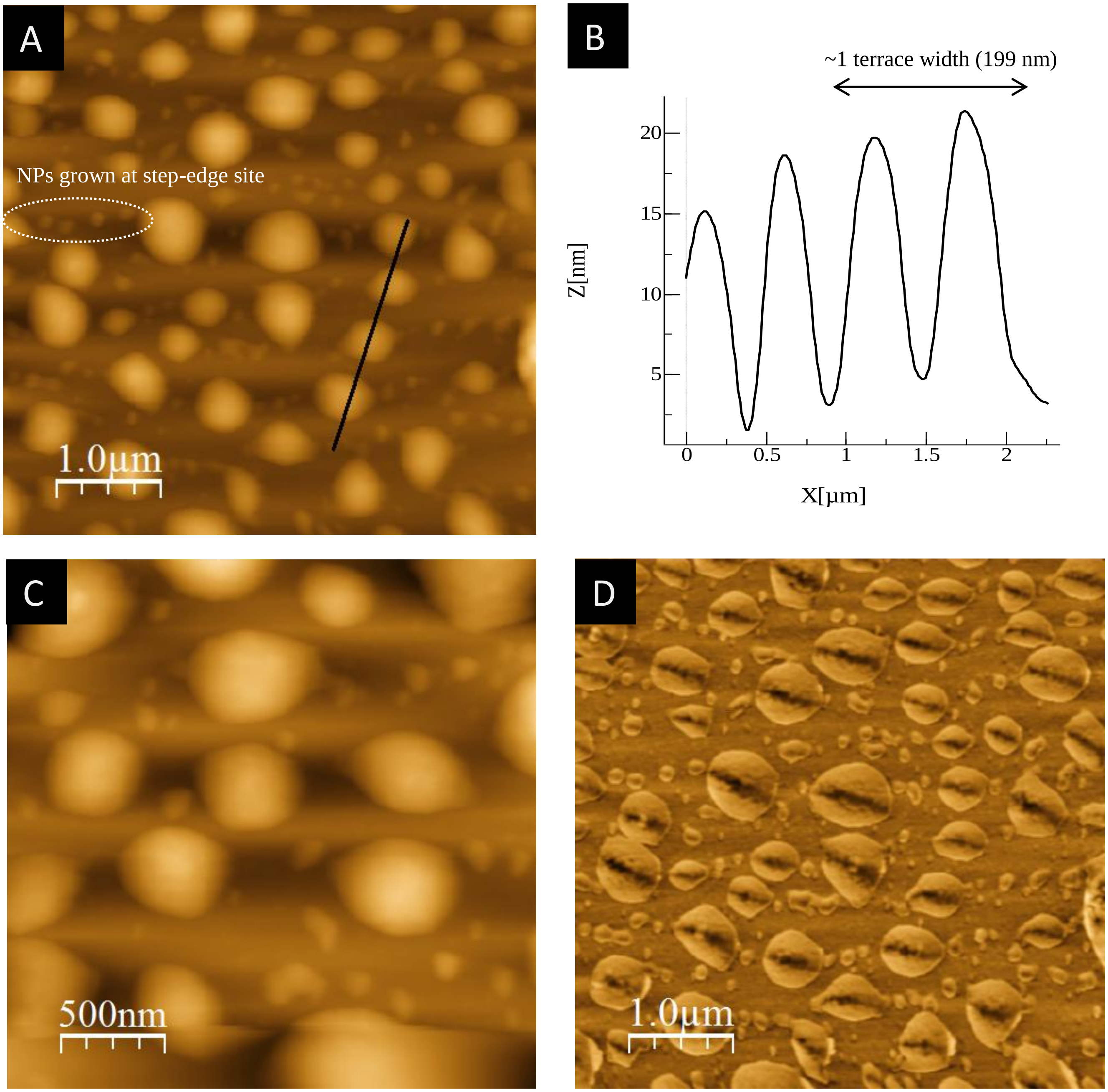

**Figure 4.3.1** High resolution NC-AFM image set of the $Co_2O_3$ nanostructures grown onto the clean YSZ(100) single-crystal surface. A) Large-scale 5.0 μm x 5.0 μm image of post-growth $Co_2O_3$-YSZ(100) system B) Height profile of the $Co_2O_3$NPs corresponding to the line profile traced in A) C) 2.5 μm x 2.5 μm image taken at different sample area to confirm consistent growth behavior of nanostructures on the support D) 5.0 μm x 5.0 μm phase image corresponding to A) that was used for grain analysis. Image shows high contrast at particle edges of the spherical $Co_2O_3$NPs as well as nanoclusters forming atop terrace features. (avg. scan speed=0.3 line /s)

In contrast to the YSZ(111) support, substrate-directed growth of the $Co_2O_3$ nanostructures on YSZ(100) support occurred predominantly in the terrace step sites and nucleation was much more abundant. Again, grain analysis shows that particle morphology is in agreement with initial experiments on NCD, but instead of the favourable nucleation at carboxyl residue sites,[16,146] it occurs at a number of defect sites found on the clean YSZ(100) Type 1 surface discussed previously. Since the resolution of the nanoclusters was not ideal from topographic AFM images, the phase images (Figure 4.3.1d) were often used for the purposes of grain analysis due to the enhanced contrast of the particle edges with respect to the support. The truncated spherical nanostructures and (larger) 3D-nanoclusters (NCs) had a lateral size of 161.7 ± 21.1 nm and 500.9 ± 54.9 nm, respectively, with an average height of 24.5 ± 7.9 nm.. As mentioned, nanostructures showed a largely spherical morphology, this time with sharper edges whose size is undoubtedly a function of the geometry of the defect site. Alignment of the $Co_2O_3$NPs occurred along what is assumed to be the step edge sites separating the parallel terraces (Figure 4.3.1a), but due to the sensitive NC-AFM image mode, it was difficult to resolve the support features completely without compromising resolution of the surface species. It is clear from the phase image that some elongation of the nanoclusters occurs, suggesting growth within the step edges and a possible mechanism for diffusion of the NPs as they formed nanoclusters. Interestingly, the nanoclusters are often found to be separated by the average terrace width (see line profile in Figure 4.3.1b) that was calculated previously, which would further serve to confirm that the array of nanoclusters formed within the defects (kink sites, etc) of the step structures. Particle density (note: 3D-NCs that were not resolvable into the component NPs were factored in as ~3NPs per NC) was found to be 26 NPs/$\mu m^2$ with a Co atom % of 6.3% calculated from XPS fitting (Figure 4.3.2). A higher NP density is noticeable around the larger clusters, perhaps meaning that formation of thermodynamically stable $Co_2O_3$ nanoclusters is occurring. Though not remarkably higher than that of the YSZ(111) support, the particle density and quantification of $Co_2O_3$ is in line with the expectations for this particular support given its intermediate defect concentration in relation to Type I YSZ(100) surfaces. The fit Co $2p_{3/2}$ band found at 780.3 eV is in agreement with expected values for a $Co_3O_4$ shift[6] but the GL-fit is actually not centered with respect to the target $Co_2O_3$ peak found at 779.8 eV. Many reasons could be inferred from this, but ultimately a component scan for that region would be required, especially to improve the poor signal-to-noise ratio by a factor of $1/\sqrt{N}$, to elaborate on the physical meaning. Contributions from other oxides with similar FWHM values could play a role, especially, too, due to the smaller energy gaps between $CoO_x$ oxidation states (< 3 eV). Previous studies with YSZ(100) using both spectrometers reveal that the line width for all peaks excluding carbon is smaller for the Kratos system, indicating relatively lower artificial, unphysical peak broadening by that spectrometer and therefore higher resolution (~0.08 eV on Specs and ~0.05 eV on Kratos).[12] Again, it is clear in the XP-spectrum that the fit (GL~50%) is shifted slightly downfield (left) of the maximum peak intensity for the $Co_2O_3$ 2p signal (appearing closer to 779.7 eV), but this fit was likely compensating for any residual broadening due to the Co $2p_{3/2}$ shake-up peak. Finally, the lack of shake-up peak intensity for the Co $2p_{3/2}$ is compatible with the lack of $Co_3O_4$ coverage, which effectively confirms that the major surface species of cobalt oxide for this system is $Co_2O_3$ as expected. Splitting of the Co $2p_{1/2}$ peak envelope confirms minor higher oxide contributions, but higher-resolution scans with longer dwell times may be necessary.

Indeed the novel mechanism for photochemical growth of $Co_2O_3$NPs (Figure 4.1.1) on nanostructured vicinal oxide surfaces like YSZ(100) is reported for the first time and our observations show a Volmer-Weber growth mode of the nanostructures for this particular system by observation of the formation and self-organization of $Co_2O_3$ NPs into 3D-nanoclusters within the highly defect-populated step-edge sites on the surface. This also suggests both aggregation of nucleation sites or CoNPs themselves due to magnetization and a high degree of mobility during the growth of the CoNP seeds before they are immobilized (i.e. form the oxide). Ideally, further control over the reproducibility of the clean YSZ(100) surface will lead to further control over the morphology of the $Co_2O_3$NPs and their orientation during photochemical growth.

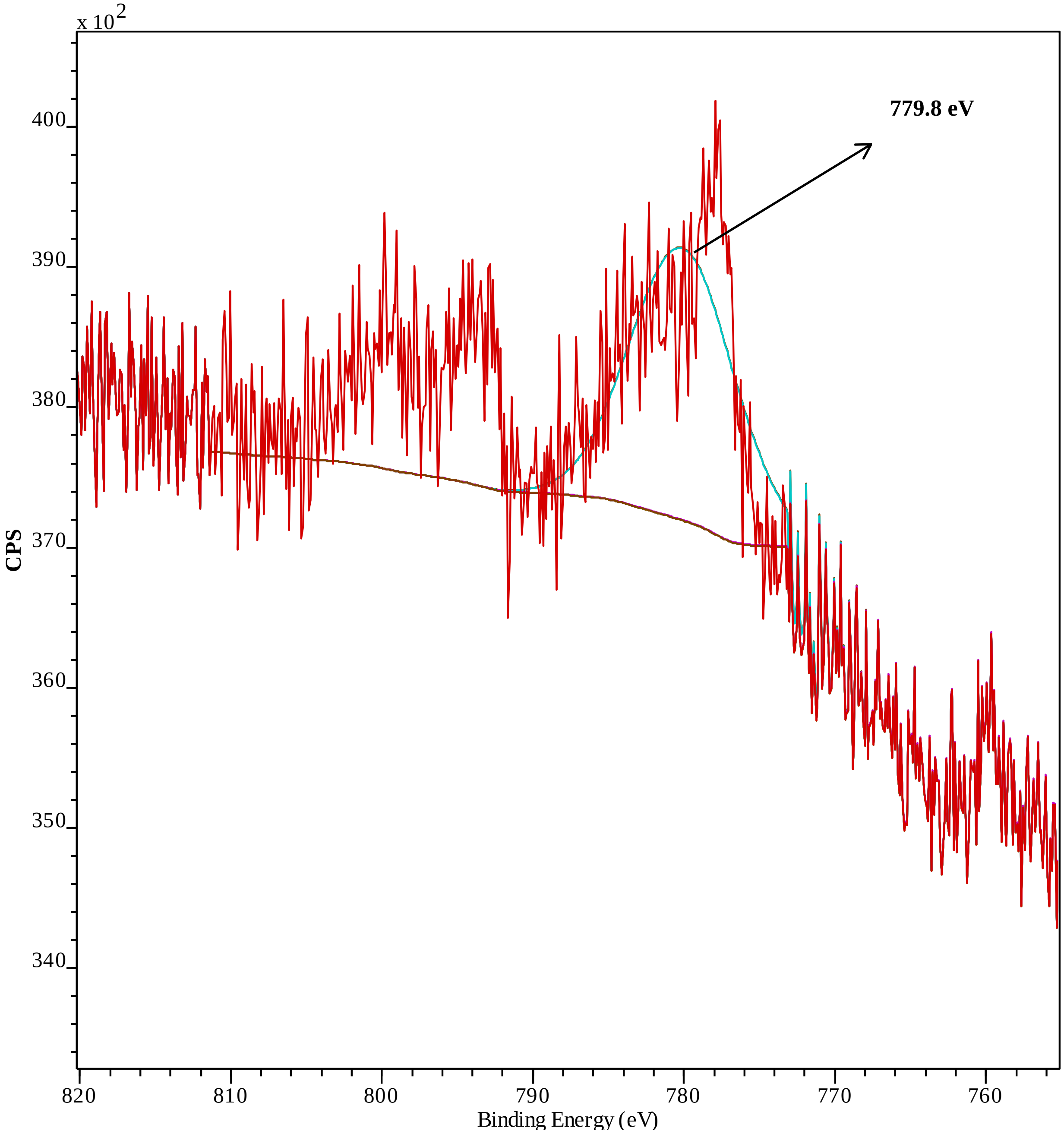

**Figure 4.3.2** XP-spectrum corresponding to the the Co 2p region of the YSZ(100)-$Co_2O_3$ surface. The Co $2p_{3/2}$ envelope was found at 779.8 eV, corresponding to $Co_2O_3$ as the principal cobalt oxide species. The Shirley background is fit clearly below the whole Co 2p envelope. Spectrum was charge-corrected with respect to the $Zr^{4+}$ $3d_{5/2}$ peak envelope which occurs at 182.6 eV. Spectrum obtained the Specs/RHK system.*Note: The Co 2p region had the highest resolution in the survey scan so this is not the fit for the component region.

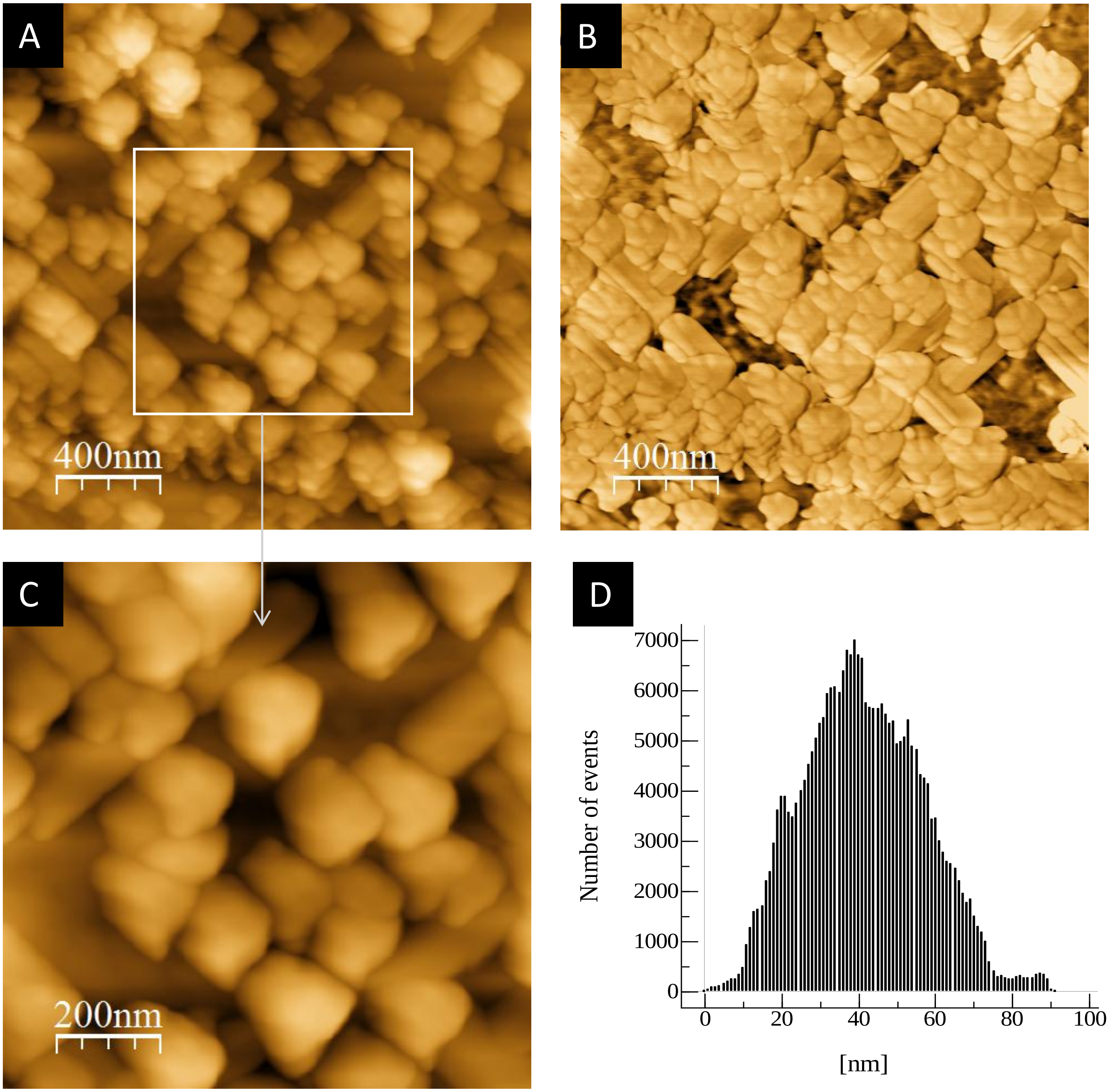

**Figure 4.4.1** High resolution NC-AFM image set of the $Co_2O_3$ nanostructures grown onto the clean MgO(100) single-crystal surface. A) 2.0 μm x 2.0 μm topographic image of post-growth $Co_2O_3$-MgO(100) system B) 2.0 μm x 2.0 μm phase image corresponding to A) showing high local contrast of the nanoclusters with the underlying surface substrate. C) 1.0 μm x 1.0 μm topographic image of the selected region in A) (scan speed=0.2-0.4 line/s, tip: SSS-NCL, RF=167 kHz).D) Height distribution histogram for the $Co_2O_3$ nanostructures.

Achieving high resolution in the NC-AFM imaging for the $Co_2O_3$-MgO(100) system was markedly challenging, and even more so than imaging of the clean MgO(100) support itself. Initial cycles of AFM imaging with the PPP-NCL tips often lead to immediate tip-blunting that evident by the linearly decreasing resolution of the image as a function of scan time, so super-sharp SSS-NCL probes had to be used for the high-resolution images reported (Figure 4.4.1). $Co_2O_3$NPs were even more abundant on the clean MgO(100) support than observed for the previous two metal-oxide supports with almost completely homogenous clustering. Phenomena like the supposed multiple-tip effect along with other potential tip-induced effects common with difficult samples was ruled out by scanning the same sample at a $90^o$ rotation and ensuring all the surface features rotated accordingly and were therefore not the result of AFM image artifacts. Grain analysis was, again, performed mostly using the phase images (Figure 4.4.1b), but unlike the other samples, the SS AFM probe allowed for resolution of the individual NPs within the nanoclusters (see topographic images in Figures 4.6a,c). A bimodal lateral size distribution was calculated with nanostructure and nanocluster sizes being 62.7 ± 13.7 nm and 229.6 ± 84.1 nm, respectively. The phase images offered a lot of interesting details including a very notable local contrast between the nanoclusters and the underlying support surface. As phase image data is often interpretable in terms of energy dissipation[147] and the phase shift itself is a function of z-position, the switchover where the tip “jumps the s-curve” from the net-attractive to the deflective force regime is likely the point where the tip is oscillating over the MgO(100) support and then contacting the $Co_2O_3$NPs. This, of course, occurs thousands of times per image and the result is sometimes excellent local contrast in the phase channel, which often indicates that an over-layer chemically distinct from its supporting under-layer may exist. The individual $Co_2O_3$NPs in the popcorn-like 3D-NCs appear to have sharper square faceting than found in the truncated spherical 3D-NCs YSZ(100)-$Co_2O_3$NP system with mean particle heights of 41.2 ± 20.0 nm (Figure 4.4.1d). Some NCs actually appear to be supported on rod-like features, indicating that the magnetic properties of the $Co_2O_3$NPs (originally as the $Co^0$ seeds) combined with their spatial localization may have contributed to this interesting geometry. Our observations, again, support the Volmer-Weber (island) type growth mode of the $Co_2O_3$NPs, with some of their growth supported by the $Co_2O_3$ "nanorod" features. The localized effect of some NCs growing out of the nanorods is actually indicative of a Stranski-Krastonov growth mechanism, but ultimately one would need information about the wetting behaviour of the NPs/NCs atop the nanorods to confirm that they are not just physically anchored or grown out of defects on other NCs.

Having the highest RMS roughness (for as-received samples) reported, the MgO(100)-$Co_2O_3$NP system also had the highest particle density of 99 NPs/$\mu m^2$ with Co atom % coverage of 14.5% determined from XPS quantification. The clean and pristine MgO(100), as shown, is both highly defective and reactive, so the high $Co_2O_3$NP density and degree of clustering supports the fact that the photochemical growth of the particles is a definitely function of high RMS roughness and/or defect concentration of the single-crystal support. Because the MgO(100) is known to be highly reactive to water, fitting of the XP-spectra was done for the the Co 2p, Mg 2p (Figure 4.4.2), and O 1s component regions (Figure 4.4.3). The Co 2p region contained the $2p_{3/2}$ band fitted at 779.8 eV (Figure 4.4.2a) along with the expected unphysical peak broadening due to shake-up satellites. The energy value obtained is in excellent agreement (0.2 eV deviation) with the expected value for $Co_2O_3$[144] and confirms that the nanostructures are predominantly this oxide species. Intensive main peaks for hydroxylated cobalt species, if they existed, would occur at higher binding energies (i.e. 780.7 eV for $Co(OH)_2$),[144] so the metastable $Co_2O_3$ oxide species was maintained and presumably the only major surface species. As previously noted, MgO(100) has an unfailing reactivity towards ambient air, meaning the support both during the liquid-phase reaction and during ambient NC-AFM imaging was fully hydroxyl-terminated, likely with carbonate residues, too.[5,148–150] Confirmation of this is found in the fitted XP-spectrum for the Mg 2p region (Figure 4.4.2b), which shows the MgO peak with contributions from the $Mg(OH)_2$ and $MgCO_3$ fittings at 49.8 eV and 50.9 eV, respectively. Since the adventitious carbon peak was found at 286.3 eV due to $CO_3^{2-}$ and OH contributions (it is normally found at 284.5 eV[151]) the whole spectrum was calibrated with respect to the metallic Mg 2s peak found at 89 eV (88.7 eV in this case, likely due to small surface charging). The measured binding energies for the $Mg(OH)_2$ and $MgCO_3$ are actually in close agreement with other experiments involving oxidation of Mg at 20 kPa $O_2$ at RT,[152]where marginal differences are attributed to local environmental factors like relative humidity, presence of the $Co_2O_3$ species, and resolution of the spectrometer. High-resolution scans of the O 1s region reveal splitting of the MgO peak into a doublet, which has been ascribed to the presence of $Mg(OH)_2$ species at 533.3 eV and $MgCO_3$ species at 534.5 eV. The broadening of the O 1s peak is expected in this case and other findings indicate that minor species like CoO(OH) and $Co(OH)_2$ may also play a role.[144] Overall, the XPS data confirms the reactivity of the MgO(100) single-crystal surface to ambient air and the chemical shift values, particularly that of the Mg 2p and O 1s peak patterns, confirm that our values are in agreement with other findings that use this support.[2,152] As such, exhaustively complex fitting procedures based on precise stoichiometry of each surface species for the individual component regions was not necessary in yielding a quantification that would be any more accurate than a GL-fit of the survey regions with the relevant RSF values.

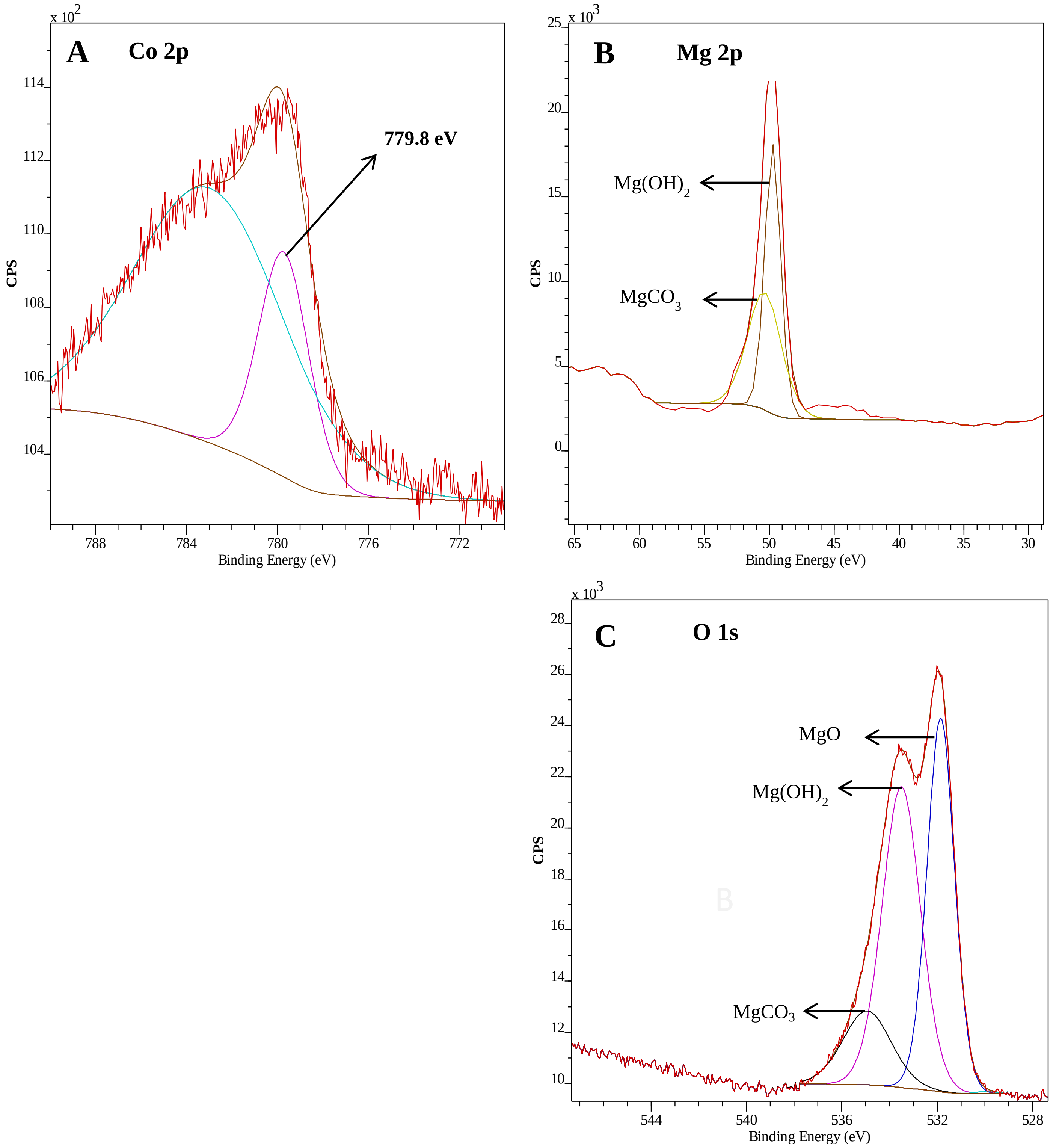

**Figure 4.4.2** XP-spectra recorded following photochemical growth of $Co_2O_3$ on the clean MgO(100) single-crystal surface. A) Co 2p envelope with peak behavior characteristic of the $Co_2O_3$ species. B) Fit Mg 2p band with broadening due to $MgCO_3$ and $Mg(OH)_2$ C) O 1s envelope showing broadening and splitting due to $MgCO_3$ and $Mg(OH)_2$ species. Spectra were collected on the Kratos Axis Ultra DLD system (CCRI), fit with Shirley backgrounds, and calibrated with respect to the Mg 2s peak (nominal position of 89 eV).

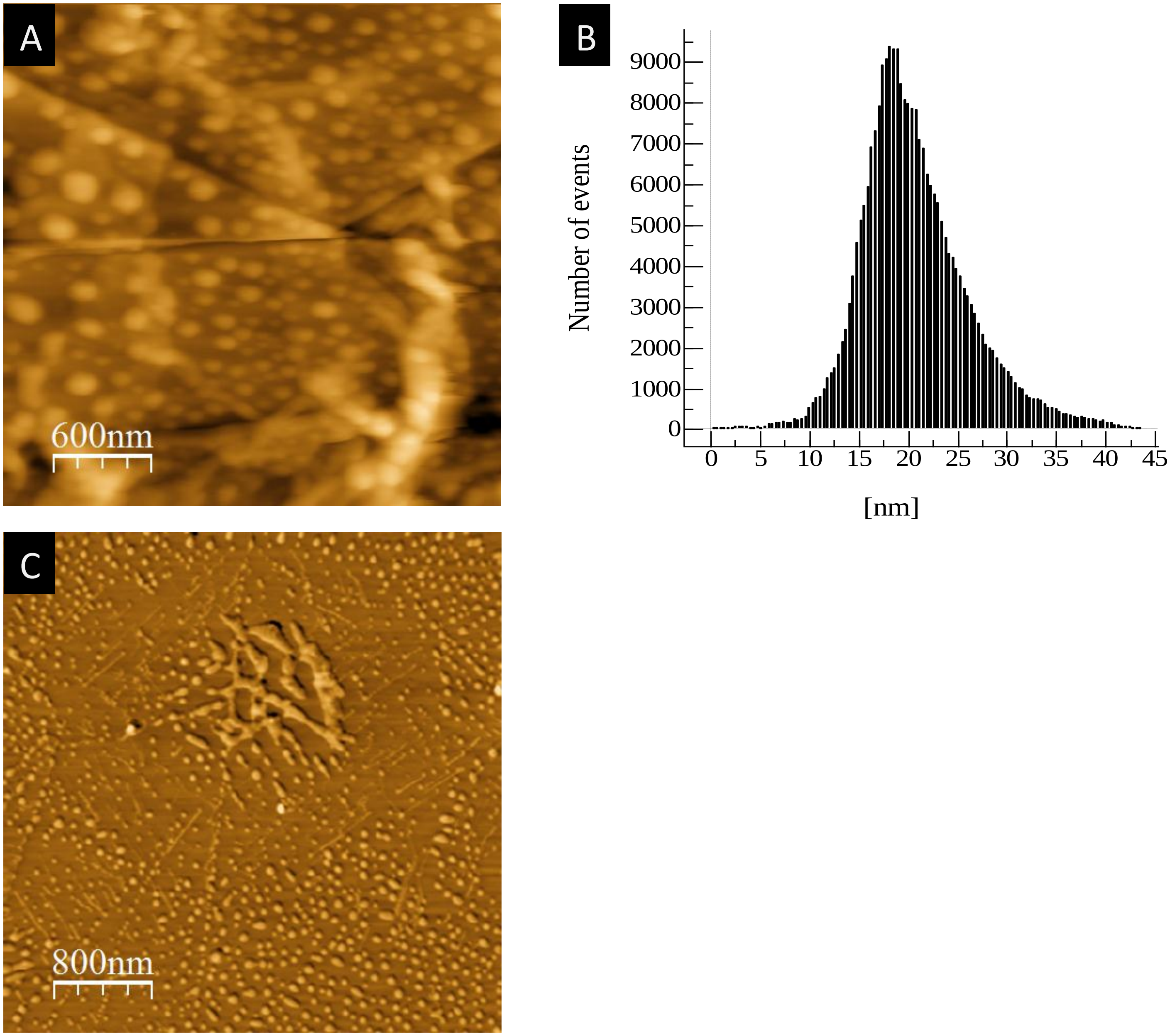

**Figure 4.5.1** NC-AFM imaging of the $Co_2O_3$-HOPG system. A) 3.0 µm x 3.0 µm topographic image of spherical $Co_2O_3$ NPs grown on the pristine HOPG surface. Clusters actually form both along cleavage steps and atop the pristine islands of freshly cleaved HOPG B) Particle height histogram for the $Co_2O_3$ nanostructures. C) 4.0 µm x 4.0 µm phase image of the nanostructures and clusters forming preferentially at cleavage steps. (scan speed=0.3-0.5, DLC tip, RF=~150 kHz).

Figure 4.5.1 shows the NC-AFM imaging of the HOPG after nucleation of the $Co_2O_3$NPs, which seems to occur randomly throughout the surface with a homogenous size distribution of 108.7 ± 38.7 nm and corresponding mean heights (Figure 4.5.1b) of ~21.0 nm. Similar to the other supports, there is evidence that assemblies of HOPG-supported $Co_2O_3$NPs represent a nanostructure with controlled arrangement. As highlighted by the large-scale phase image in panel C of Figure 4.5.1c, the growth of the nanostructures may be occur right in the cleavage step sites. The clean HOPG surfaces have well-defined defect structures that come mostly in the form of cleavage step sites (resolved by NC-AFM, Figure 4.5.1a) that get produced after mechanical exfoliation of the basal plane. This is essentially a buckling of the adjacent graphene layers (recall that HOPG is a intercalated or "tiered" structure) exposing many layers of graphene at once.[153] It is interesting to note that the NC-AFM topographic images of the HOPG under-layer show cleavage steps between 4-20 graphite planes (~14-70 Å). Flooding analysis (also performed in WSxM 5.0 software[74]) of the NC-AFM images showed that only 31.2% of the topmost layer is composed of $Co_2O_3$ nanostructures, which speaks to how stepped the clean HOPG surface actually was prior to growth experiments. Particle density was on the order of ~63 NPs/$\mu m^2$, but unfortunately the XP-spectrum (Figure 4.5.2) for this particular system was too noisy to yield an accurate quantification of Co atom %. That being said, the high-resolution scan of the Co 2p region (Figure 4.5.2) did yield the fit Co $2p_{3/2}$ band associated with $Co_2O_3$ and within 0.2 eV of the expected value.[144] The lower spectral sensitivity for the $Co_2O_3$ could simply be due to the inert nature of the HOPG crystal surface; the highly pristine surface lacks ionic nucleation points for the particles to grow out, possibly making them shallow enough to be on the order of the X-ray spectrometer skin depth (<7 nm). Angle-resolved XPS would possibly serve to improve the intensity of this region, otherwise, and more improbably, the cobalt was inserting itself into the bulk of the crystal. Peak-to-peak spin-orbit splitting distance of the Co $2p_{3/2}$-$2p_{1/2}$ was not measurable from the survey scan or high-resolution scan due to the low S/N ratio. The O 1s region contains spectral information for literally thousands of compounds, meaning XPS analysis in an absolute sense is very complex. In the case of the O 1s broadband (Figure 4.5.2), the peak envelope is fit for at least two bands that, if not unphysical, correspond to the stable and major $Co_2O_3$ oxide species at 529.8 eV with broadening contributions from cluster or support-bound OH groups at 530.9 eV. This finding bears affinity to another study that shows a 1.1 eV peak separation between the XPS fit bands for $Co_3O_4$ and CoOOH at 530.0 eV and 531.1 eV, respectively.[145] Very small discrepancies in this region would be a function of instrument error or due to unknown surface compositions that include the photoemission of higher cobalt oxides (that is, > 3+).[154] Though the spectrum was not calibrated, the C 1s peak (not shown) was found to be shifted 2 eV from the peak position at 284.4 eV diagnostic for the $sp^2$-carbon framework of pristine, air exfoliated HOPG.[6] Instrument error and other discrepancies concerning unknown sample composition could also shift this peak; for instance the photoemission of higher oxides (> 3+). Furthermore, atypically high baseline noise was observed and effectively buried the Co 2p signal quite substantially, so this may also serve to confirm, at least qualitatively, that local charging effects were very much present, making the major metal-oxide peak nearly indistinguishable. For the purposes of this work, it is sufficient to report and confirm the presence of the target cobalt oxide species; however, a proper correction for the intrinsic signal of the C 1s region for this sample is possible if the line-broadening functions of the spectrometer, properly correlated to instrumental parameters, are estimated.[155] As noted, the possibility of contamination (OH, hydrocarbon, etc) during the period between growth and XPS analysis may have also caused the diminished Co 2p signal and, finally, the weak adsorbate-substrate interactions commonly associated with the VB growth regime[156], particularly where a lack of support-cluster wetting causes the nanostructures to be weakly tethered to the surface, is another strong consideration to be made when interpreting XP-spectra for this particular system.

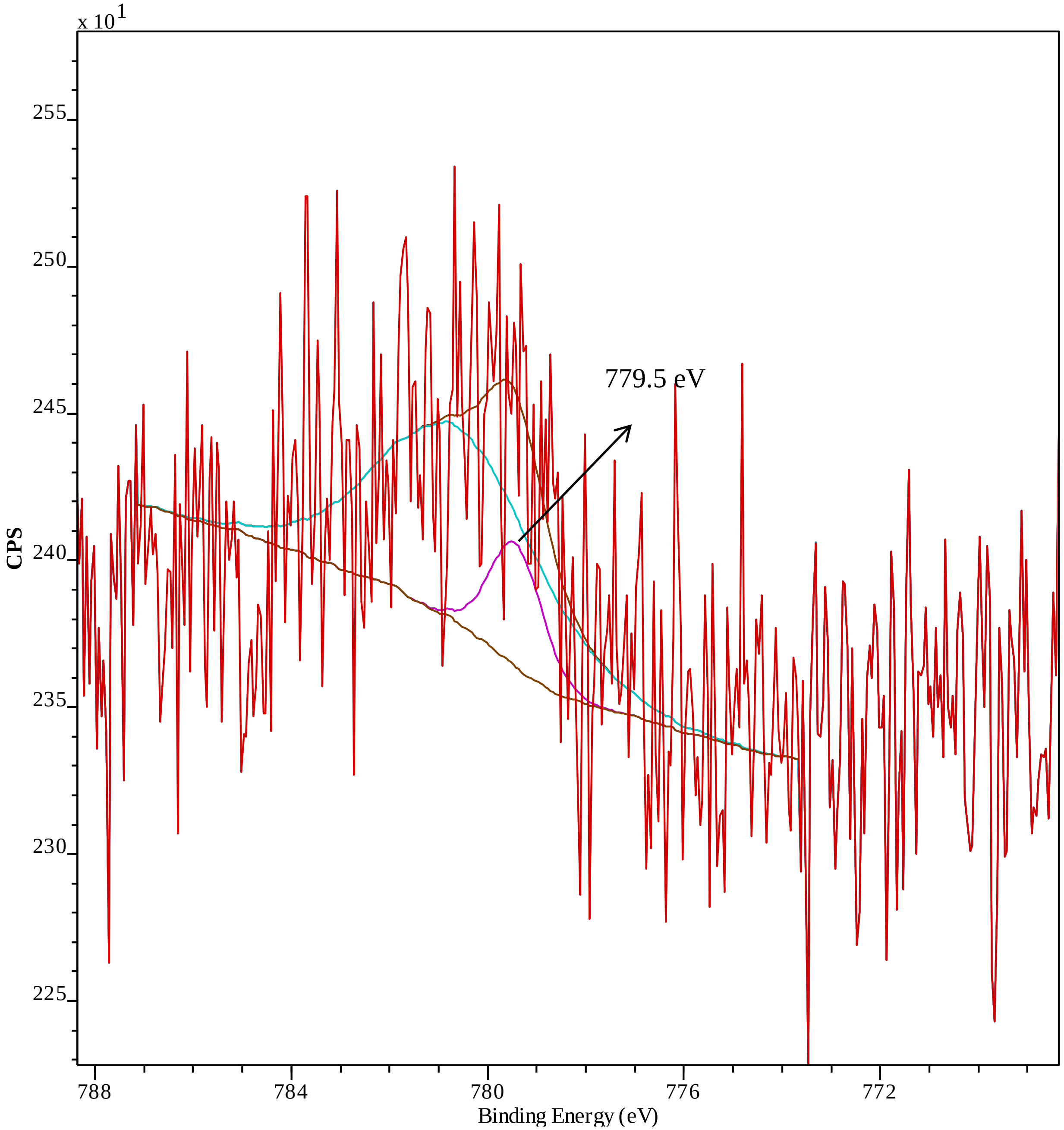

**Figure 4.5.3** XP-spectrum of the Co 2p region of the Co $2p_{3/2}$ band fit for $Co_2O_3$NP species (major) at 779.5 eV. CoO was also a likely minor species causing the broadband. There is likely a higher contribution from the surface contamination of other cobalt oxides on this support, evident from the smaller fit for $Co_2O_3$. Spectrum was acquired on the Kratos Axis DLD Ultra system (CCRI).

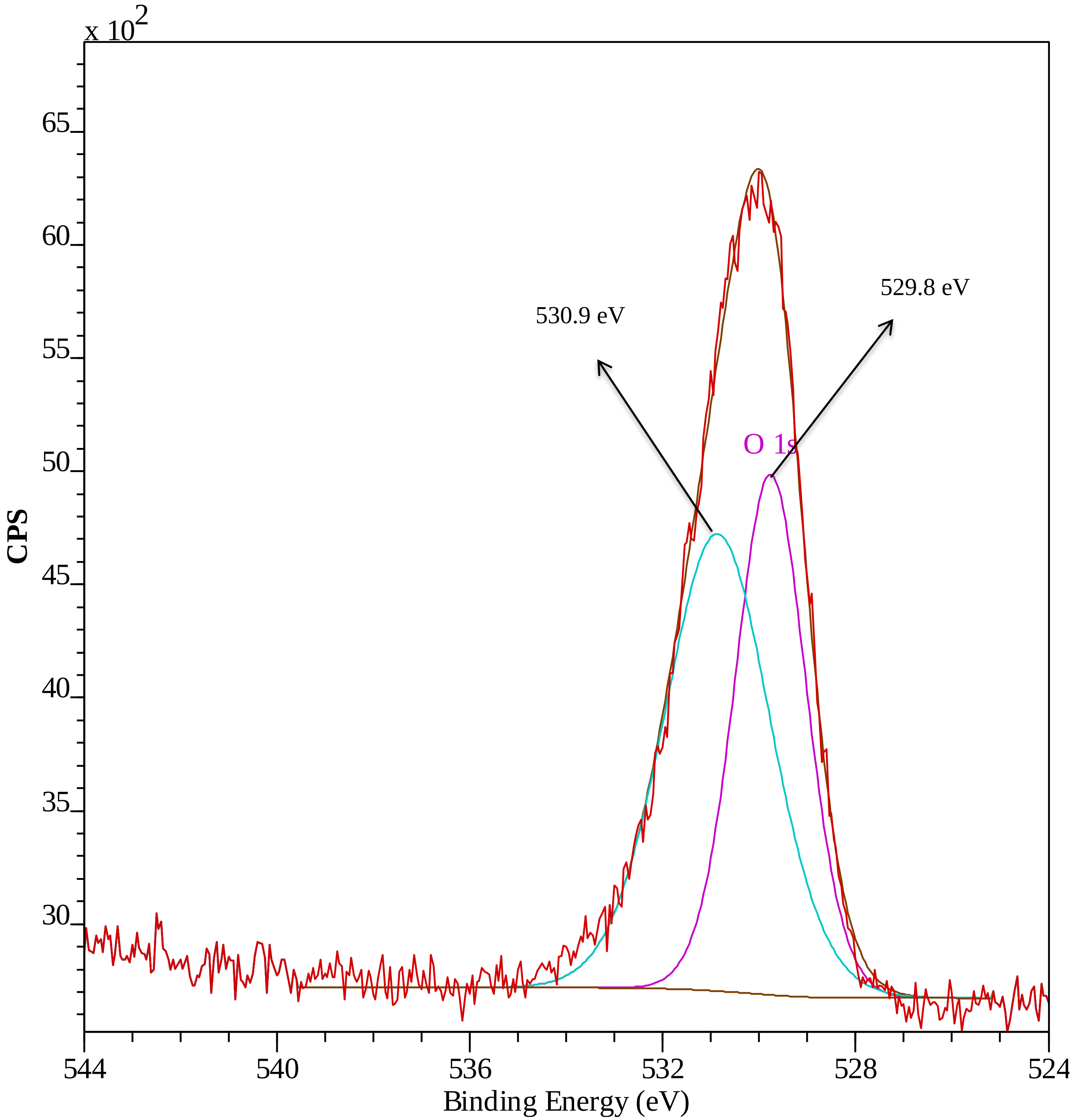

**Figure 4.5.4** Fit component XP-spectrum of O 1s envelope corresponding to the $Co_2O_3$ and other surface species attributed to hydroxyl contaminants ($Co(OH)_2$/CoO(OH)) and the presence of higher oxides of cobalt.[75] Spectrum acquired on Kratos Axis DLD Ultra (CCRI).

# CHAPTER 5 CONCLUDING REMARKS AND FUTURE OUTLOOK

The novel photochemically induced growth of $Co_2O_3$ nanostructures on four carefully selected single-crystal supports were thoroughly and successfully studied by the methods of NC-AFM and XPS. The growth behavior of the nanostructures (summarized below) is in accordance with NC-AFM analysis of the local surface features for the supports, especially the RMS roughness parameter, which indicates that an enhancement of the surface defect concentration leads to higher particle densities, At% Co, and this is particularly true of the insulating supports. Grain sizes are related to the nucleation and aggregation of either CoNPs during growth or aggregation of nucleation sites themselves due to potential surface modifications while in solution. Ease of NC-AFM imaging was also proportional to the roughness parameters of the clean samples and the growth modes on the supports provided a range of interesting and complex nanostructures. If more control with respect to ensuring reactivity of the support does not impinge particle growth of the target oxide, there is a precedent for the development of a model catalyst system.

**Table 5.0** Summary of some of the $Co_2O_3$NP-support data

| Support | RMS, $R_q$(Å) | [a]Type-Polarity | $Co_2O_3$ At %, $X_i$ | Particle density (NP/µm$^2$) | Morphology | Grain (NP) Size(nm) | Growth Mode |
|---|---|---|---|---|---|---|---|
| YSZ(111) | 1.006 | 2-Relatively Nonpolar | 1.65 | 8.75 | Flat/Pancake | 110.9 | Lateral/VW |
| YSZ(100) | 1.180 | 1-Polar | 6.3 | 26 | Truncated Spherical | 161.7 | VW |
| MgO(100) | 1.890 | 3- Nonpolar Ionic | 14.5 | 99 | Popcorn-like Sharp/Faceted | 61.7 | VW and local SK |
| [b]HOPG | 3.462nm | Nonpolar | N/A | 63 | Spherical | 108.7 | Weak VW |

[a]Surfaces are classified with respect to the Tasker's rules for surface types[10]
[b] This surface is not a metal oxide and therefore is not in accordance with Tasker's rules for surface types.

XP-spectra ra showed the $Co_2O_3$ species at the expected BE values of 779.7, 779.8, 779.8, and 779.5 eV for YSZ(111), YSZ(100), MgO(100), and HOPG, respectively. The work presented in chapters 3 and 4 is a detailed insight into the mechanism of growth of the $Co_2O_3$NPs on nanostructured surfaces carefully selected based on a range of fundamental surface properties. Acknowledging Table 5.0, the density, growth mode, morphology, grain size, and composition of the surfaces play a key role in controlling the heteroepitaxial growth of $Co_2O_3$NPs nanostructures on single-crystal surface nanostructures formed during cleaning processes. Future experiments involving these novel $Co_2O_3$NP-support systems should seek to determine if the CoNPs have surface mobility both before and after they undergo oxidation. *In situ* NC-AFM may be employed in conjunction with a temperature ramp program to probe the thermal stability and mobility of the particles along various surface features. While numerous attempts have been made to synthesize the Cobalt (III) Oxide in the literature,[157–161] this work reports a completely novel photochemical synthesis of the $Co_2O_3$NPs as major species on single-crystal surfaces with only minor contributions from other $CoO_x$ species, as confirmed by XPS. Future work perhaps probing catalytic viability of these systems may seek to first determine presence of $Co_2O_3$ species by Temperature Programmed Reduction (TPR) and Attenuated Total Reflectance Infrared Spectroscopy (ATR-IR) data revealing the reduction temperature at 480 K and sharp Co-O IR bands at 580 $cm^{-1}$($\nu_1$) and 665 $cm^{-1}$($\nu_2$) , respectively. These values are well-documented and would allow the researcher to explore $Co_2O_3$ as a catalytic candidate for water-splitting[16] or CO oxidation experiments[162].

# APPENDIX A: XRD ANALYSIS OF $Co_2O_3$-SUPPORTS

The following XR-diffractogram is typical of the $Co_2O_3$-YSZ(100) and similar for other supports where the blue ticks indicate peaks for the $Co_2O_3$, but we attribute the low S/N ratio to the low cobalt loading amounts used in the study (018µg $Co_2O_3$ or 1.05 nmol $Co_2O_3$ in the film with a $CoCl_2$ loading concentration of 0.05 mM) for all samples. In the interest of consistency and reproducibility, the loading concentrations were not altered.

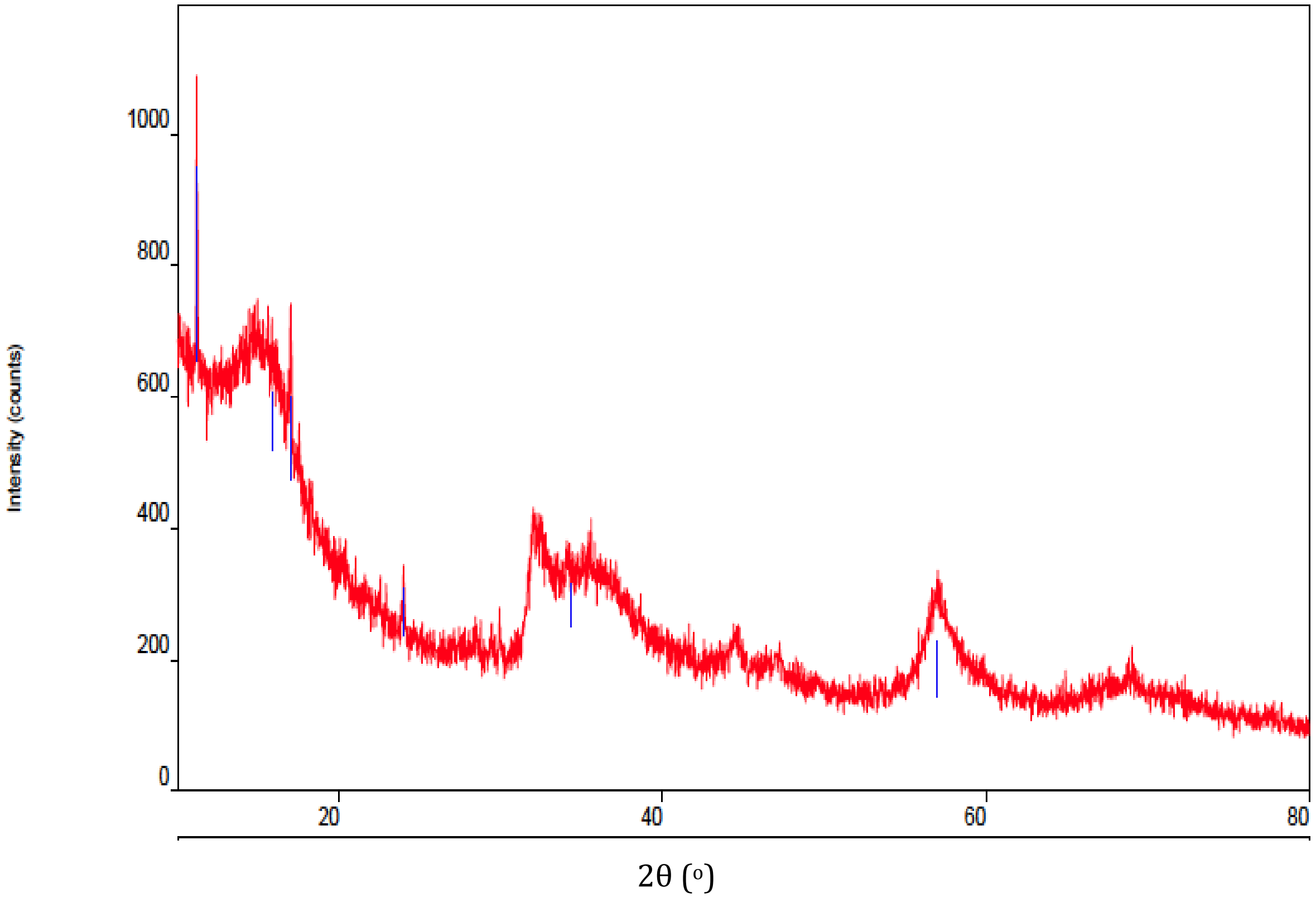

# APPENDIX B: NOVEL WORK

Chapter 3: I performed all thermal treatments and extensive NC-AFM imaging on the single-crystal samples (all purchased from MTI Corp except for HOPG, which came from NanoScience). Additionally, all image processing and analysis was performed by me using a host various soft/freeware programs, but most image filtering was performed in WSxM 5.0 Develop 4.1. XPS was performed on our custom built Specs/RHK system for the YSZ(100) and this work originally was done to develop reproducible cleaning procedures via exposure of $O_2$ to the sample (base pressure of 5 x $10^{-6}$ torr) and $Ar^+$ sputtering followed by STM imaging.

Chapter 4: I collaborated with Post-Doctoral student Tse-Luen (Erika) Wee in the Scaiano group at the University of Ottawa Department Of Chemistry to photochemically grow $Co_2O_3$ on the cleaned single-crystal samples from Chapter 3. I am indebted to Erika for use of her materials and instrumentation to do the growth experiments and later modify them. She performed all work in preparing these samples. XPS analysis was performed by myself on our Specs/RHK system for the $Co_2O_3$-YSZ(100) sample and was attempted on the $Co_2O_3$-YSZ(111) sample, but the rest of the samples were analyzed on the CCRI's Kratos Axis Ultra XPS system by Sander Mommers.